\documentclass[fleqn]{article}
\usepackage{graphicx}
\usepackage[figuresright]{rotating}
\usepackage{axodraw}
\usepackage[T1]{fontenc}
\begin{document} 
{\centering
{\Large {\bf
Overview of Studies on the \\

SPIROC Chip Characterisation}

\vspace{0.3cm}  
R.\ Fabbri$^1$, B.\ Lutz$^1$, W.\ Shen$^2$\\

\vspace{0.3cm}      
{\large \bf
$^1$ DESY, Hamburg \\
$^2$ University of Heidelberg} \\
\vspace{0.5cm}
      \hspace{3.7cm} 
\centering{\today}
}
}
  \tableofcontents
  \vspace{-20cm} \hspace{6cm} 
\newpage  
\section{Introduction}
%
The requirements imposed by the high precision physics forseen 
at the International Linear Collider (ILC) set high demands on 
calorimetry. The ambitious required jet energy resolution of
$\sigma_{\tiny E} /E \approx 0.3/ \sqrt{E}$~\cite{ILC}  
could be achieved by combining the potentials of the particle flow 
approach~\cite{ILC} with extremely segmented (in both transverse and 
longitudinal directions) electro-magnetic (ECAL) and hadronic (HCAL) 
calorimeters.

The ECAL design is based on a Si-W sampling structure with sensitive
silicon pads of $1$ cm$^2$ surface. The prototype will consist of 
approximately $10000$ channels.
A group of the CALICE collaboration has developed the analog option for
the hadronic calorimeter (AHCAL) based on a sampling structure with 
scintillating tiles (of smallest size $3$x$3$x$0.5$ cm$^3$) individually 
readout by Silicon Photo-Multiplier (SiPM) mounted on each tile.

The SiPM is a pixelated avalanche photo-diode operated in limited Geiger
mode. The detector surface typically of few mm$^2$ is divided into 
hundreds to thousands pixels. 
The analog output is obtained by adding the response of all pixels firing 
as independent digital counters. SiPMs are operated at $2$-$3$ volts 
overvoltage.
Given an internal pixel capacitance $C_{pixel}$ of typically $50$ fF, 
the charge collected for one photoelectron signal is approximately $1.2$ pC 
(equivalent to  $\approx 7.5 \cdot 10^5$ electrons). The SiPM offers a 
very fast response with a typical rise time of $3$-$5$ ns. 
The dynamic range is determined by the finite number of pixels and 
reaches $\approx 92$ pC.

In order to test the feasibility of the particle flow approach prototypes 
of both ECAL and HCAL are being built by the CALICE collaboration and
being tested in a combined test beam experiment.
To meet the needs of the analog HCAL prototype it was decided to adapt
the design developed for the ECAL front-end electronics to read the SiPM
signal. This ILC-SiPM ASIC has been used during the test-beam runs
at CERN and FNAL in 2006-07 and 2008-09, respectively.
SPIROC is an evolution of the ILC-SiPM to be used for the next 
generation of ILC AHCAL prototype.
In the following the main features of the new ASIC, version SPIROC 1B,
proposed for the readout are described and its properties analysed. 
More specificately, the results on the analogue part of the chip 
are presented in this work. When reading out signals from 
photon detectors, SiPMs from MEPhy/Pulsar, identical to the 
devices used in the AHCAL test beam operations, were used.  
These SiPMs have a $1$x$1$ mm$^2$ surface divided into $1156$ pixels. 

The on going investigation of the digital component of the ASIC, 
implemented in the version SPIROC 2, will be presented in a separate 
future note.

\section{SPIROC ASIC Description and \mbox{Properties}}
%
The SPIROC chip is a dedicated very front-end electronics for an ILC 
prototype hadronic calorimeter with silicon photomultiplier readout. 
This ASIC has been designed and developed by OMEGA at Orsay~\cite{ORSAY}. 

SPIROC was submitted in Fall 2007 and being tested since then.
It embeds cutting edge features that fulfil ILC final detector requirements. 
It has been realised using AMS $0.35$ $\mu m$ SiGe technology, 
and developed to match the requirements of large dynamic range, 
low noise, low power consumption, high precision and large number of 
readout channels needed.
SPIROC is an auto-triggered, bi-gain, $36$-channel ASIC which
allows to measure on each channel the charge from one to $2000$ 
photoelectrons and the signal timing with a \mbox{$100$ ps}
accuracy TDC. The integrated ASIC components allow $16$ selectable 
pre-amplification gain factors (with output gain values from  
$3$ to $100$ mV/pC), and seven CR(RC)$^2$ shaping times from  
$25$ to $175$ ns. After shaping, the signal is held at its maximum 
amplitude with a track and hold method.
For each channel, an analogue $16$ slots memory array is used to store 
the time information and the charge measurement, and a $12$-bit 
Wilkinson ADC has been embedded to digitise the analogue 
memory content (time and charge for the two gain modes). 
The data are then stored in a $4$ kbytes RAM. A very complex digital 
part has been integrated to manage all theses features and to transfer
the data to the data acquisition system.
An adjustable $8$-bit DAC ($0$-$4.5$ V) in the ASIC allows individual 
adjustment of the SiPM forward voltage for each one of the $36$ channels.

\begin{figure*}[t!]
    \begin{minipage}{5.8cm}
    \hspace{-1.3cm}
      \includegraphics[height=8cm,width=14.cm]{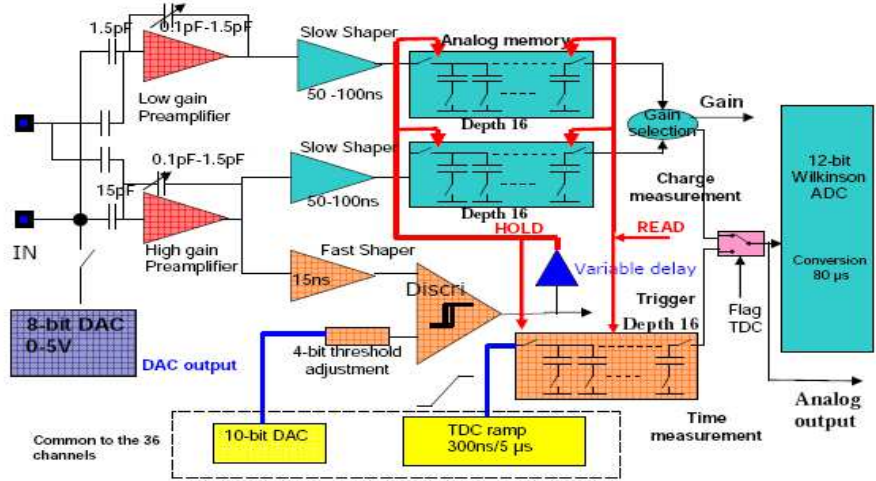}
   \end{minipage}
  
  \caption{SPIROC: one analog channel structure.} 
  \label{fig:SPIROC_schematics}
\end{figure*}

A schematic view of a single channel of the ASIC chip is given in 
Fig.~\ref{fig:SPIROC_schematics}. 
%
%

\section{Test-Bench Description}
\begin{figure}[t!]
 \begin{center}
   \rotatebox{-90}{
   \includegraphics[height=8.5cm,width=5.5cm]{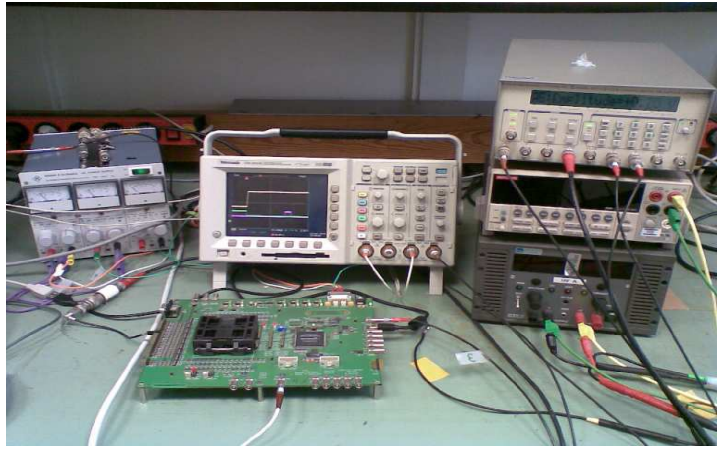}}

   \vspace{0.8cm}
   \includegraphics[height=5.5cm,width=8.5cm]{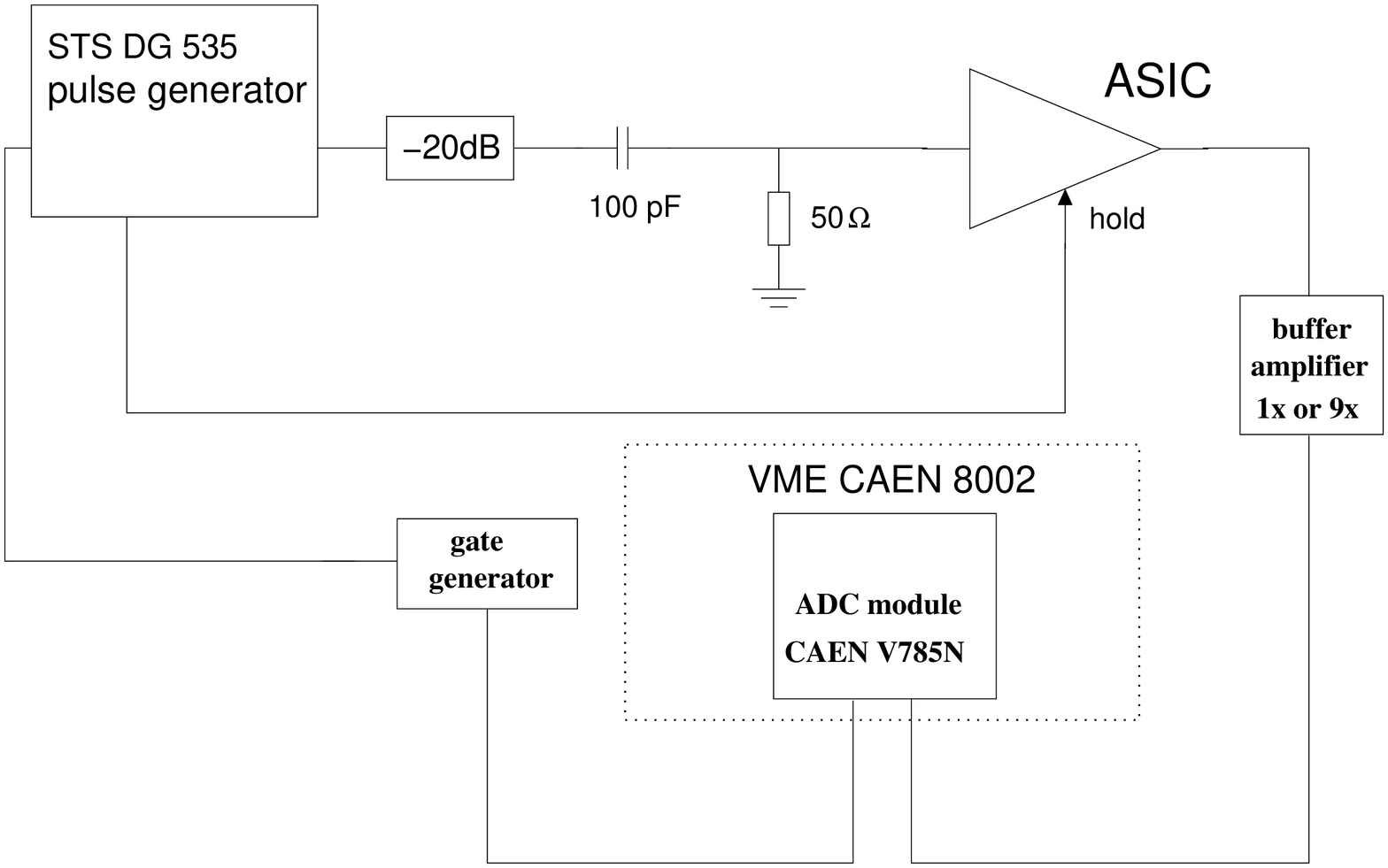}
   \caption{Typical test bench set up for the commissioning of the 
            SPIROC chip at DESY.}
   \label{fig:TestBench}
 \end{center}
\end{figure}

The characterisation measurements of the SPIROC chip were
performed at DESY. The typical set up of the test bench 
is shown in Fig.~\ref{fig:TestBench}.
A digital pulse voltage generator (STS DG535) provides a rectangular 
signal at $10$ kHz
rate and with variable amplitude up to $4$ V. Additionally, a $20$ dB 
attenuator ($10$ factor attenuation) is inserted in the 
signal line. To simulate a real current signal from SiPMs, the generated 
squared signal is driven through a coupling capacitor (\mbox{$C=100$ pF}) 
and a resistor ($R = 50$ $\Omega$) located before the input channels in 
the SPIROC board, resulting in a characteristic decay time \mbox{$\tau = 5$ ns}
for the current signal.  

The signal is then injected in one of the $36$ input channels, 
and the processed output signal can be measured either at the 
oscilloscope (Tektronix TDS $3034B$, running at $300$ MHz bandwidth 
with $2.5$ Gs/s sampling rate) or directly via an external ADC module 
(CAEN $V785N$), accommodated in a VME create (CAEN $8002$). The data 
acquisition is performed via a LINUX machine connected to the 
VME crate.
The pulse generator is also connected to the LINUX machine, and is 
driven by scripts, allowing, together with the acquisition, 
for automatic, flexible, and systematic measurements in a large 
parameters space using SHELL scripts.
Via an USB connection the chip is driven by a LabView user interface
running under WINDOWS because the USB driver to steer the 
chip is not available for LINUX systems. This restricts the
automation and flexibility of the measurements performing 
the scan of the ASIC related parameters.
The porting of this software to a LINUX platform is ongoing. 

A buffer amplifier, built at DESY, was often used (before the ADC
input channels) to improve the precision of the measurement, as in 
the noise studies.

\section{Electrical Noise Investigation}
%
The noise affecting the processed signal out of the board
was measured at different working conditions of the chip. 
Additional sources can superimpose to the chip noise during the 
measurement, as the USB-connection to the computer driving the 
board, the clock on the board which handles the input/output 
communications with the computer, and the connections to the 
pulse generator for the input and hold signals.

\begin{figure}[t!]
  \includegraphics[height=7.5cm,width=10.5cm]{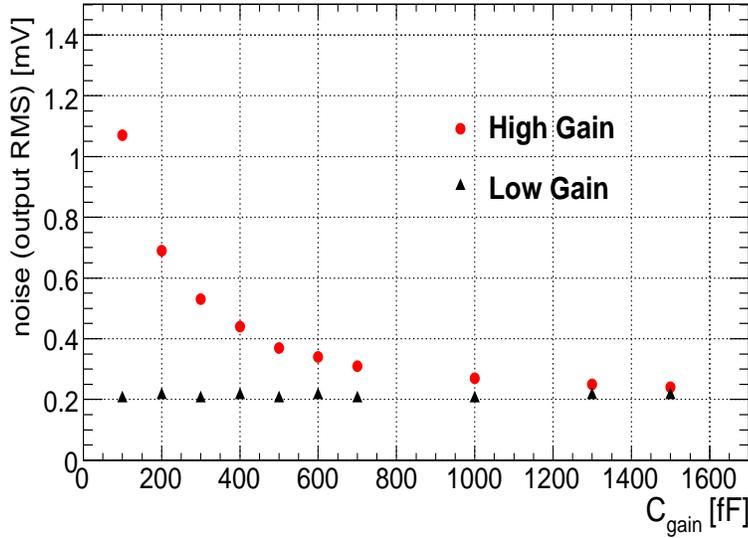}
  \vspace{-0.6cm}
  \caption{The chip noise is measured versus the variable capacitance
           in the preamplifier stage, for $50$ ns shaping time.}
  \label{fig:NoiseVsFc}
\end{figure}

After disconnecting these additional noise source from the board, 
the noise was measured as the RMS of the analogue output signal. 
The dependence of the noise on the variable capacitance in the 
preamplifier stage of the chip is shown in Fig.~\ref{fig:NoiseVsFc},
for both high and low gain operation modes.
The expected $1/C_{gain}$ dependence in the high gain mode is observed. 
The noise reaches its asymptotic value for the largest feedback 
capacitance investigated value. For the low gain mode, where the 
amplification is one order smaller than for the high gain mode, no 
dependence is visible, being the noise dominated by the either the chip 
or the experimental board (the noise from the external ADC module was 
measured to be up to $0.1$ mV).

\begin{figure}[t!]
  \includegraphics[height=7.5cm,width=10.5cm]{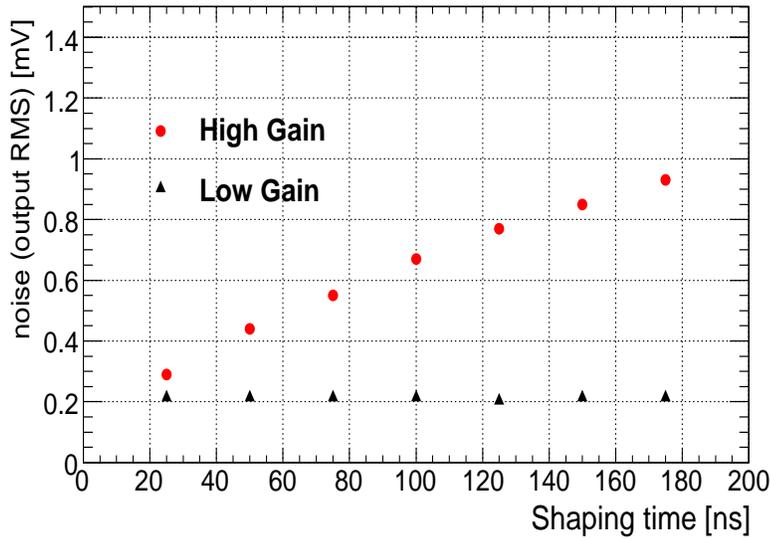}
  \vspace{-0.6cm}
  \caption{The chip noise is measured versus the shaping time
           for $400$ fF variable capacitance in the preamplifier stage.}
  \label{fig:NoiseVsSt}
\end{figure}

The dependence of the chip noise on the shaping time values is 
presented in Fig.~\ref{fig:NoiseVsSt}. 
An almost linear dependence in the high gain mode is observed,
while for the low gain mode no dependence is visible, similarly to 
what observed for the dependence on the feedback capacitance.

\begin{figure}[t!]
  \includegraphics[height=7.cm,width=10.5cm]{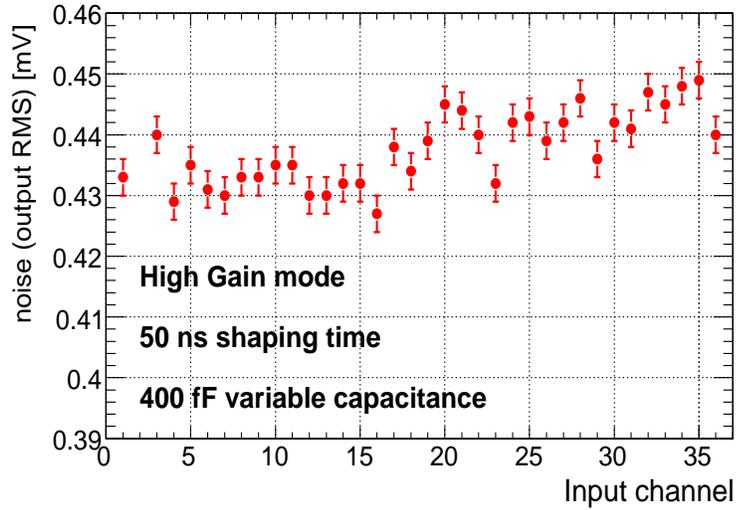}
  \vspace{-0.6cm}
  \caption{The noise is measured independently for all the $36$ input 
           channels of the chip in high gain mode.}
  \label{fig:NoiseVsCh}
\end{figure}

\begin{figure}[t!]
   \includegraphics[height=7.cm, width=10.5cm]
         {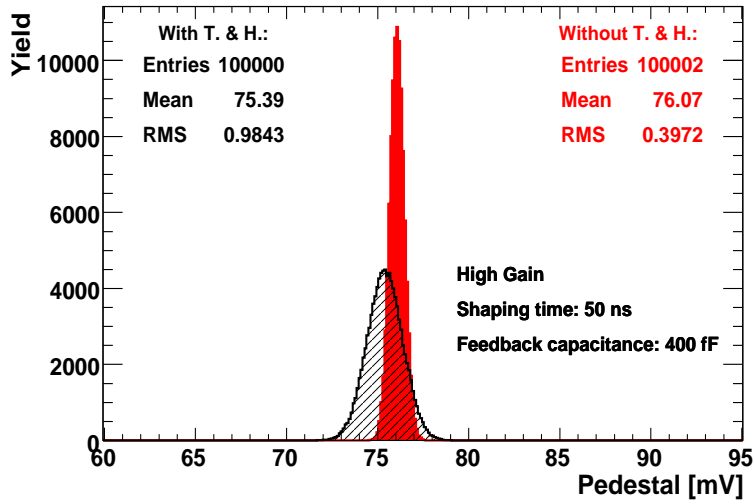}
  \vspace{-0.2cm}
    \caption{The chip pedestal and noise are measured in high gain mode
           for $50$ ns shaping time and $400$ fF feedback capacitance. 
           The distributions were obtained separately  while keeping the 
           track and hold component switched on and off (hatched and 
           filled histograms, separately). The USB connection to the 
           driving PC, and the internal clock were left on during these
           measurements.}
  \label{fig:ENC_EFFECTS_FROM_TH}
\end{figure}

\begin{figure}[t!]
  \includegraphics[height=7.cm,width=10.5cm]{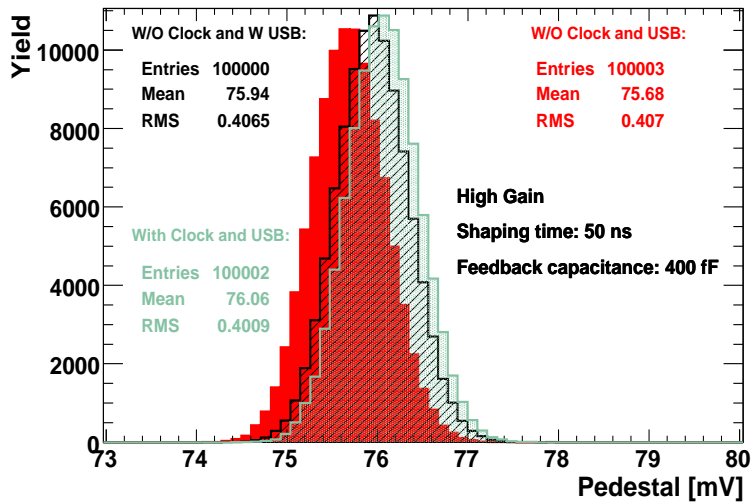}
  \vspace{-0.5cm}
  \caption{The chip pedestal and noise are measured while keeping the
           track and hold component switched off (filled histogram), and
           in sequence plugging the USB connector to the board and 
           switching on the internal clock (striped and
           point histograms, separately).}
  \label{fig:SourcesOfNoise}
\end{figure}

\begin{figure}[t!]
   \includegraphics[height=7.cm, width=10.5cm]
         {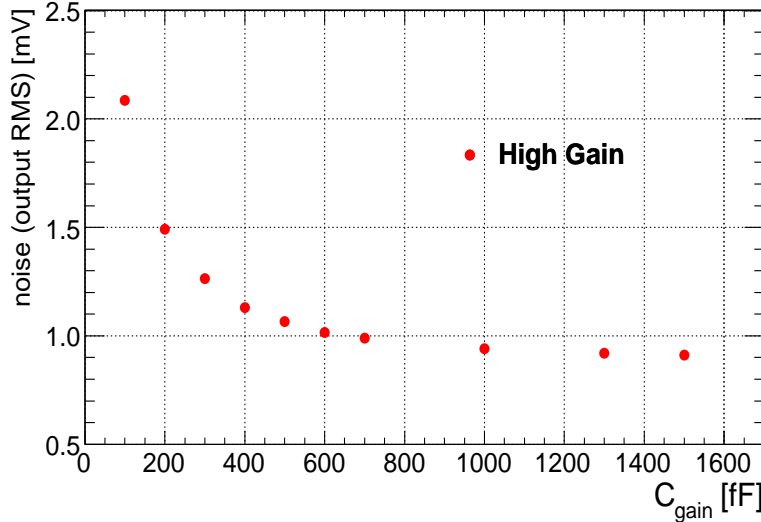}
  \vspace{-0.5cm}
    \caption{The chip noise is measured versus the variable capacitance
           in the preamplifier stage keeping the track and hold component
           switched off. The chip was operated in high mode at 
           $50$ ns shaping time.}
  \label{fig:ENC_VS_FC_WITH_TH}
\end{figure}

The uniformity of the noise in the SPIROC $36$ input channels 
was also investigated, and the result of the measurement
is presented in Fig.~\ref{fig:NoiseVsCh}, for the chip operating
in high gain mode, at $50$ ns shaping time, and $400$ fF
variable capacitance in the pre-amplifier stage. The measured 
noise is uniform within fractions of millivolts, well enough 
for the chip purposes. Although the upper half of channels appears 
to have a higher noise, this feature was not furtherly investigated
being the effect negligible.
The largest effect superimposed to the electrical noise is given 
by using the track and hold switch to hold the pre-amplified 
and shaped signal at its peaking amplitude. An increase of the 
noise up to a factor $2$-$3$ is observed in the output signal, 
Fig.~\ref{fig:ENC_EFFECTS_FROM_TH}.
With the used setup we could not distinguish between an increase of 
the noise or of the sensitivity to the noise (due to the increased
bandwidth induced by the track and hold switch).
Instead, a negligible effect is observed by the switching on the internal 
clock for input/output communication with an external driving interface
(via USB port), and by plugging the USB connector into the board, 
Fig.~\ref{fig:SourcesOfNoise}. 
In normal data taking conditions the ASIC is supposed to process
the incoming signals via the track and hold component switched on. 
Therefore it is important also to measure its impact on the 
measured noise dependence on the preamplifier gain, 
Fig.~\ref{fig:ENC_VS_FC_WITH_TH}.
Consistently with what shown in Fig.~\ref{fig:ENC_EFFECTS_FROM_TH}, 
an increase of the noise up to approximately a factor $3$ is observed
at $C_{gain} = 1500$ fF. 

\section{ENC and Input Detector Capacitance}
%
Sofar the electrical noise of the stand-alone ASIC was investigated, 
without any input signal line connected to the board. Connecting 
a SiPM to the readout system is expected to increase the noise, 
depending on the detector capacitance.

In principle, a charge-sensitive preamplifier would be 
favoured~\cite{LEO}, providing small sensitivity to changes of the 
parasitic capacitance at its input, as in the case of SiPMs whose 
capacitance changes with temperature.
Nevertheless, in order to cover the large dynamic range of SiPMs, a 
voltage-sensitive preamplifier was adopted for the chip. 
It is crucial therefore to quantify the noise sensitivity to an 
external variable capacitance coupled via the connected detector.

Since the signal from the SiPM appears as electric charge, electronic
noise will be quantified by giving its 'equivalent noise charge' (ENC), 
defined as the input charge which would be necessary to generate 
a signal equivalent in amplitude to the measured noise output of the ASIC. 
This value can be obtained by normalising the noise to a reference signal. 
As an example, the signal originating from one photo-electron in the SiPM, 
and its corresponding output are considered. 
Assuming a SiPM gain $G_{SiPM}$ of 
order $5\cdot 10^{5}$ would result in an input charge $Q_{in}$ approximately 
of $G_{SiPM} \cdot e = 80$ fC. The corresponding output signal $V_{out}$ for 
the highest gain was found to be $\approx 7.6$ mV (see 
Sec.~\ref{sec:THRESHOLD_CUTS}), obtaining an ENC of order
\begin{equation}
   ENC = V_{RMS} \cdot \frac{Q^{1pxl}_{in}}{V^{1pxl}_{out}} =
   V_{RMS} \cdot \frac{80}{7.6} \approx V_{RMS} \cdot 10.5 \ \ \ 
   [ fC \cdot mV^{-1}].
\end{equation}
Here $V_{RMS}$ is the average voltage noise level (in millivolt) 
appearing at the output. 
Dividing it by the the electron charge $e$ (in femtocoulomb units) gives 
the ENC in number of electrons
\begin{equation}
   ENC = V_{RMS} \cdot \frac{10.5}{1.6 \cdot 10^{-4}} \approx
   V_{RMS} \cdot 7 \cdot 10^4 \ \ \  [ \mathit{electrons} \cdot mV^{-1}].
\end{equation}

\begin{figure}[t!]
   \includegraphics[height=6.75cm, width=10.5cm]
         {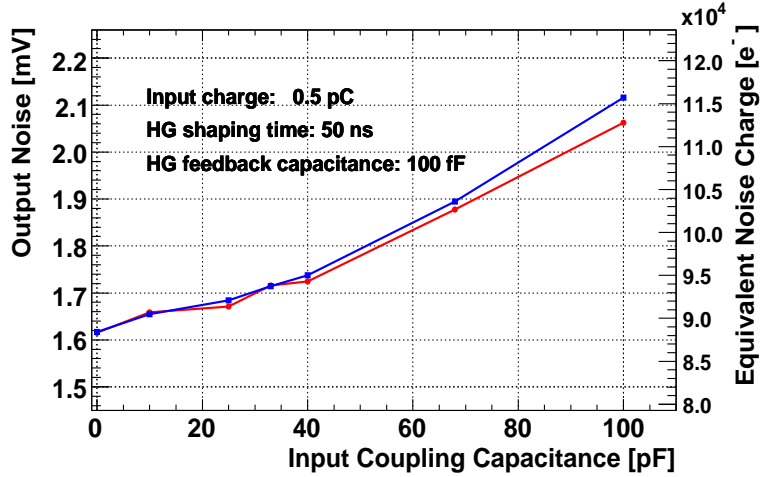}
  \vspace{-0.55cm}
    \caption{Equivalent noise charge measured while varying 
             the coupling capacitance at the input signal line.}
  \label{fig:ENC_VS_CC}
\end{figure}

Using the external pulse generator, the output noise was measured 
and converted in number of equivalent electrons following the above 
formula (2). During the measurement, the coupling capacitance value was 
varied to simulate the change of the SiPM internal capacitor. 
The results of two scans performed in sequence, without any change 
in the setup apart replacing the capacitor, are presented in 
Fig.~\ref{fig:ENC_VS_CC}, 
showing a noise $V_{RMS}$ increase up to $20\%$ in the range 
\mbox{$0$-$100$} pF, 
from $1.6$ to $2.1$ mV (corresponding to an ENC variation from $9\cdot 10^4$ 
to $11.5\cdot 10^4$ electrons).
According to these measurements, the signal over noise ratio achievable 
by the SPIROC can be calculated, ranging approximately between $4.7$ and $3.6$
assuming a SiPM of $5\cdot 10^{5}$.

\begin{figure}[t!]
 \begin{minipage}{6.75cm}
   \includegraphics[height=5.5cm, width=7.cm]
         {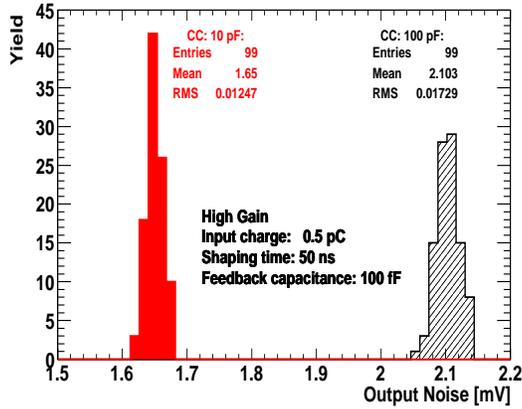}
 \end{minipage}
 \begin{minipage}{5.cm}
    \caption{Distribution of noise measurement performed at two different 
             coupling capacitance values.}
  \label{fig:ENC_SCAN}
 \end{minipage}
\end{figure}

To verify that the observed non-reproducibility of the measurements 
is not given 
by any feature of the ASIC, a series of measurements was taken in 
sequence at a fixed coupling capacitance to simulate the detector 
capacitance. No additional change in the experimental apparatus was 
introduced. 
The results for two values of the coupling capacitance, 
Fig.~\ref{fig:ENC_SCAN}, shows a spread in the measurement which is of the 
order of a fraction of millivolt. This is well below the maximum spread 
observed in the two consecutive scans, thus suggesting a noise source 
external to the ASIC, possibly due to the experimental setup.

\begin{figure}[t!]
\begin{center}
  \includegraphics[height=7.5cm,width=10.5cm]{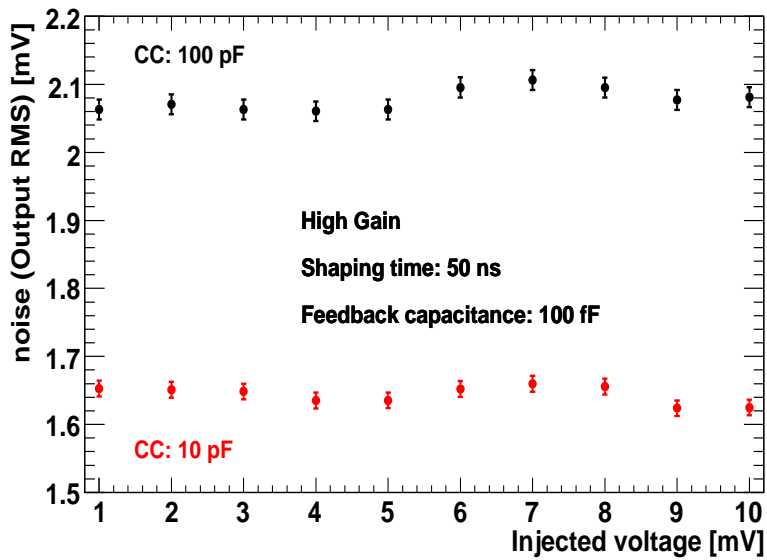}
  \vspace{-0.25cm}
  \caption{The signal noise is measured for different values of the 
           injected voltage from the pulse source, and for two 
           extreme values of the coupling \mbox{capacitance (CC)}.}
  \label{fig:ENC_VS_CHARGE}
\end{center}
\end{figure}

The measurement of the coupling capacitance dependence of the ENC was 
performed injecting $0.5$ pC into the ASIC (except for the measurement
at zero capacitance, when the line was disconnected from the board). 
To validate 
the results it should be verified the reasonable assumption that the 
measured noise is not influenced by the amount of injected charge.
A measurement of the noise was taken at different values of 
injected charge, and for two values of the coupling capacitance, 
showing, as expected, no dependence, Fig.~\ref{fig:ENC_VS_CHARGE}. 
%

\section{Pedestals Uniformity}
\begin{figure}[t!]
\begin{center}
  \includegraphics[height=7.5cm,width=11.5cm]{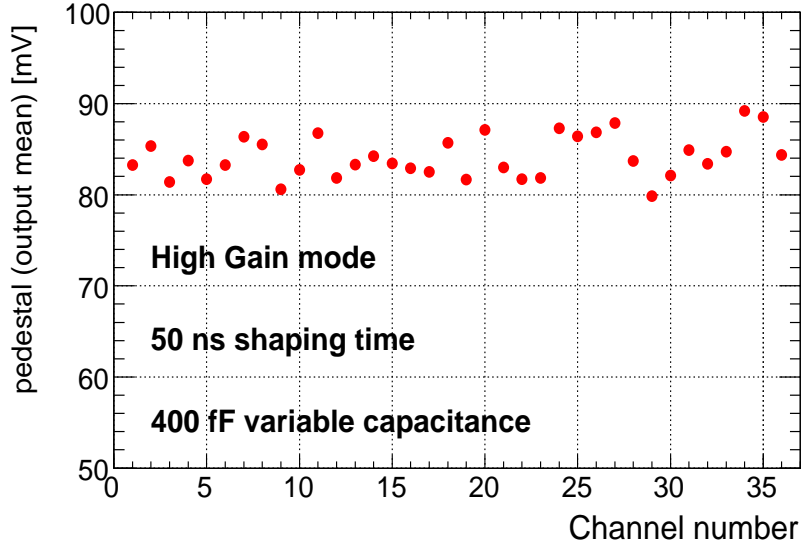}
  \vspace{-0.5cm}
  \caption{Pedestals measured for all the $36$ input
           channels of the chip, using the external ADC module.}
  \label{fig:PedestalVsCh}
\end{center}
\vspace{-0.3cm}
\end{figure}

%
Pedestals were investigated for all the the $36$ input channels 
of the chip, to verify their uniformity, and are presented in
Fig.~\ref{fig:PedestalVsCh}, using the external ADC module.
Note that the mean values reported here are not the absolute 
pedestal values, due to an offset introduced for convenience 
in the amplifier before the ADC. 

\begin{figure}[t!]
 \begin{minipage}{7.5cm}
 \includegraphics[height=6.5cm,width=7.5cm]{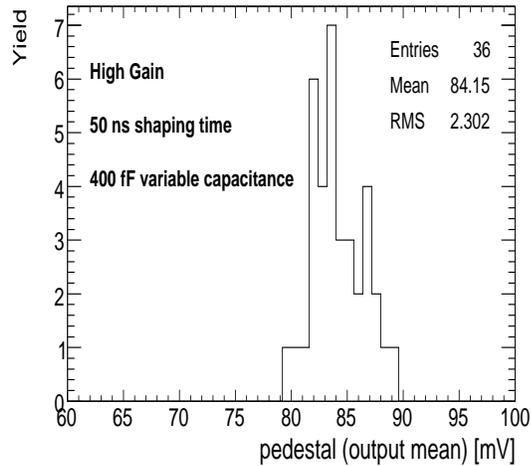}
  \vspace{-0.8cm}
 \end{minipage}
 \begin{minipage}{4.cm}
  \caption{Pedestal distribution for all the $36$ input 
           channels.}
  \label{fig:PedestalSpread}
 \end{minipage}
\end{figure}

The measured spread of pedestals is about $2.3$ mV, 
Fig.~\ref{fig:PedestalSpread}, a value 
slightly larger than what reported in~\cite{ORSAY}.
%

\section{Trigger Discriminator and Efficiency}
\label{sec:TRIGGER_EFF}
The chip is designed to operate in the so called 'auto-trigger' mode. 
The input signals are first pre-amplified according to both 
low and high gain line settings. In the high gain section a dedicated
$15$ ns fast shaping line is also present in parallel to the line to 
the analogue memory. There the signal enters a discriminator with a 
threshold common to all $36$ input lines, and a $120$ ns wide trigger 
is generated whenever the voltage in the line 
is above the threshold value set, Fig.~\ref{fig:SPIROC_schematics}. 
The common threshold is tunable by a $10$ bits DAC. On top of the 
common threshold value, each channel threshold can be individually
tuned via a $4$-bit DAC. 

In normal ILC operations the generated trigger is forseen to subsequently 
hold the pre-amplified signal, processed in the meanwhile by slower shapers, 
at its amplitude peaking value. It is therefore crucial to investigate the
trigger efficiency and homogeneity for the $36$ channels.

\begin{figure}[t!]
\begin{center}
  \includegraphics[height=6.3cm,width=10.0cm]{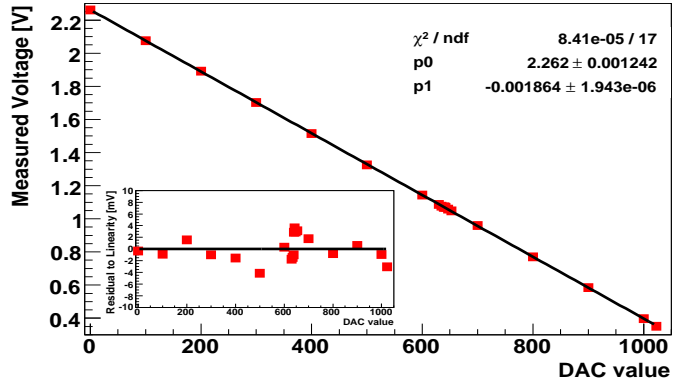}
  \vspace{-0.8cm}
  \caption{The calibration of the common signal threshold $10$-bit 
           DAC values into volt units.}
  \label{fig:DAC_calib}
\end{center}
\end{figure}

The calibration of the common DAC values was performed, and is shown
in Fig.~\ref{fig:DAC_calib}. The threshold is found to range from $0.35$ V 
up to $2.26$ V. The residuals to linearity, found to be within 
few millivolts (a fraction of a photon-electron signal), are presented 
in the small insert panel.

\begin{figure}[b!]
  \hspace{.75cm}
  \includegraphics[height=6.3cm,width=10.0cm]{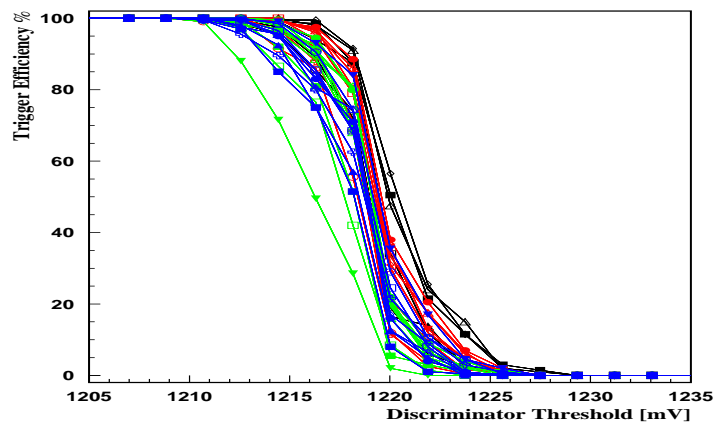} 
  \vspace{-0.5cm}
  \caption{The trigger efficiency spread for all the $36$ input 
           channels in the ASIC, relative to an injected charge of
           $100$ fC.}
  \label{fig:S_curve}
\end{figure}

After having performed the discriminator DAC calibration, the trigger 
efficiency can be investigated channel by channel. This is done using 
a built-in procedure in the LabView user interface. The program increases 
the threshold level in sequence of DAC units, remains at a fixed DAC 
value for $200$ cycles, and counts the number of triggers generated
by the pedestal of the investigated channel.  
It provides thus a measurement of the trigger efficiency, which is 
presented in Fig.~\ref{fig:S_curve} 
for an injected charge of
$100$ fC, separately for each of the $36$ channels.

The increase of the trigger counting is smoothed by the noise amplitude
in the line, which was typically found to be around $2$-$3$ mV during
that specific measurement, and was obtained considering the threshold value 
variation needed to increase the trigger efficiency from $10\%$ to $90\%$.
Also, the maximum efficiency is reached at threshold level values 
different from channel to channel, with a spread of around $5$ mV, 
resulting in a negligible pedestal spread  between all the $36$ channels. 
During forseen data taking operations, the observed spread can, 
in principle, be cured tuning the discriminator level channel by 
channel using the forseen $4$-bit threshold finer adjustment. At 
the moment, this feature is not properly working in the released
versions of the chip (SPIROC 1 and \mbox{SPIROC 2}), and could not be 
investigated. 

\section{Trigger Time Walk and Jitter}
\begin{figure}[t!]
 \begin{minipage}{6.5cm} 
   \includegraphics[height=2.4cm, width=4.5cm, viewport=0 0 350 245]
         {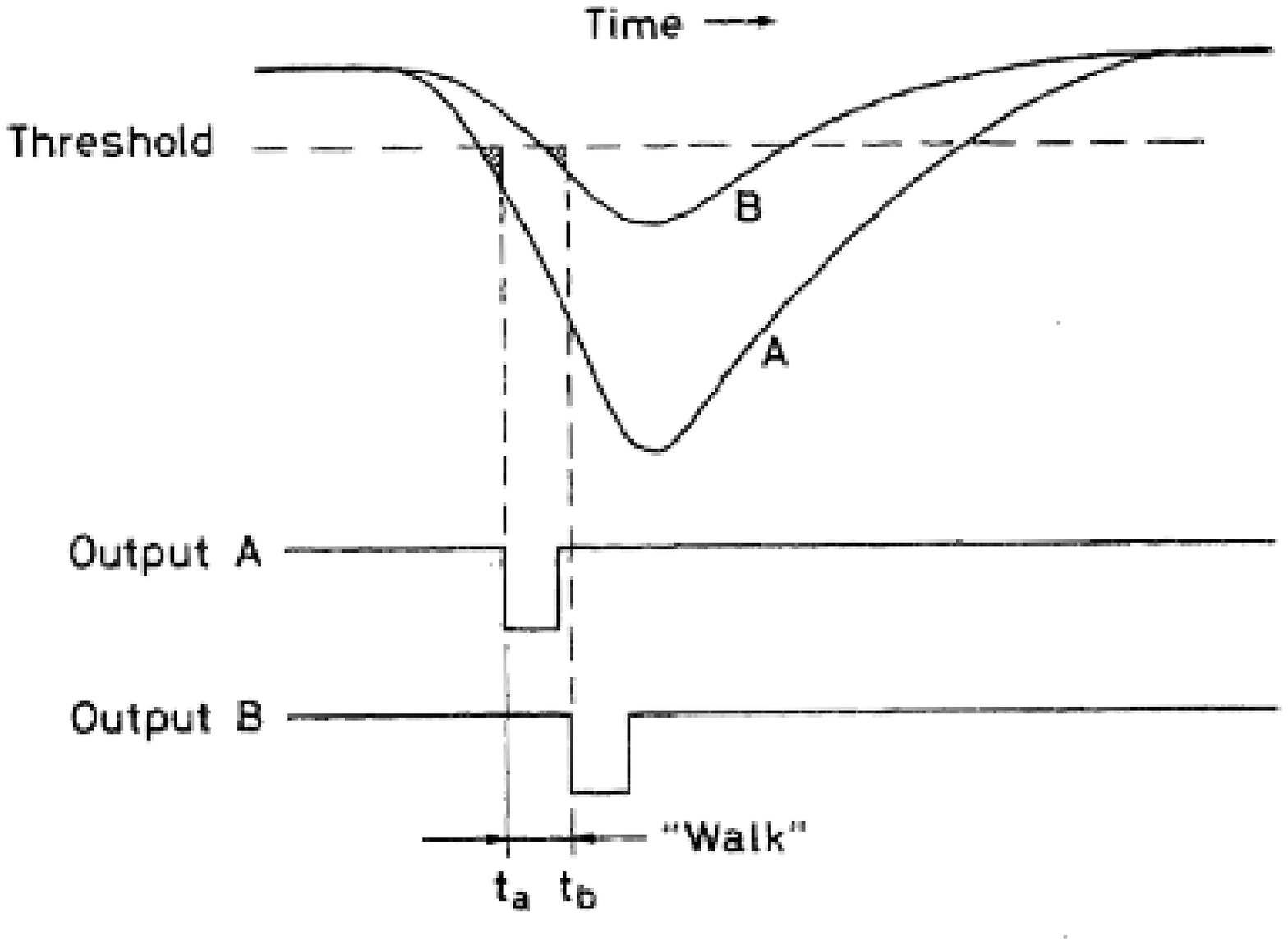}
 \end{minipage}
 \begin{minipage}{6.5cm}
   \hspace{-0.5cm} 
   \includegraphics[height=4.3cm, width=5.8cm] {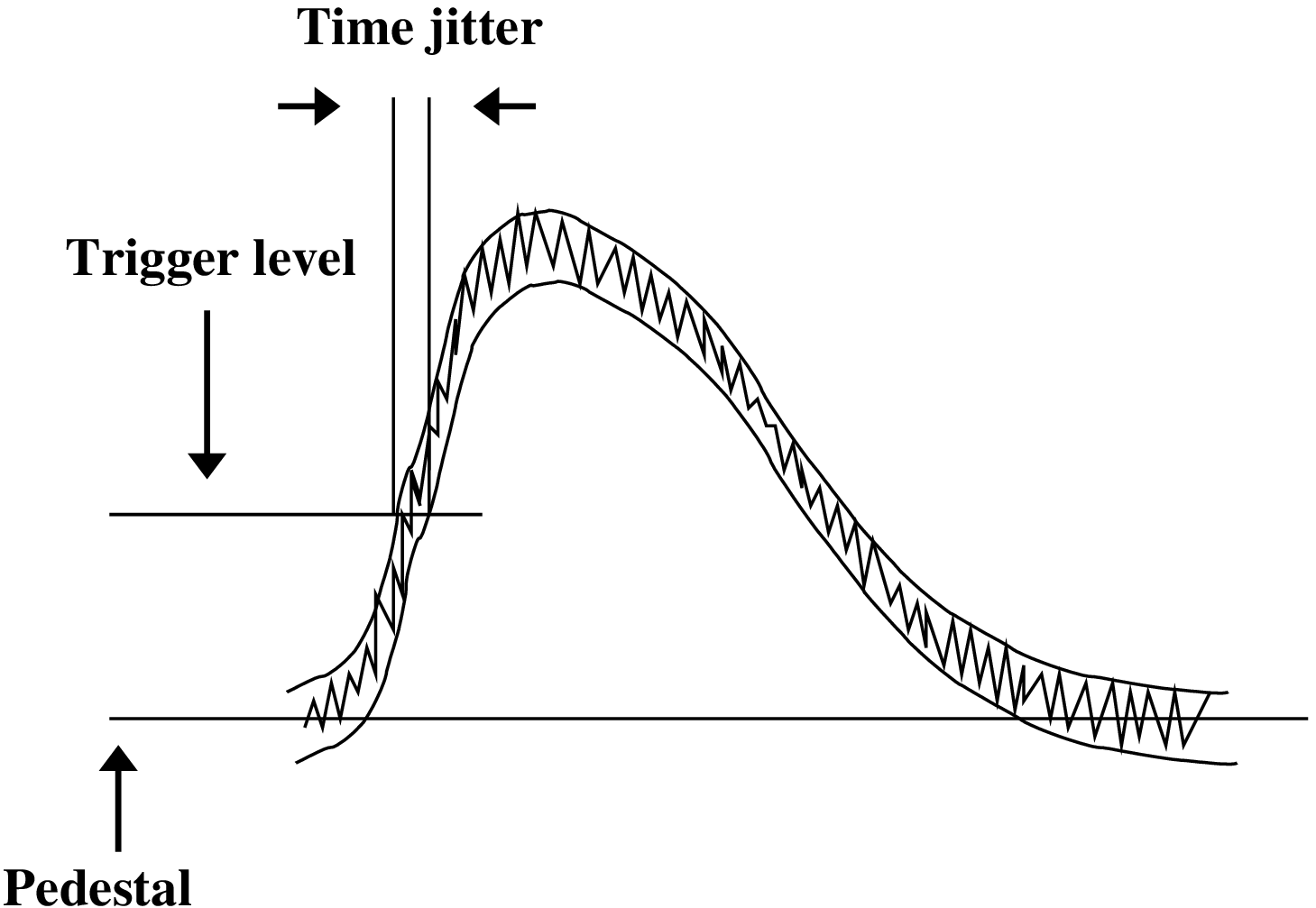}
 \end{minipage}
 \vspace{0.2cm}
 \caption{Examples of trigger time walk (left panel) and jitter
           (right panel).}
  \label{fig:trigger_uncertainties}
\end{figure}

%
\label{sec:trigger}
As mentioned in Sec.~\ref{sec:TRIGGER_EFF},
SPIROC is forseen to be used in auto-trigger mode during 
ILC running conditions.
It is crucial therefore to investigate the size of the 
main uncertainties which can affect the trigger timing. 
Signals with different amplitudes (and same peaking time)
cross the discriminator level at different times, resulting 
in a time shift (trigger walk) of the generated processing trigger, 
left panel of Fig.~\ref{fig:trigger_uncertainties}.
On top of this uncertainty, the crossing of the discriminator 
level is varied in time (trigger jitter) by the noise in the analogue signal, 
right panel of Fig.~\ref{fig:trigger_uncertainties}.

\begin{figure}[t!]
 \begin{center}
   \includegraphics[height=6.75cm,width=10.5cm]{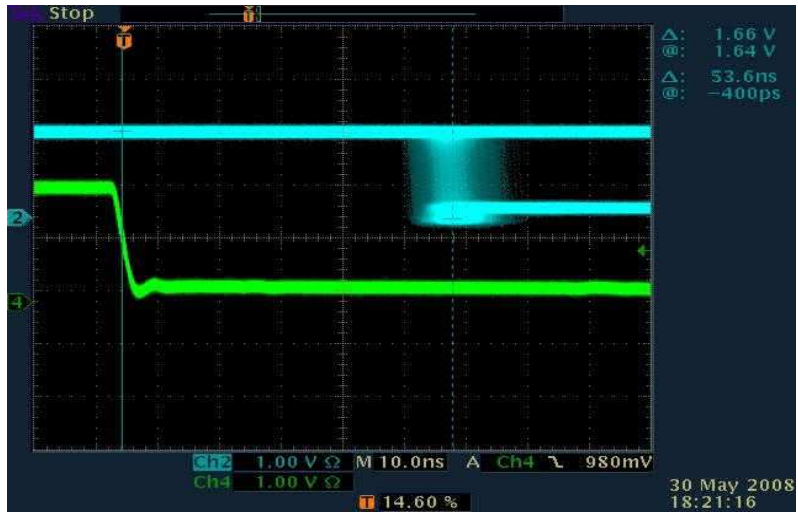}
   \caption{The trigger signal (light blue) is measured in the oscilloscope 
           in coincidence with the pulse generator trigger (green).
           The shift of the mean and the RMS value of the trigger distribution 
           provide the measurement of the trigger time walk and jitter,
           respectively.}
   \label{fig:oscilloscope_jitter}
 \end{center}
\end{figure}

\begin{figure*}[t!]
  \includegraphics[height=6.cm,width=6.8cm]{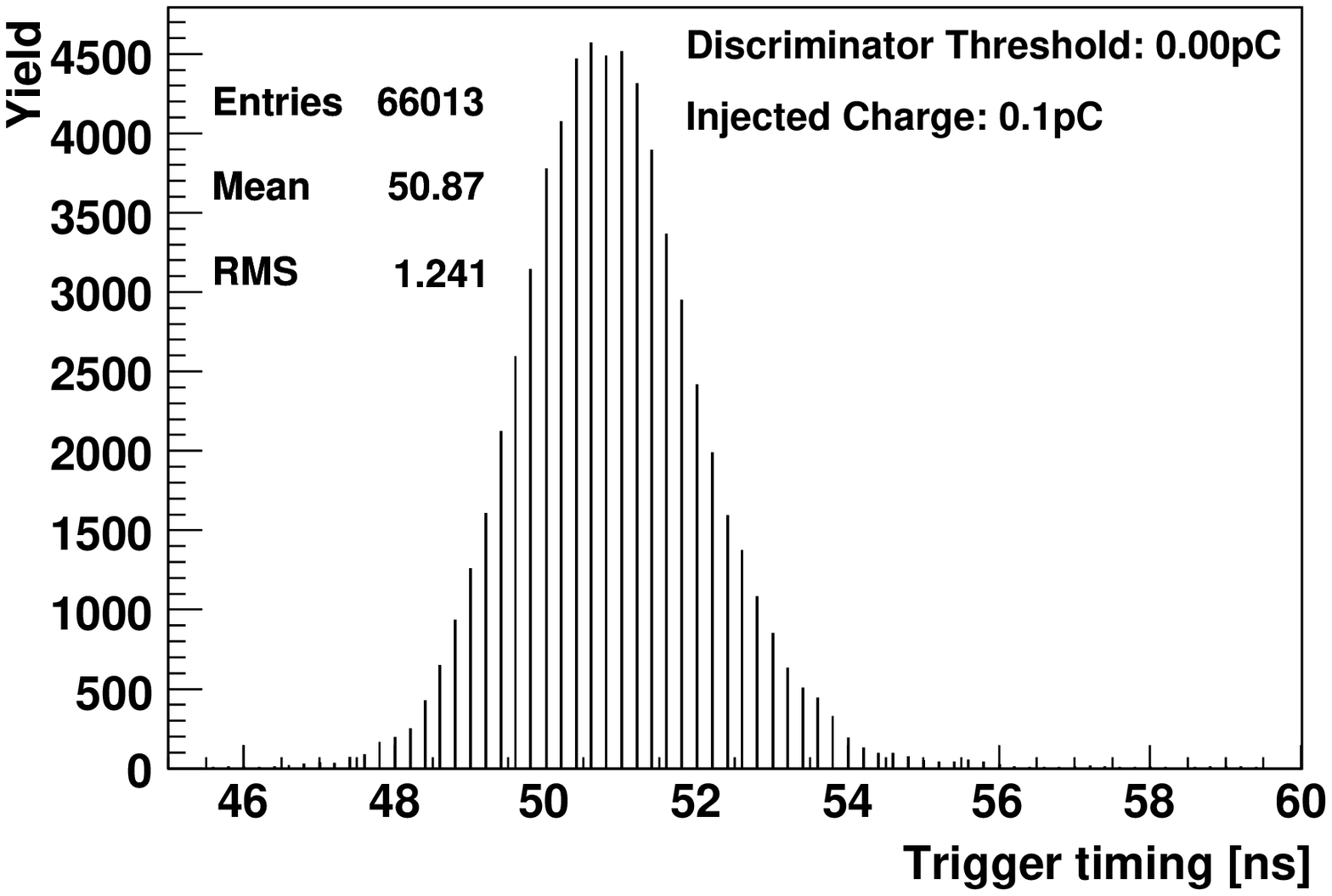} 
  \hspace{-0.8cm}
  \includegraphics[height=6.cm,width=6.8cm]{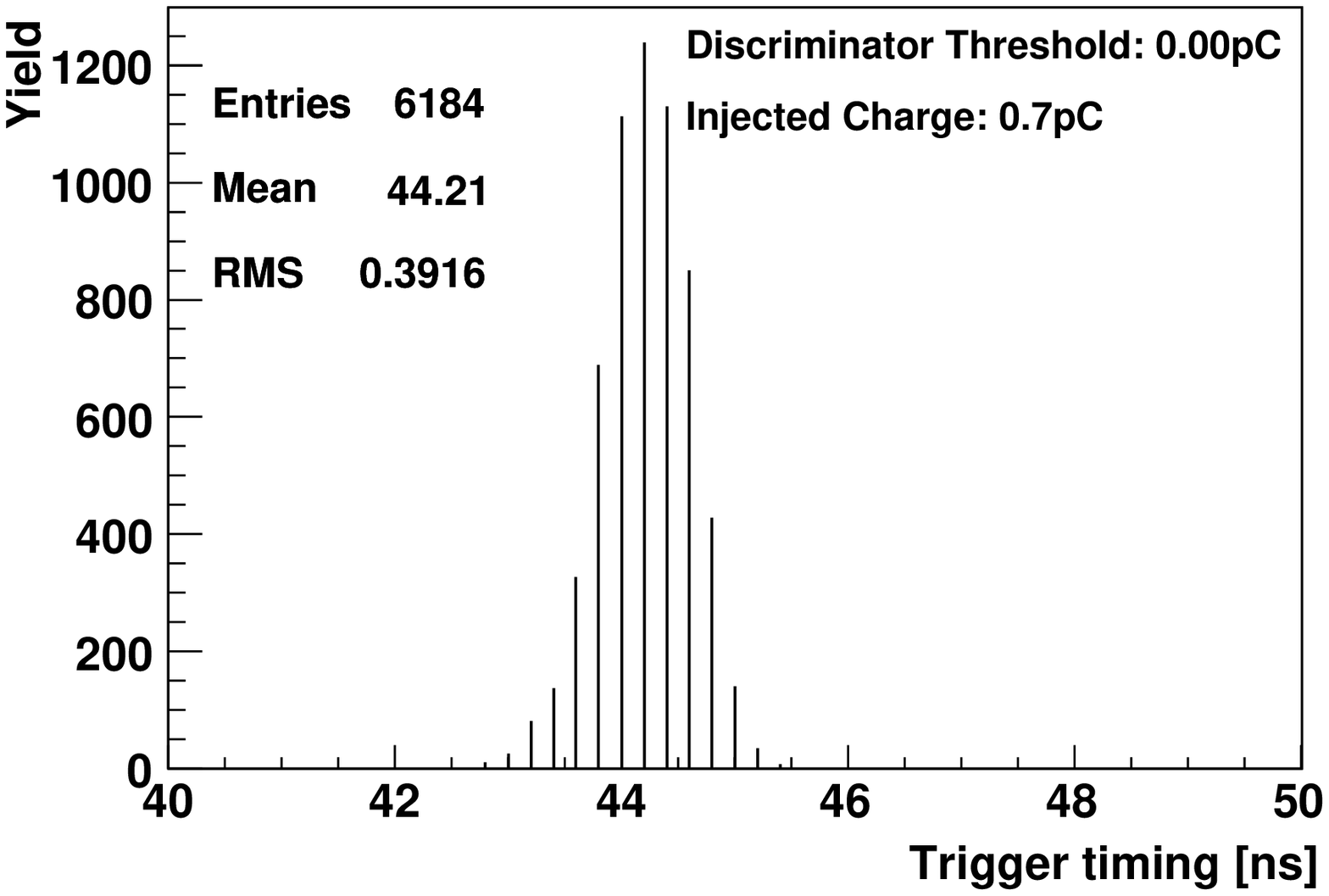}\\
  \vspace{-0.5cm}
  \caption{The trigger timing distribution is presented for the discriminator
           threshold value set above the pedestal, and for 
           different values of the injected charge in the SPIROC board.}
  \label{fig:trigger_jitter_distrib}
\end{figure*}

The trigger time walk and jitter were investigated injecting a signal 
from the pulse generator, and then measuring at the oscilloscope the 
timing of the coincidence of the generated 
trigger in the chip with the main pulse generator trigger. 
Different threshold values for the analogue signals were set 
(via the LabView interface) at the $10$-bit DAC discriminator 
(common to the $36$ input channels) in the fast shaping line, 
and different charge values were injected,  
thus allowing for the study of the  time walk and jitter dependence 
on the trigger threshold level and on the amplitude of the input signal.
The shift of the mean and the RMS value of the measured trigger distribution,
which is shown for a typical setting in Fig.~\ref{fig:oscilloscope_jitter},
provide the measurement of the trigger time walk and jitter,
respectively.
Typical distributions of the trigger timing above described are presented in 
Fig.~\ref{fig:trigger_jitter_distrib}.

\begin{figure}[t!]
  \includegraphics[height=7.35cm,width=10.5cm]{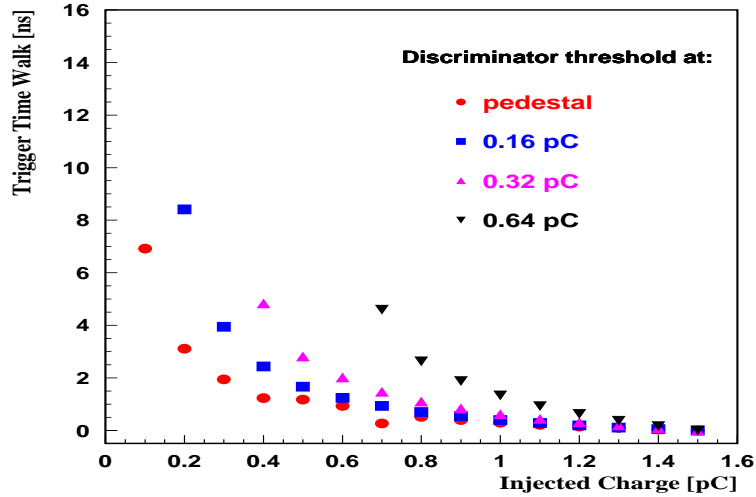}
  \vspace{-0.7cm}
  \caption{The trigger time walk is presented for different 
           threshold values of the $10$-bit DAC discriminator, and 
           for increasing values of the injected charge in the 
           SPIROC board.}
  \label{fig:trigger_timewalk}
\end{figure}

The results of the time walk measurements are shown in 
Fig.~\ref{fig:trigger_timewalk}, for increasing values of the 
injected charge, from $0.08$ up to $1.60$ pC (equivalent to one pixel 
and $20$ firing pixels in an AHCAL SiPM, respectively), 
and of the discriminator threshold level. The time walk values are
presented with respect to the trigger timing corresponding to the 
largest injected charge.
When setting the threshold value above the pedestal the time walk 
appears to be up to $7$ ns only for the first pixel peaks. For a larger 
number of firing pixels it is within one nanosecond; this feature is
observed also for a threshold at about half a mip 
(SiPMs are configured such that on average $15$ pixels 
should fire due to a mip energy deposition~\cite{Groll}, thus 
corresponding to $1.2$ pC input charge).

\begin{figure}[t!]
  \includegraphics[height=7.35cm,width=10.5cm]{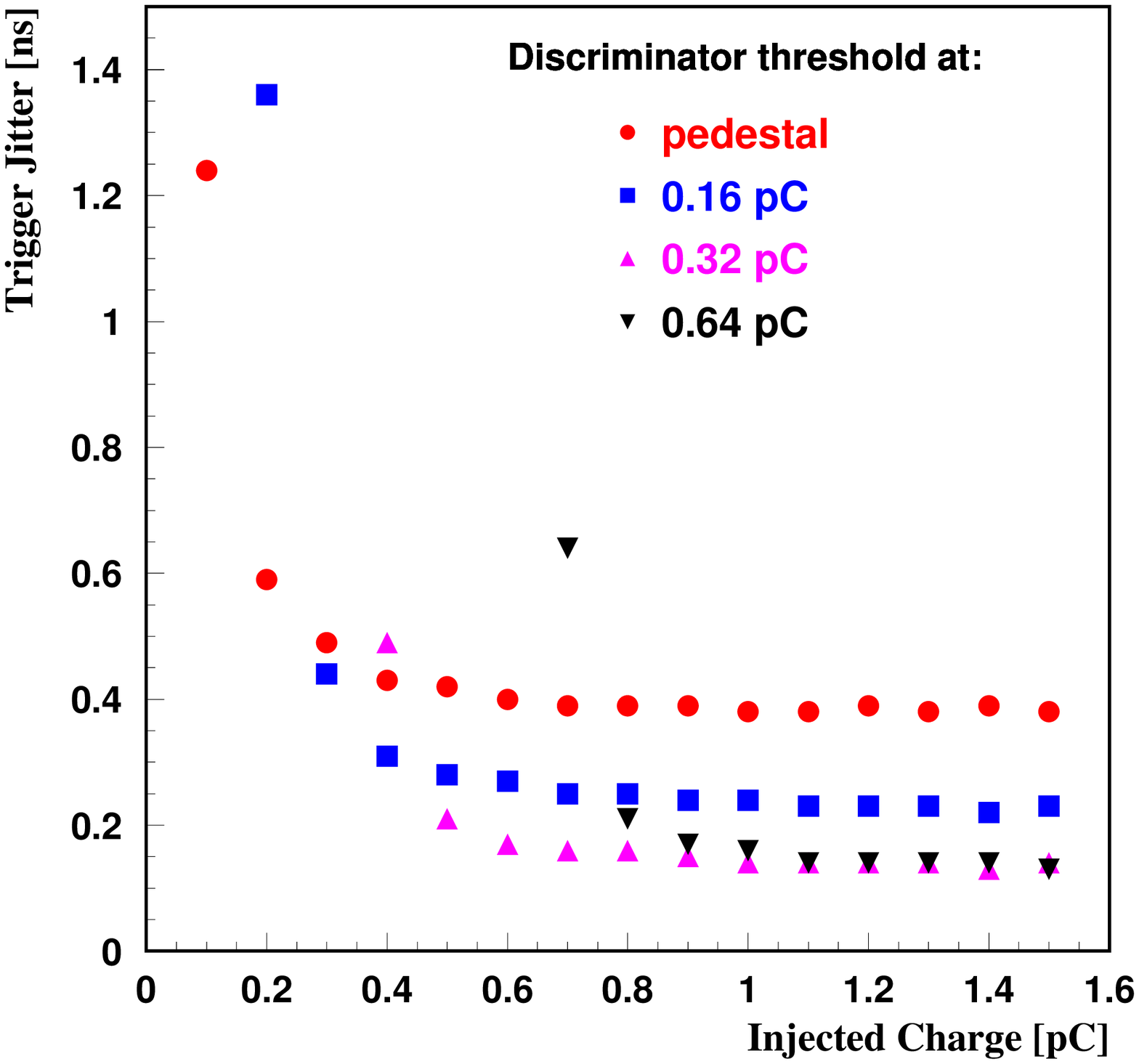}
  \vspace{-0.7cm}
  \caption{The trigger jitter is presented for different 
           threshold values of the $10$-bit DAC discriminator, and 
           for increasing values of the injected charge in the 
           SPIROC board.}
  \label{fig:trigger_jitter}
\end{figure}

The trigger jitter dependence on the injected charge, for different 
values of the discriminator threshold, is presented in 
Fig.~\ref{fig:trigger_jitter}.
The jitter appears to be up to $1.4$ ns, at the smallest threshold 
values and for small values of injected charge, while it is within 
$1.0$ ns at a threshold of about half a mip.

Concerning the forseen operation of the chip in auto-trigger mode, 
resolving single-pixel structure in the SiPM spectra (i.e., 
in calibration mode) appears to be not infeasible, due to the 
reasonably small measured trigger jitter and time walk
At the moment, the SiPM calibration is expected to be performed via an 
LED system providing an external trigger to the ASIC readout.
This issue will be extensively presented in 
Sec.~\ref{sec:single_pixel_spectra}. 
Instead, the small time walk and jitter values observed at thresholds
and input charges above half a mip, should definitely 
allow the chip to properly operate in the physics mode.  
The influence of largely varying SiPM gain values, light generation 
and collection efficiency has not been studied in this work.

\section{SiPM Voltage Adjustment}
\label{sec:InputDAC}
\begin{figure}[t!]
\begin{center}
   \includegraphics[height=5.cm,width=10.cm]{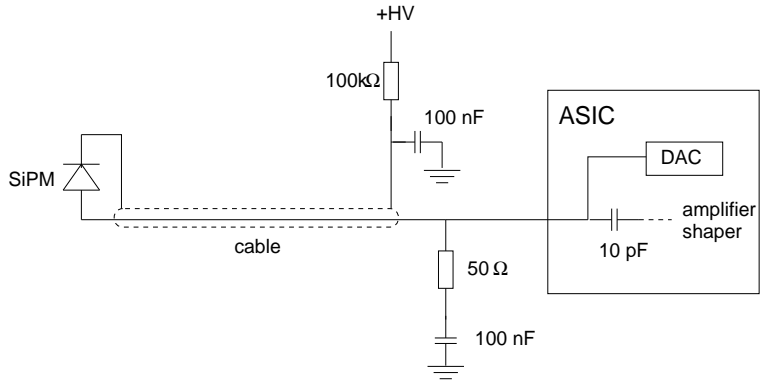}
   \vspace{-0.3cm}
   \caption{Diagram of the SiPM connection to the ASIC board. The same 
            line is used for both power and signal. While a common  
            voltage bias is provided by an external power supply to all $36$ 
            SiPMs connected to the ASIC, each voltage is tunable 
            via a separate \mbox{$8$-bit} DAC HV adjustment.}
   \label{fig:InputHvDAC}
 \end{center}
\end{figure}

The chip is forseen to be connected to an external power supply and 
to provide a common high voltage bias to all connected $36$ SiPMs.
Tuning of the applied voltage, channel by channel, is achieved
using an $8$-bit DAC HV adjustment dedicated to each
detector power line, Fig.~\ref{fig:InputHvDAC}.

\begin{figure}[b!]
 \begin{minipage}{6.5cm}
    \includegraphics[height=5.5cm, width=6.5cm]{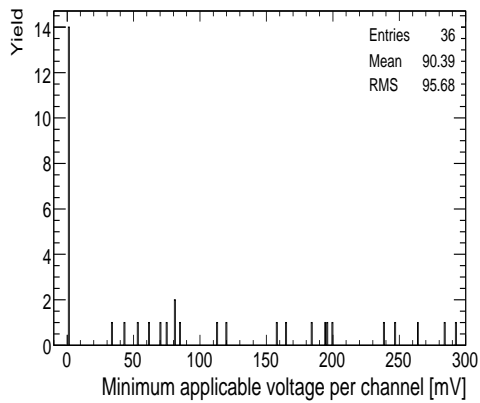} 
 \end{minipage}
 \begin{minipage}{5.5cm}
    \caption{Distribution of the minimum applicable voltage 
             via the \mbox{$8$-bit} DAC HV adjustment. Each entry represents 
             the minimum voltage value measured in one SPIROC channel.}
    \label{fig:InputHvDAC_MinVoltage}
 \end{minipage}
\end{figure}

The calibration of the input HV DAC for all channels is at the moment 
performed via dedicated LabView interface routines. The applied 
voltage ranges approximately from $0$ V to $4.5$ V, varying from 
channel to channel, Fig.~\ref{fig:InputHvDAC_MinVoltage}. 

\begin{figure}[t!]
  \includegraphics[height=5.cm, width=6.0cm]{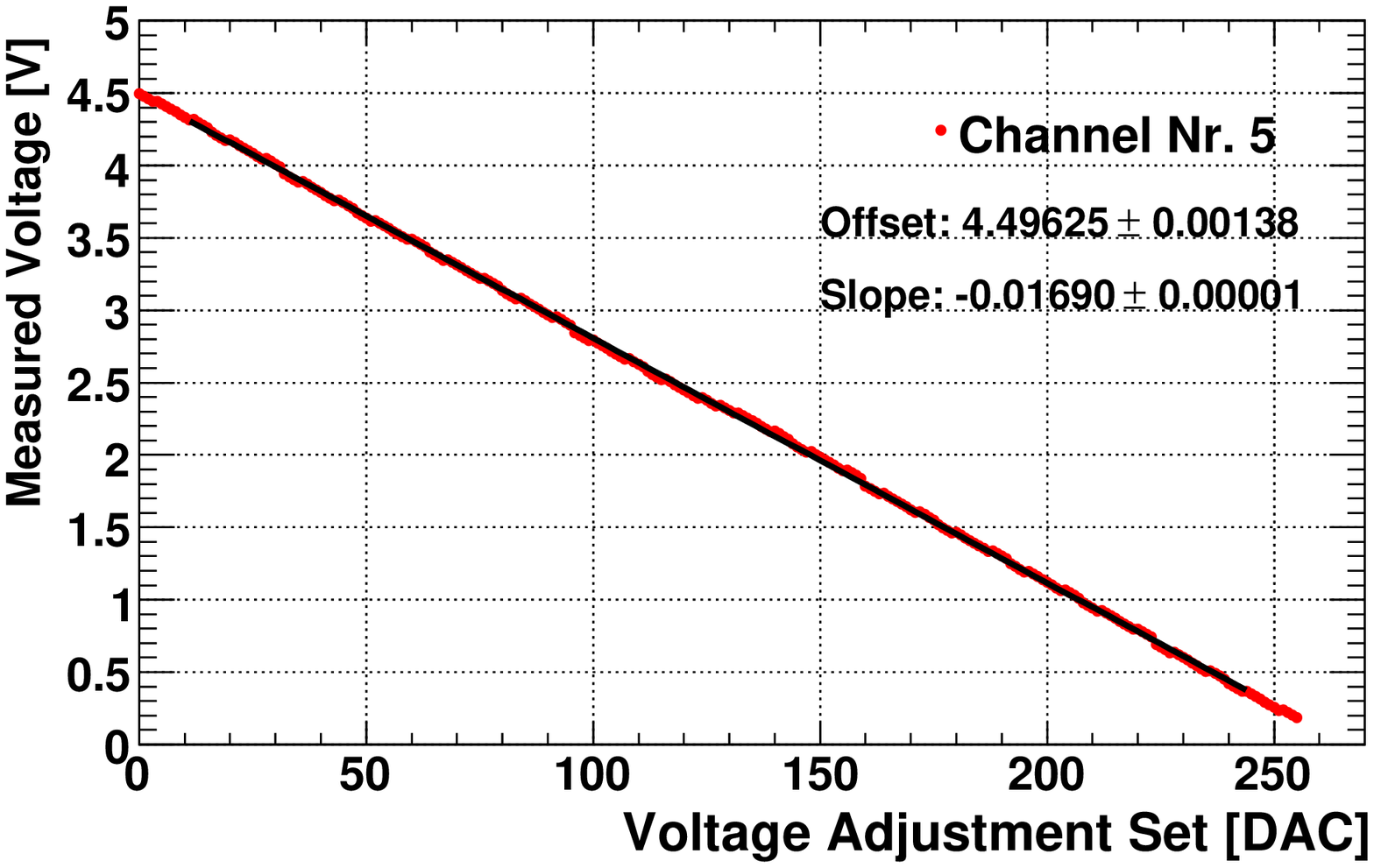} 
  \includegraphics[height=5.cm, width=6.0cm]{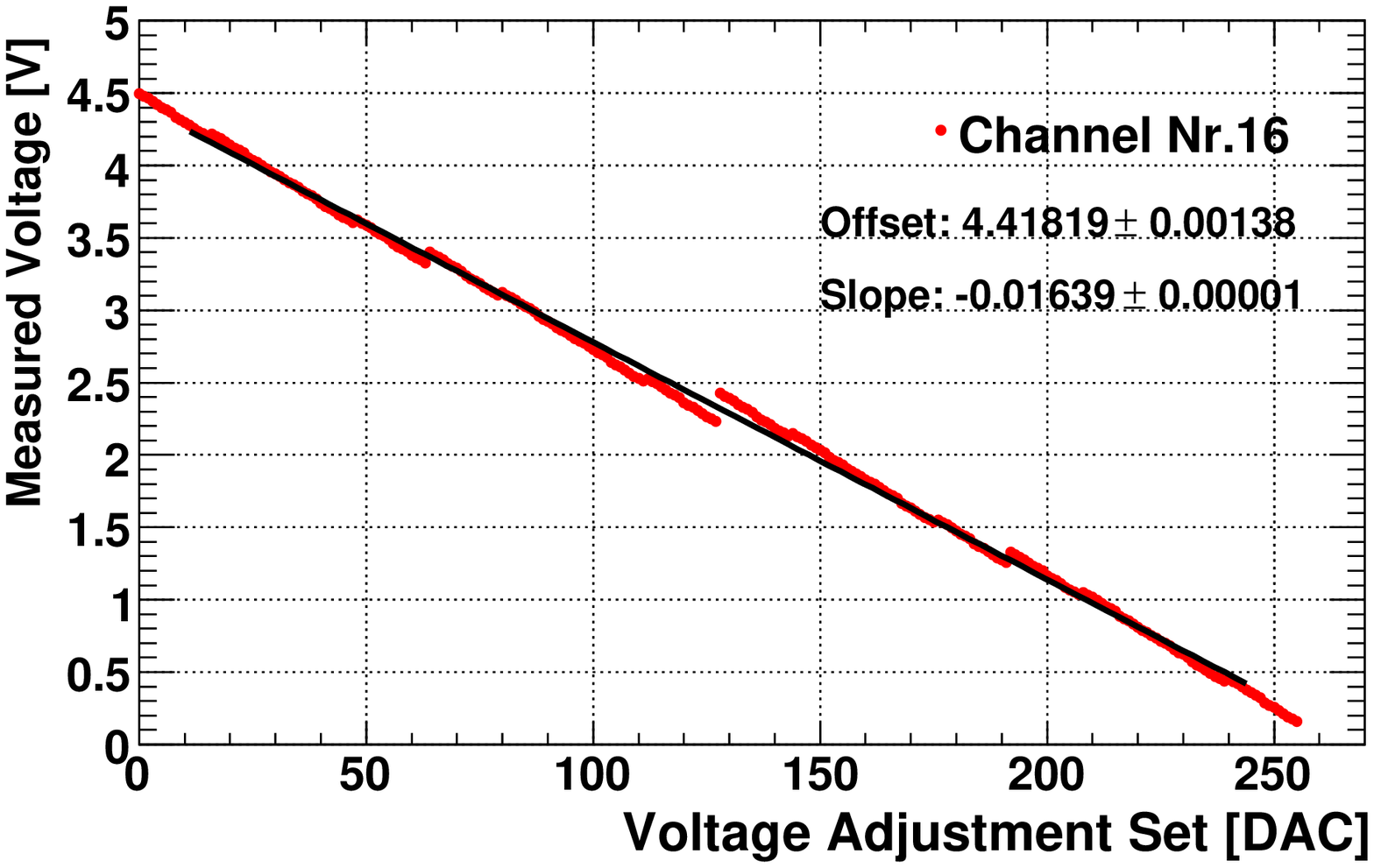} \\
  \includegraphics[height=5.cm, width=6.0cm]{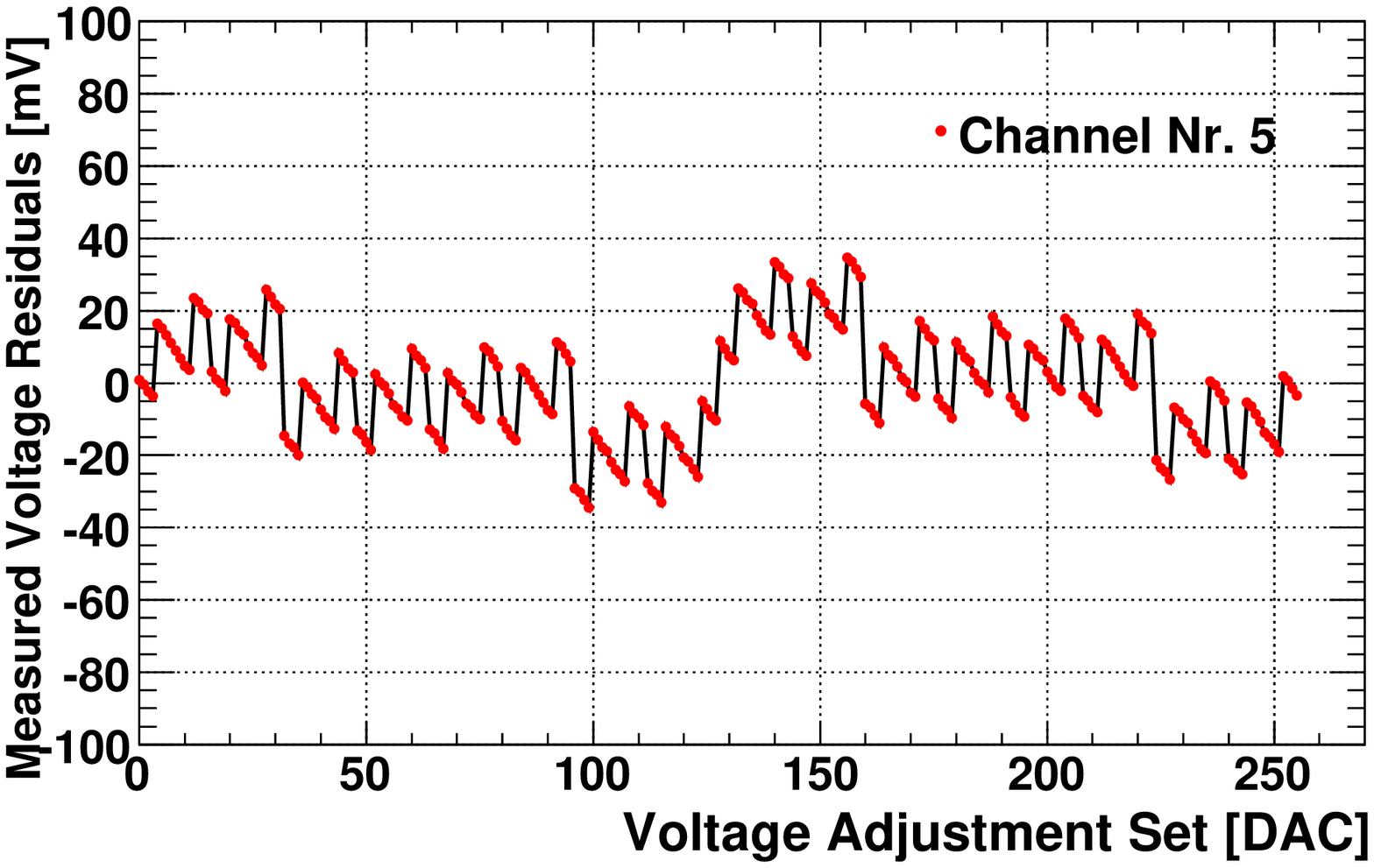} 
  \includegraphics[height=5.cm, width=6.0cm]{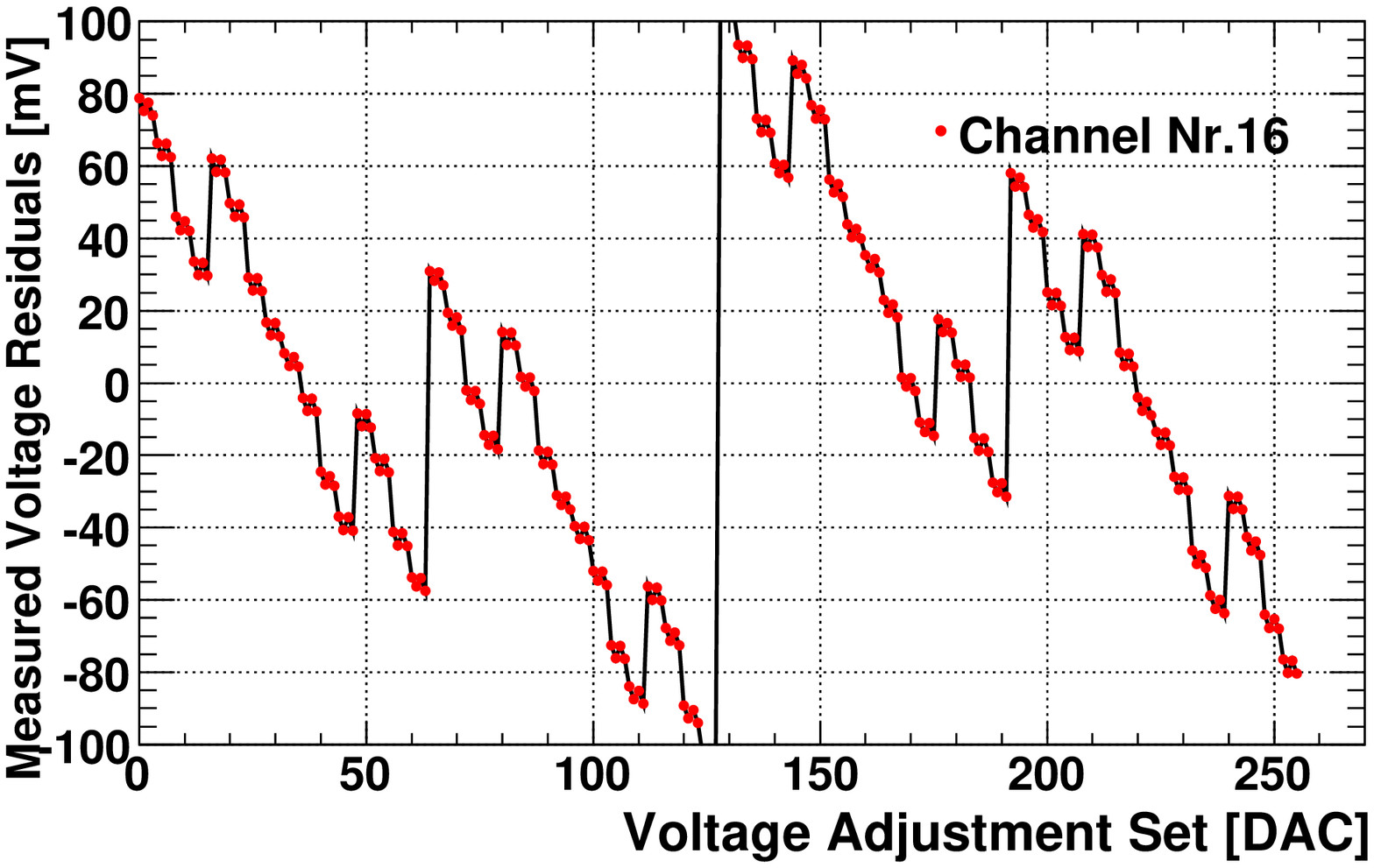} \\
  \caption{Upper Panels: Calibration of the high voltage DAC adjustment.
           Lower Panels: The residuals of the measurements with 
           respect to the function obtained by a linear fit.}
  \label{fig:InputHvDAC_Scan}
\end{figure}

The measurements are linearly fitted in the range 10-245 DAC units, 
and the residual of each measurement to the resulted linear function is
calculated. 
A large channel by channel variation in the residuals size 
is observed, and two extreme cases are reported 
in Fig.~\ref{fig:InputHvDAC_Scan}. 

\begin{figure}[t!]
 \begin{minipage}{6.5cm}
    \includegraphics[height=5.5cm, width=6.5cm]{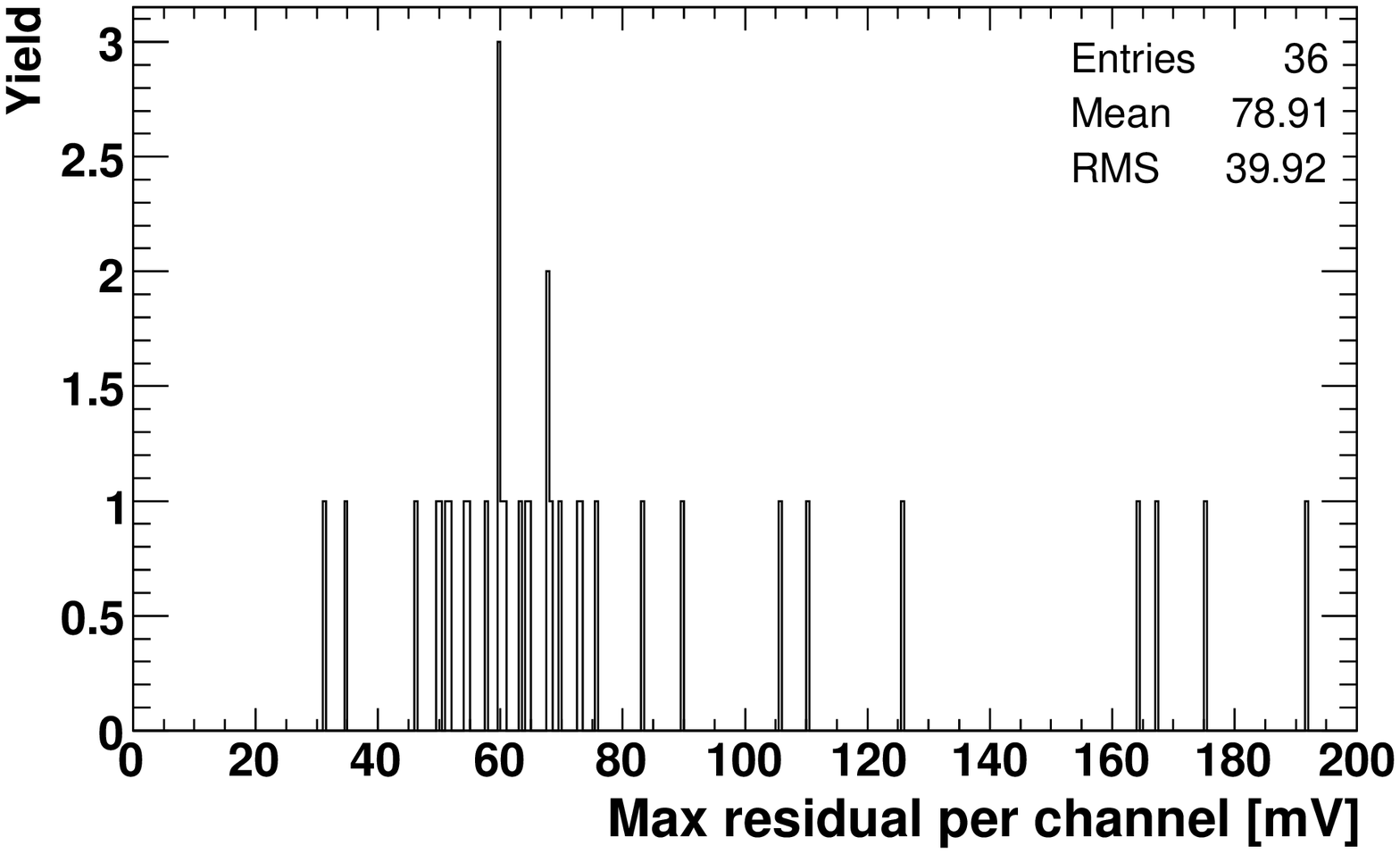} 
 \end{minipage}
 \begin{minipage}{5.5cm}
    \caption{Distribution of the deviation from linearity 
             of the high voltage adjustment. Each entry represents 
             the maximum deviation observed in one channel.}
    \label{fig:InputHvDAC_ScanResiduals}
 \end{minipage}
\end{figure}

The results show a differential non-linearity on average up to $70$ mV, and 
for some channels up to $200$ mV, 
Fig.~\ref{fig:InputHvDAC_ScanResiduals}, resulting in a relative gain 
change of approximatively $1.8\%$ and $5.2\%$~\cite{NILS}, respectively, and 
thus providing a potential systematic uncertainty of similar size to 
the energy calibration of the detector, 
in case the observed deviation cannot be systematically reproduced.

\begin{figure}[t!]
    \includegraphics[height=7.cm, width=10.5cm]
              {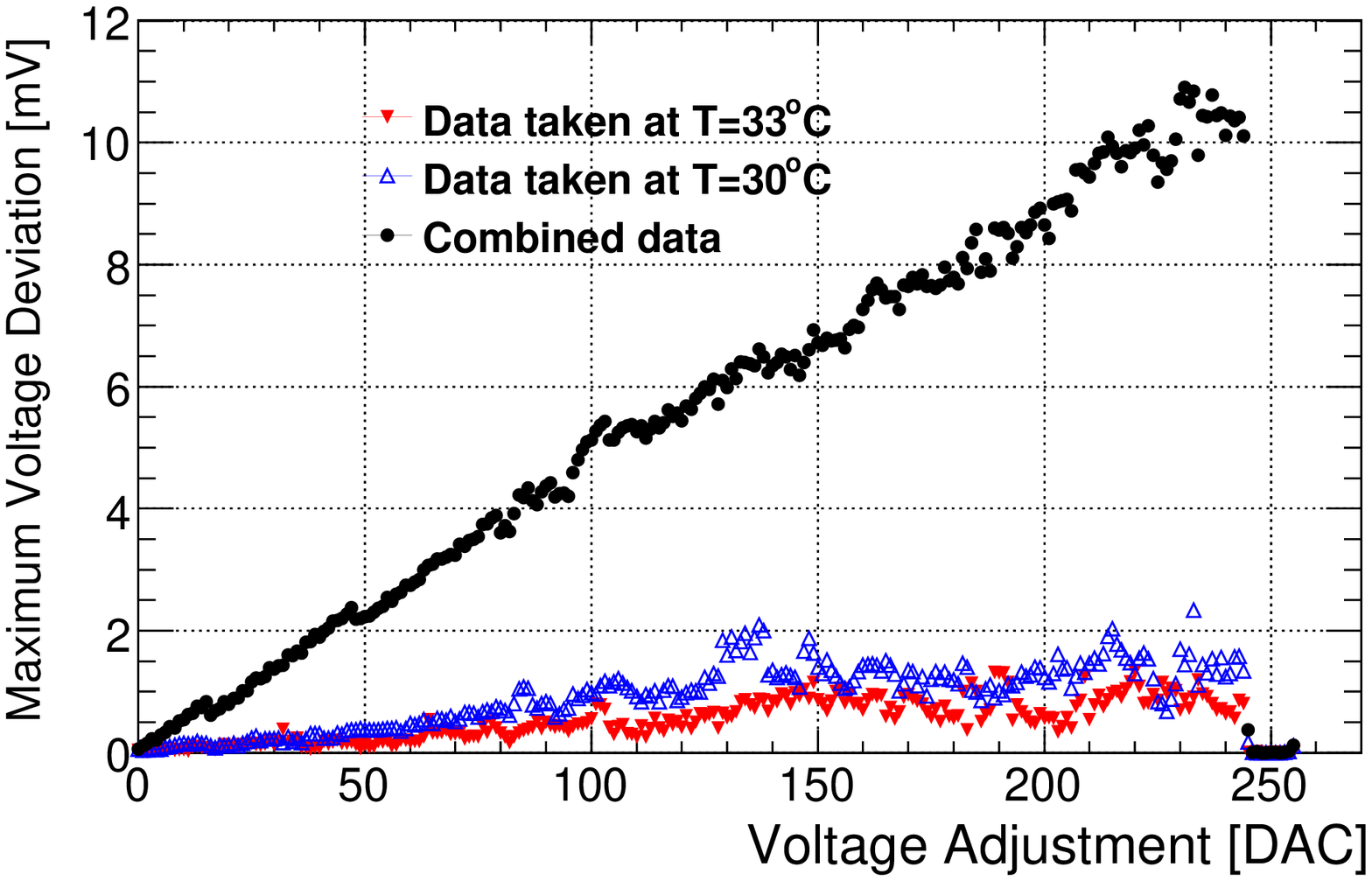} 
    \vspace{-0.5cm}
    \caption{The measured maximum deviation of the voltage adjustment 
             applied to the SPIROC channels is presented for two separate 
             DAC scans performed at different temperatures. When combining 
             the measurements
             of the two data sets the spread increases up to $10$ mV.}
    \label{fig:InputHvDAC_SystematicScan}
\end{figure}

To investigate the reproducibility of the voltage tuning applied 
to the ASIC channels, two series of DAC scans (five scan per series) 
were performed for one channel in separate periods characterised by 
a temperature variation of three degrees Celsius. 
Within each data set, the maximum deviation between the measured 
voltage values was calculated for each DAC value, and is 
presented in Fig.~\ref{fig:InputHvDAC_SystematicScan}.
As expected, for each series of measurements the spread increases with 
the DAC value, i.e. with the amount of current in the DAC switch, 
resulting in a maximum systematical deviation up to one millivolt. 
As a consequence, the observed large differential non-linearity 
can be corrected via a calibration of channel by channel HV  
adjustment DAC.

Although a negligible deviation is observed among the 
measurements, when combining the two data sets the maximum spread 
of the measurements increases up to $10$ mV. This can be possibly 
interpreted as due to the large variation of the temperature 
(up to three Celsius degrees) measured during the systematical 
studies, resulting in a larger applied voltage value observed at a lower 
temperature. In this case, the observed change of voltage adjustment value 
could be corrected via a proper calibration.
The calibration might not be needed in case of small temperature 
gradient during data taking, since the observed 
$10$ mV systematical spread is still small compared to the overvoltage 
of SiPMs, typically of $2$-$4$ V for the devices used in current
AHCAL test beam operations~\cite{NILS}. Thus, this spread would correspond 
to a gain change below $1$\%. 
%

\section{Low Gain - High Gain Coupling}
\begin{figure}[t!]
 \begin{minipage}{6.5cm}
    \includegraphics[height=5.5cm, width=6.5cm]{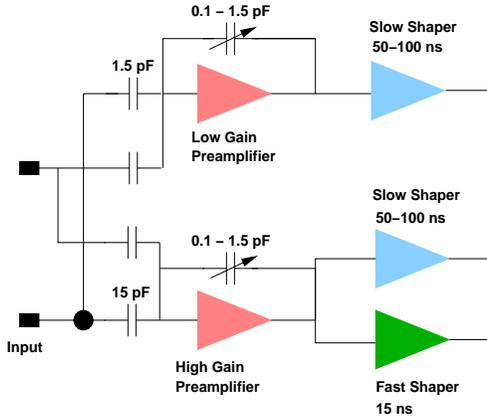} 
 \end{minipage}
 \begin{minipage}{5.cm}
    \caption{Electrical diagram of the high gain and low gain paths 
             per single channel in the ASIC chip.}
    \label{fig:HG_LH_Coupling}
 \end{minipage}
\end{figure}

In each input channel of the ASIC the low and high 
gain paths are electrically coupled, Fig~\ref{fig:HG_LH_Coupling}.
As a result, the charge sharing between the two paths is influenced 
by the feedback capacitance value set. 

\begin{figure}[t!]
  \includegraphics[height=5.5cm, width=6.0cm]{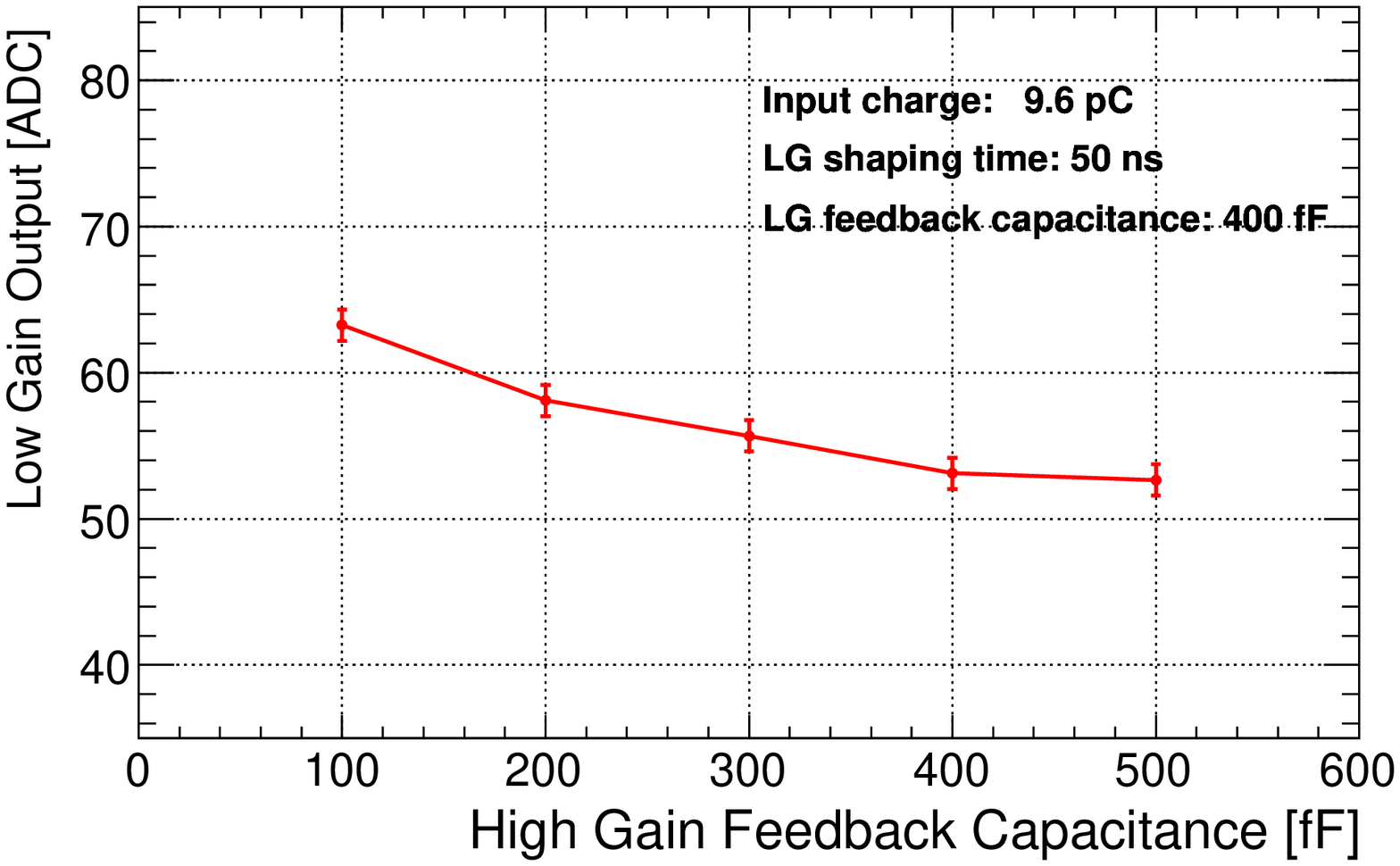} 
  \includegraphics[height=5.5cm, width=6.0cm]{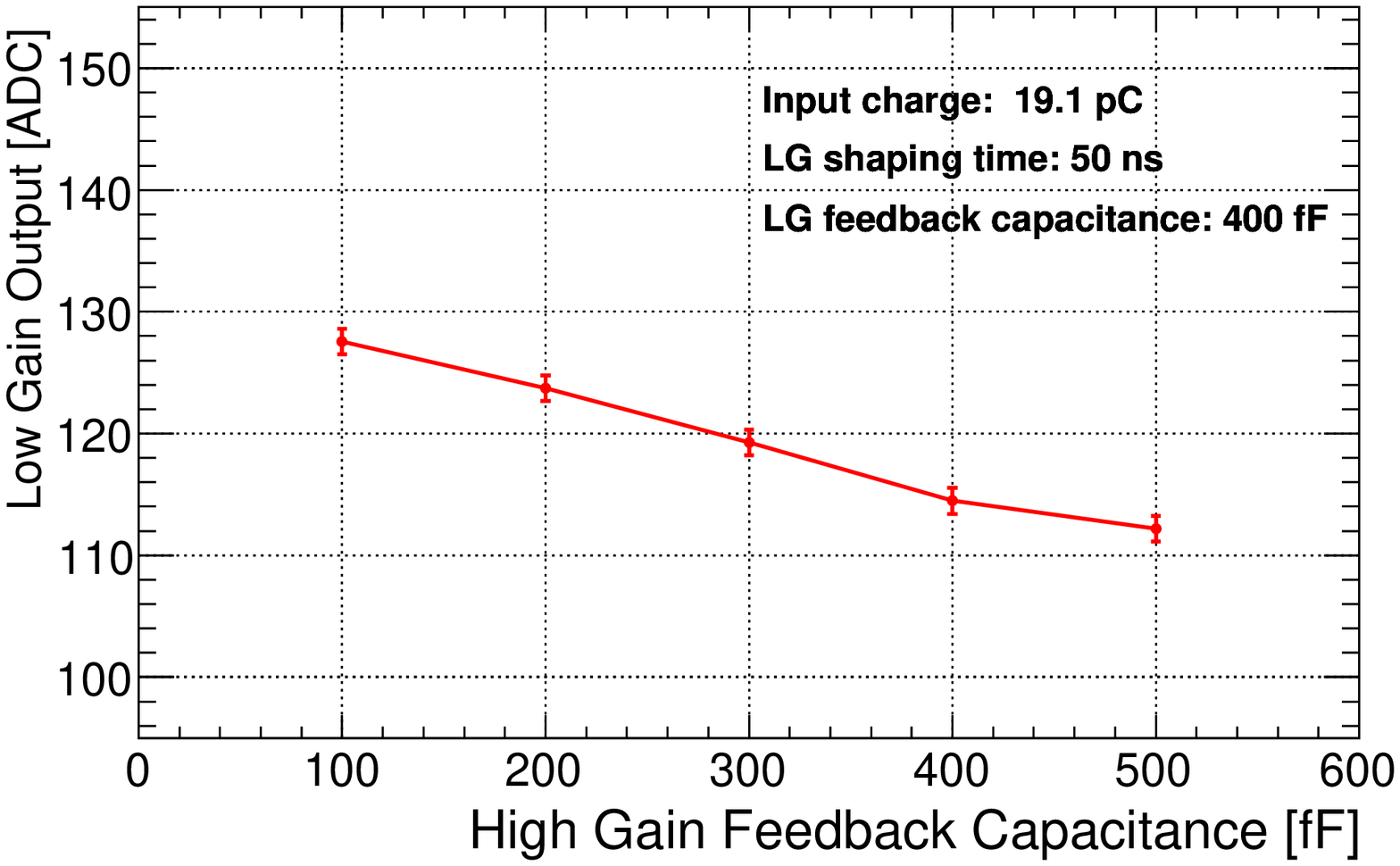} \\
  \vspace{-0.75cm}
  \caption{The output signal (in ADC units) is measured in the low 
          gain path while changing the feedback capacitance value
          for the high gain amplifier. Here presented are two 
          measurements obtained using different values of input 
          charge in the ASIC.}
  \label{fig:HG_LG_Effect}
\end{figure}

In order to estimate the effects of the high-low gain coupling, 
the output signal in one path was measured for different 
capacitance values in the other preamplifier line. 
No significant effect is observed in the output signal of the 
high gain line while changing the amplification in the other 
line. Instead, a sizable effect (up to $10\%$) is visible when 
measuring the output signal in physics mode and changing the 
amplification in the high gain line. An example of the observed
correlation is presented in Fig.~\ref{fig:HG_LG_Effect} for different
values of the charge injected in the ASIC (using a $6.4$ dB 
attenuator at the pulse generator output).
%

\section{Cross Talk between Input Channels}
%
The chip will handle $36$ incoming signal lines. According to 
previous measurements~\cite{ORSAY} the estimated cross-talk 
among the neighbouring channels is below $0.3\%$ for an injected charge 
of $15$ pC. The measurement in~\cite{ORSAY} was presented for 
few injected charge values only. This investigation 
of the cross-talk was done performing a wider scan of 
injected charge.
While increasing the injected charge in steps of $1$ pC 
from one to $40$ pC, corresponding approximately to the range 
$1-33$ mips, the output signal 
(pedestal subtracted) was measured in the neighbouring channels. 
The ASIC was run in low gain mode (physics mode) at $50$ ns of 
shaping time, and $400$ fF of feedback capacitance. 

\begin{figure}[t!]
  \includegraphics[height=8.cm, width=11.0cm]{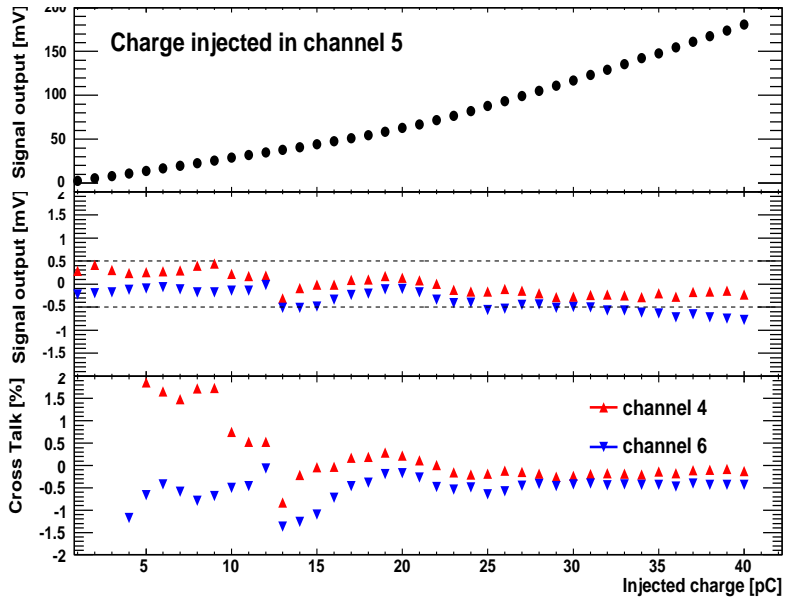} 
  \caption{The measurement of the cross-talk is presented for the 
           low gain mode at $50$ ns of shaping time, and $400$ fF of 
           feedback capacitance.}
  \label{fig:CROSS_TALK}
\end{figure}

An example of the performed measurements is presented in 
Fig.~\ref{fig:CROSS_TALK}. The output signal in the line
where the charge was injected is shown in the upper panel. 
The reason of the observed non-linearity was found in the 
track and hold switch (see Sec.~\ref{sec:TH}). 
In the neighbouring lines the measured signal is typically 
within $\pm 0.5$ mV in the analysed input charge range (middle panel), 
confirming 
a negligible charge leakage between the channels.
A small trend is visible, within the $\pm 0.5$ mV range, moving 
from small to large injected charge.
For each charge value the cross-talk (bottom panel) in one channel 
was calculated dividing its signal (held at its maximum amplitude) 
to the corresponding signal in the line where the charge was injected. 
As expected by using this method, the cross-talk inflates at small 
input charge values. At the $15$ pC of injected charge, the measurement 
is close to what reported in~\cite{ORSAY}.
%

\section{Track and Hold Switch}
\begin{figure}[t!]
  \begin{minipage}{6.5cm}
   \includegraphics[height=5.5cm,width=6.5cm]{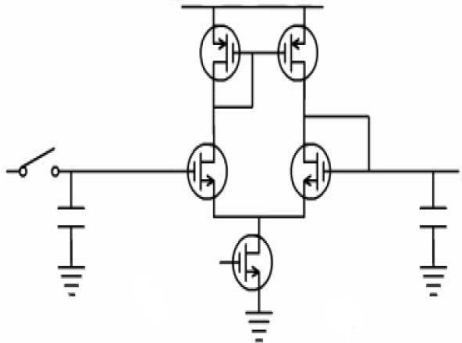}
  \end{minipage}
  \begin{minipage}{5.5cm}
   \caption{Electrical diagram of the Track and Hold switch.}
   \label{fig:TH}
  \end{minipage}
\end{figure}

%
\label{sec:TH}
The chip has to save the amplitude of the pre-amplified and shaped 
signal at its peaking time (arrival time of signal maximum). This 
is achieved using the track and hold switch, Fig.~\ref{fig:TH}, 
which holds the signal at the amplitude corresponding to a provided
holding-trigger arrival, to be tuned to hold the signal at its 
maximum amplitude. 

\begin{figure}[t!]
\begin{center}
  \includegraphics[height=8.5cm, width=12.0cm]{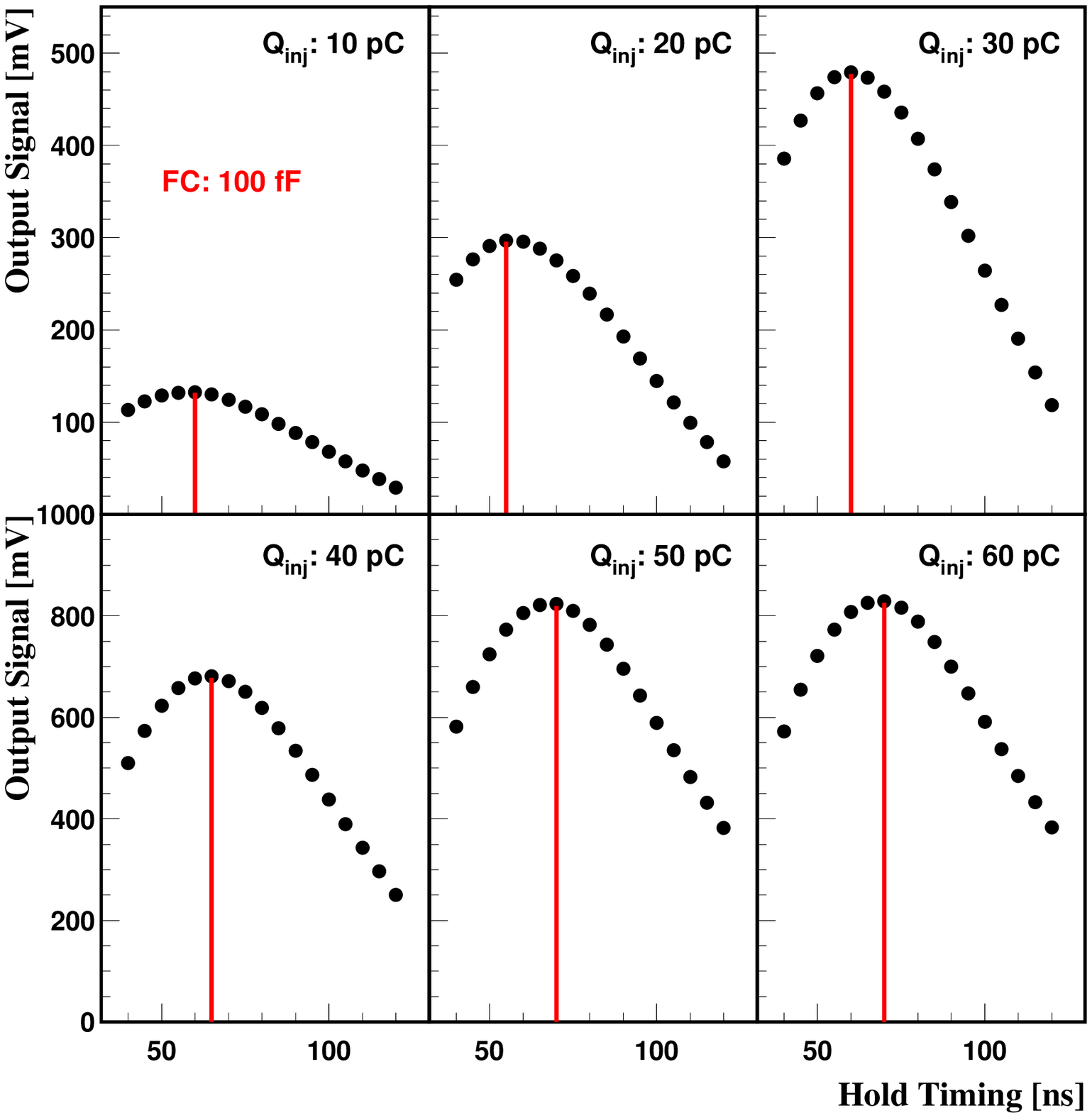} \\
  \vspace{-0.35cm}                    
  \includegraphics[height=8.5cm, width=12.0cm]{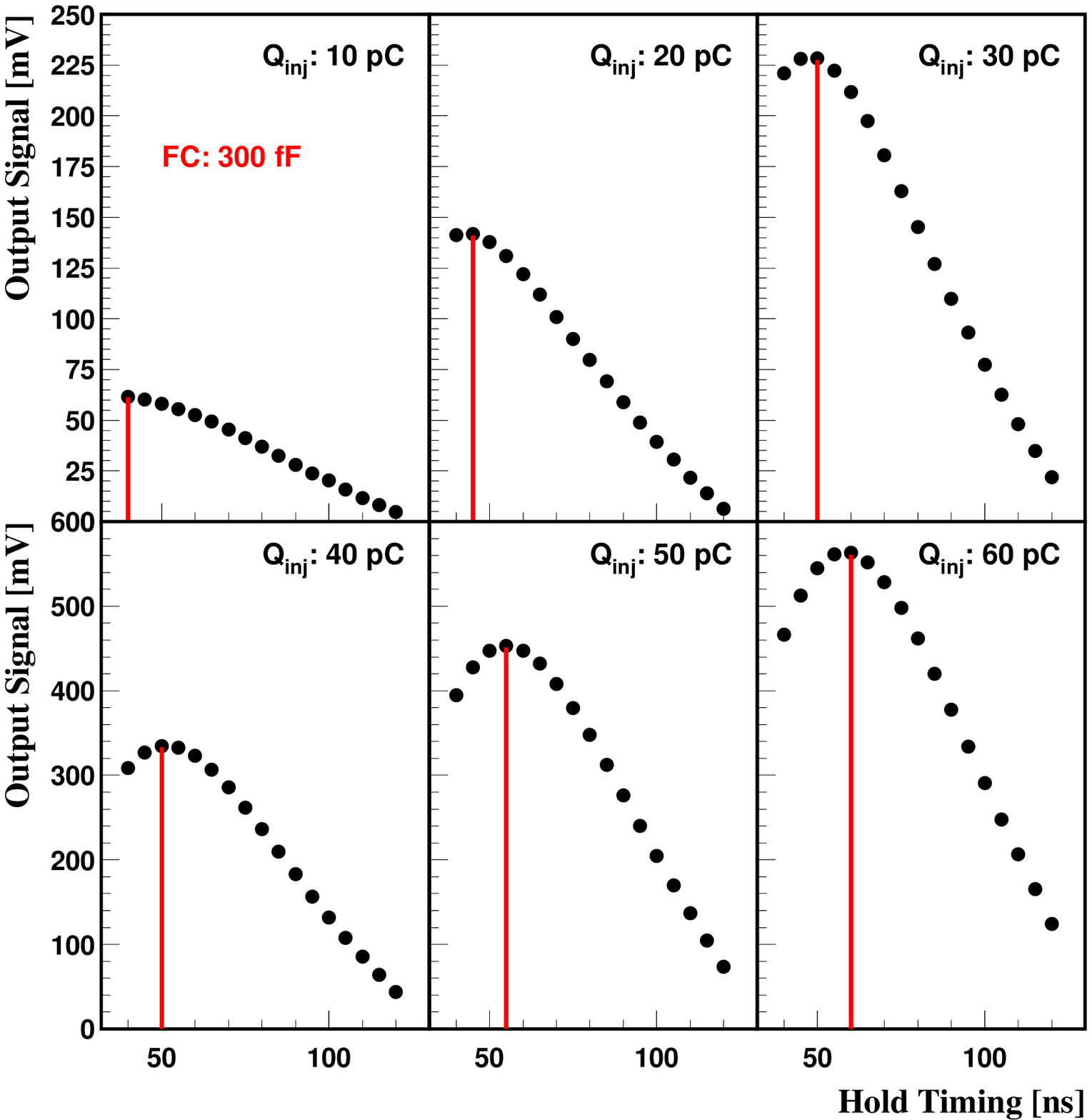} \\
  \vspace{-0.25cm}
  \caption{The output signal from the ASIC is measured in low gain mode 
           for $50$ ns shaping time and for different input charge values, 
           while performing a scan in the hold arrival time. No attenuator 
           on the input line was used.}
  \label{fig:TH_SCAN_100_300}
\end{center}
\end{figure}

The peaking time is not expected to depend on 
the amount of injected charge (mainly depending instead on the shaper 
type~\cite{KOWALSKI}), and the choice of the holding-trigger should 
hold for all values of input charge. Surprisingly, it was observed 
that the peaking time varies with the charge. The signal amplitude
was measured with the external ADC while performing a scan of the 
holding-trigger arrival time, in both low and high gain modes, for $50$ 
ns shaping time, and for all possible values of the feedback capacitance. 
An example of these measurements is shown in  
Fig.~\ref{fig:TH_SCAN_100_300}. In these 
test-bench measurements the hold signal is generated by the external 
voltage source. Its arrival time can be arbitrarily varied to anticipate 
or delay it with respect to peaking amplitude of the injected charge.
The measurements show that the peaking time (displayed with a vertical
line in each plot) increases with increasing
injected charge (while remaining constant only after reaching saturation),  
thus suggesting that either the peaking time is charge dependent 
(contrary to the expectations) or that the track and hold switch 
somehow interferes with the signal developing in time. 

After investigation by the Orsay group, a possible explanation of 
the observed feature in the small charge region was proposed~\cite{LUDOVIC}.  
Due to the small bias transistor size in the track and hold buffer 
(bottom transistor in Fig.~\ref{fig:TH}) and 
to the large parasitic capacitance at its output, small input charge 
will make the buffer enter the non-linear working region. The 
investigation with simulations is on-going by the Orsay group.

\section{Dynamic Range, Linearity, and Gain of the ASIC}
%
During normal AHCAL data taking (low gain) the dynamic range of the ASIC 
should cover the region between one mip and $\approx 77$ mips 
($1156$ pixels per SiPM / $15$ pixels firing per mip), considering 
saturation effects in the SiPM due to the limit number of 
pixels and to the pixel recovery time. Assuming a SiPM gain of
the order of $5\cdot 10^5$ (corresponding to $0.08$ pC charge 
generated per pixel), the covered dynamic mips range corresponds
to the region between $1.2$ and \mbox{$92.5$ pC}. 
It should be noted that this upper charge limit depends
on the specific photodetector gain (MEPhy/Pulsar devices, considered
here, have a pixel gain variation between $0.25 \cdot 10^6$ and 
$1.00 \cdot 10^6$).

The chip preamplifier gain should be chosen such to provide a linear 
response in the whole energy range. On this purpose, a scan was 
performed over all feedback capacitance values possible in low gain 
mode, while increasing the amount of injected charge. 
For each input charge, it was chosen the hold-trigger timing corresponding 
to the maximum amplitude for the output signal measured at the 
external ADC. This choice 
was motivated by the impossibility of performing such a time 
consuming multi-parameter scan extracting the proper 
holding-time via the oscilloscope, without an automatised procedure
(each single point of the scan was systematically measured several 
times). Systematic effects arising by the track and hold switch,  
as discussed in Sec.~\ref{sec:TH}, might affects the results. 

\begin{figure}[t!]
\begin{center}
  \includegraphics[height=8.cm, width=11.0cm]{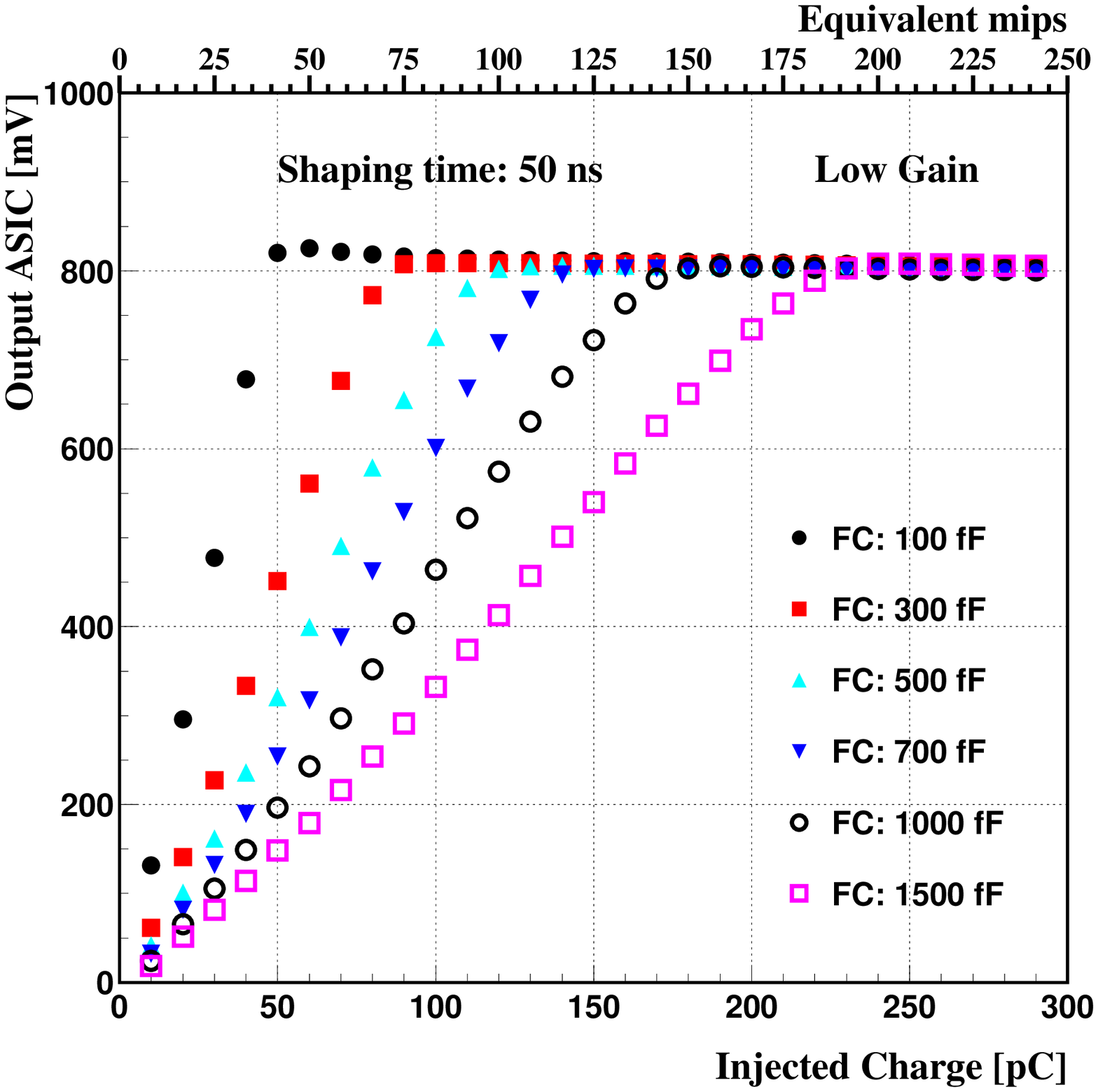} \\
  \vspace{-0.2cm}
  \includegraphics[height=8.cm, width=11.0cm]{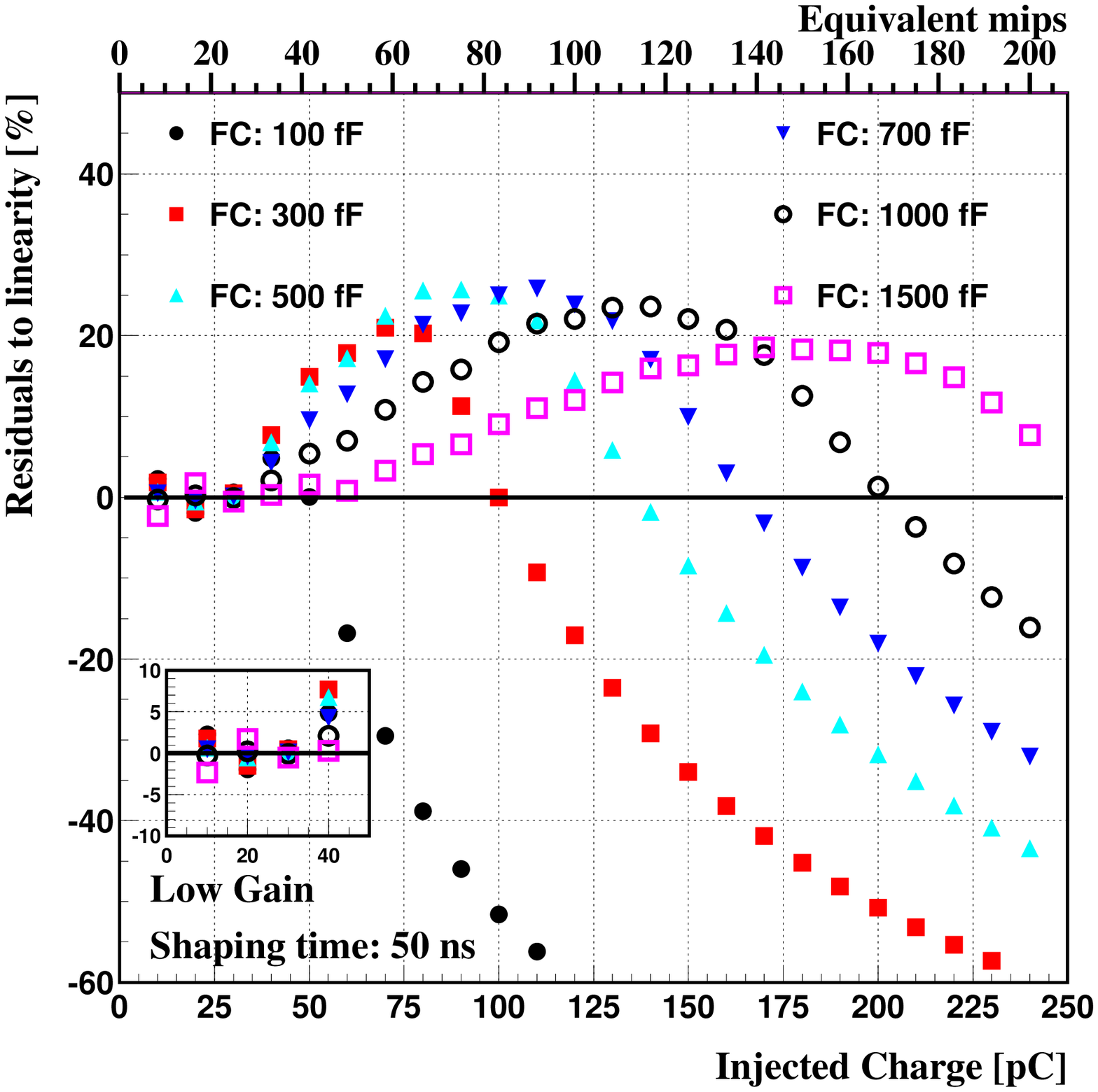} 
  \vspace{-0.6cm}
  \caption{The output signal from the ASIC is measured in low gain mode 
           for $50$ ns shaping time and for different input charge values.  
           Upper panel: Results obtained using different feedback 
           capacitance values are compared. Bottom Panel: Residuals to 
           linearity calculated as described in the text.}
  \label{fig:LOW_GAIN_SCAN}
\end{center}
\end{figure}

The results of this scan, presented in the upper panel of 
Fig.~\ref{fig:LOW_GAIN_SCAN} for $50$ ns shaping time, indicate 
that the minimum feedback capacitance value needed to cover the 
required SiPM dynamic range ($77$ mips) is $400$ fF. Above this value the 
maximum expected mip signal can be processed before the ASIC runs 
into saturation. As a reference, the values of the injected charge 
are also reported in terms of equivalent AHCAL mips values.

The sets of data were linearly fit in the range $10$-$30$ pC
and the residuals to the linearity were calculated dividing the deviation 
of the measured points from the fit results over the fit values. 
A non linearity 
up to $20\%$ is observed with increasing injected 
charge, bottom panel of Fig.~\ref{fig:LOW_GAIN_SCAN}. 
The region where the ASIC saturation appears (anyway not relevant for the
forseen operation of the chip) shows a consistent drop of the residuals.

\begin{figure}[t!]
\begin{center}
  \includegraphics[height=8.cm, width=11.0cm]{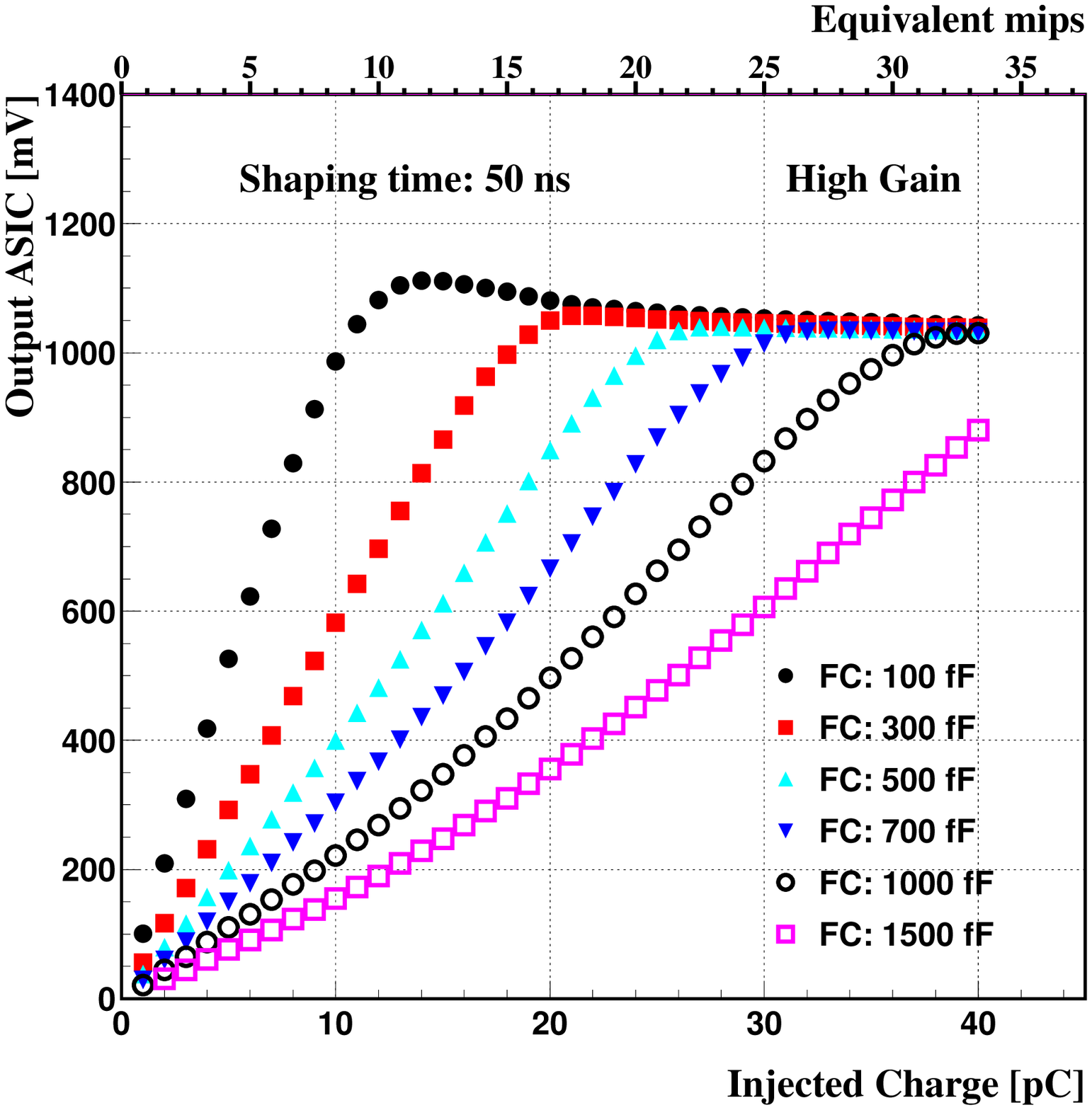} \\
  \vspace{-0.2cm}
  \includegraphics[height=8.cm, width=11.0cm]{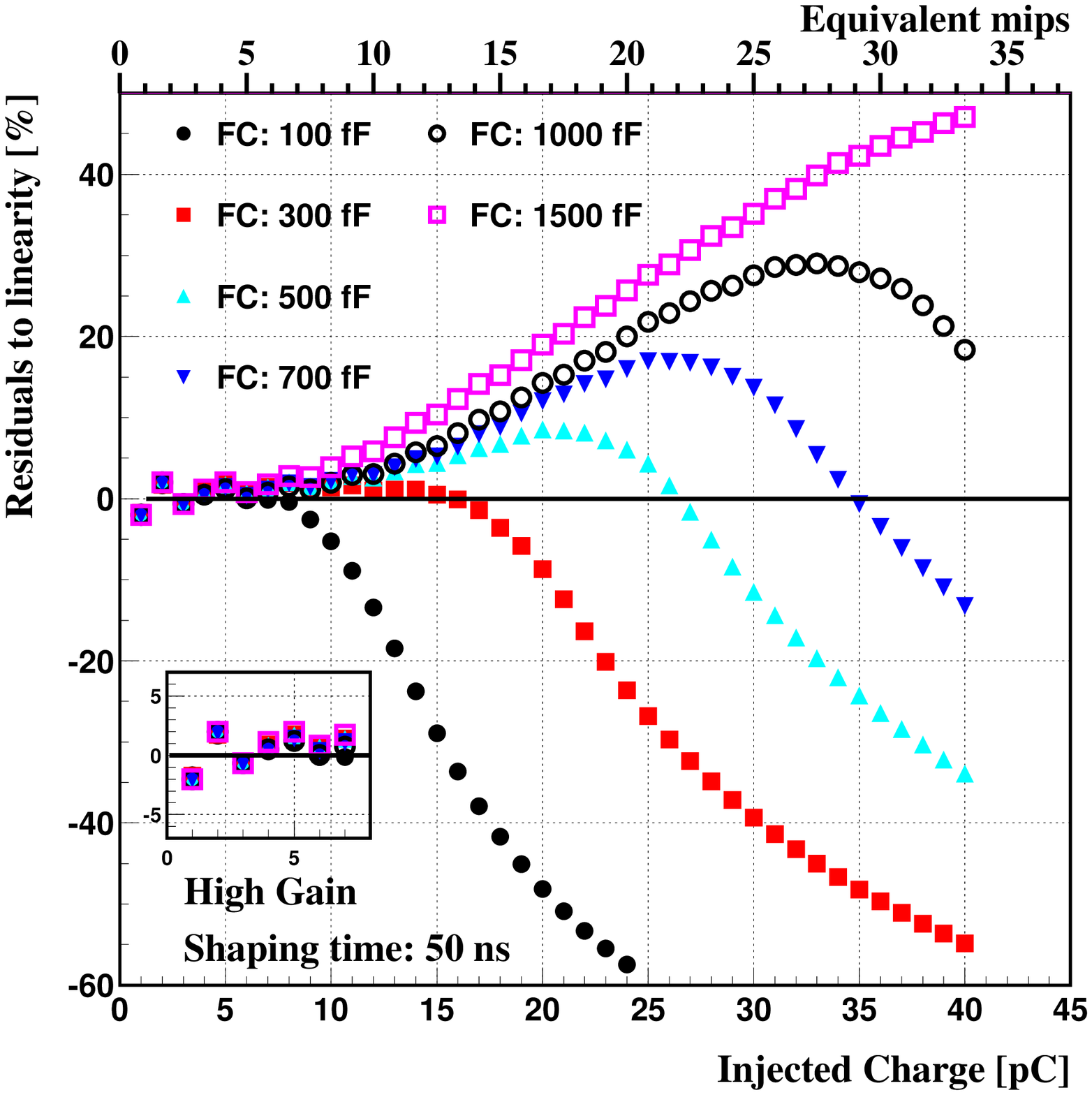} 
  \vspace{-0.6cm}
  \caption{The output signal from the ASIC is measured in high gain mode 
           for $50$ ns shaping time and for different input charge values.  
           Upper panel: Results obtained using different feedback 
           capacitance values are compared. Bottom Panel: Residuals to 
           linearity calculated as described in the text.
           A $20$ dB attenuator at the input line was used during 
           these measurements.}
  \label{fig:HIGH_GAIN_SCAN}
\end{center}
\end{figure}

Similar measurements were performed operating the chip in high gain 
mode, Fig.~\ref{fig:HIGH_GAIN_SCAN}. 
In this mode, used for calibration, it is preferable to use the 
largest (smaller) gain (feedback capacitance) to investigate the 
SiPM single-pixel spectra.  
The results show that for $100$ fF of feedback capacitance the 
saturation is reached for values of input charge above $10$ pC, 
corresponding to a dynamic range up to $\approx 8$ mips, enough 
for performing the calibrations. Note that here the linear fit
was performed in the range $1$-$3$ pC. 

The observed non linearity is mainly due to the mentioned features 
of the track and hold switch, which are under investigation and will be 
possibly cured in the next generation of the chip. In addition to the 
observed injected charge dependence of the peaking time, it was also 
observed that the peaking amplitude obtained via the track and hold
switch is typically larger than what observed when the signal is not 
held.

\begin{figure}[t!]
\begin{center}
  \includegraphics[height=8.cm, width=11.0cm]{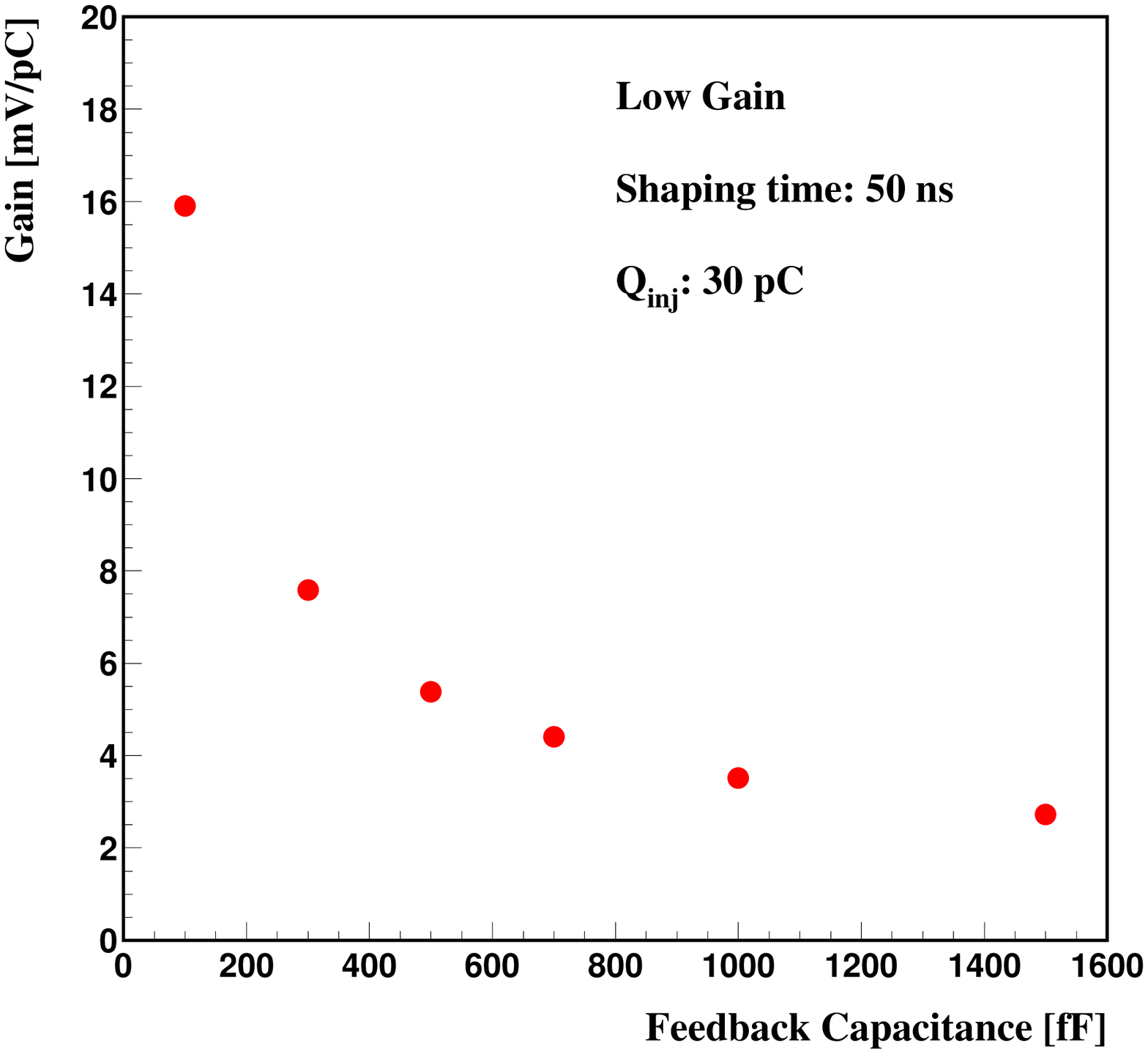} \\
  \vspace{-0.9cm}
  \includegraphics[height=8.cm, width=11.0cm]{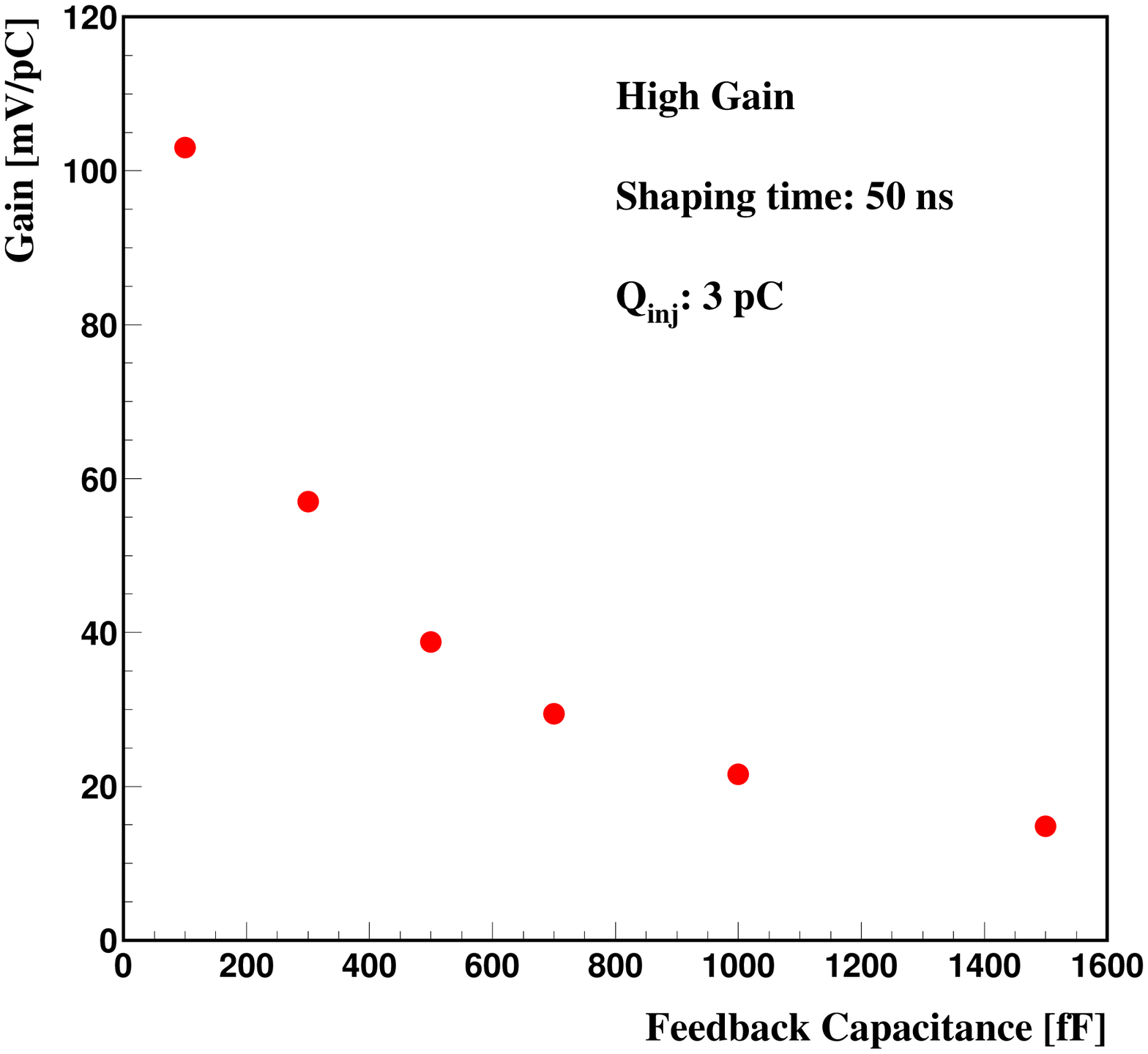} 
  \vspace{-0.7cm}
  \caption{The gain of the ASIC is measured in both low and high gain modes 
           for $50$ ns shaping time and for $30$ pC and $3$ pC of 
           injected charge, respectively.}
  \label{fig:GAIN_FACTOR}
\end{center}
\end{figure}

The gain of the chip was calculated out of the same data according 
to the formula $G_{ASIC} = \frac{V_{output}}{Q_{input}}$ (in $\frac{mV}{pC}$
units), and is presented in Fig.~\ref{fig:GAIN_FACTOR} for different 
values of the feedback capacitance $C_{FC}$ in both low and high gain modes at 
$50$ ns shaping time.

In the low (high) gain mode the chip amplification factor was measured 
for $30$ pC ($3$ pC) of injected charge, a value not sizably 
affected by the non-linearity induced by the track and hold switch.
The expected $1/C_{FC}$ functional dependence of the gain~\cite{LEO} 
is observed, quantitatively in agreement with the 
measurements presented in ~\cite{ORSAY} and~\cite{LUDOVIC_TALK}.

\section{Amplitude Dependence on Shaping Time}
\begin{figure}[t!]
\begin{center}
  \includegraphics[height=8.cm, width=11.0cm]{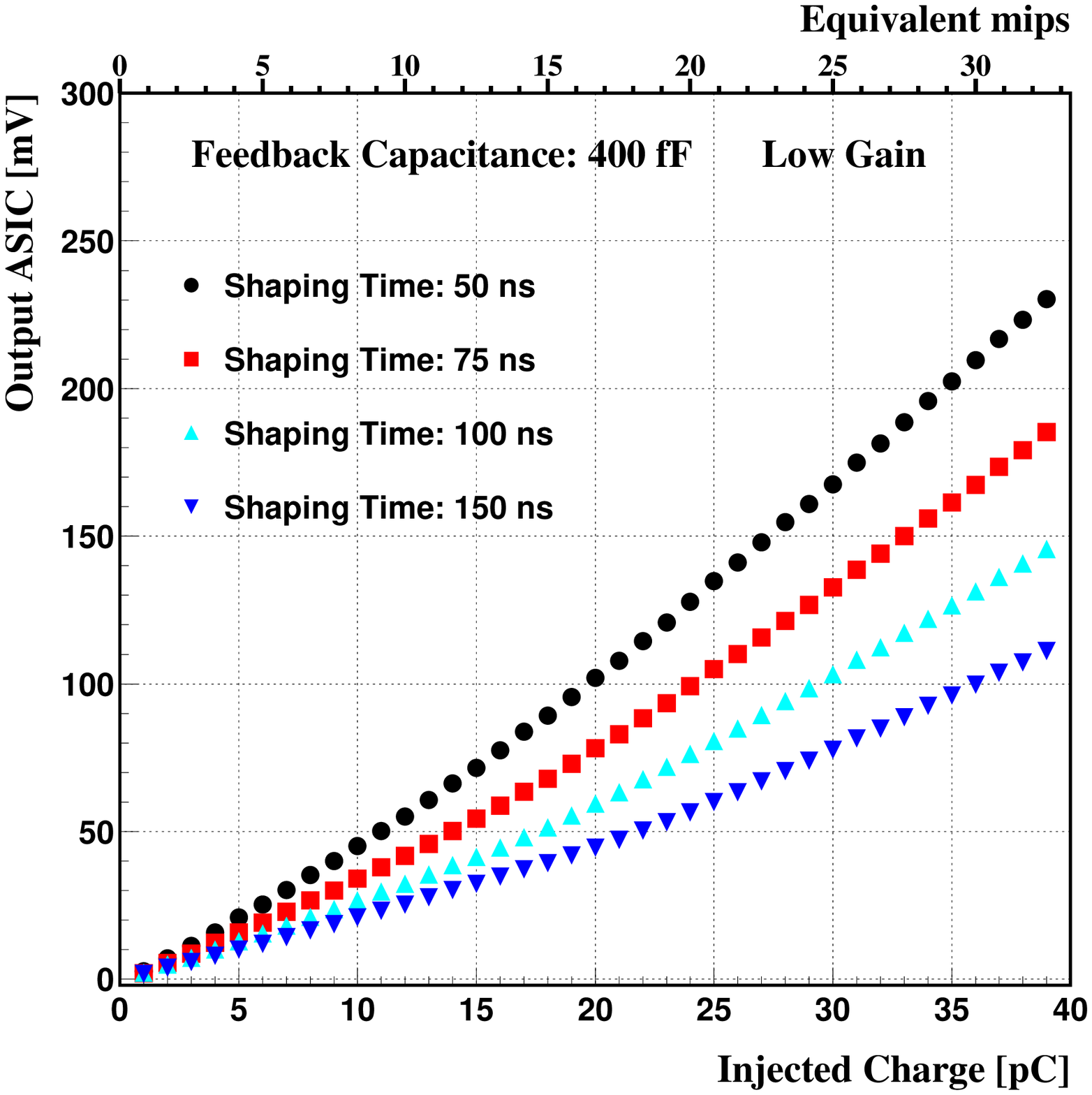} \\
  \includegraphics[height=8.cm,width=11.0cm]{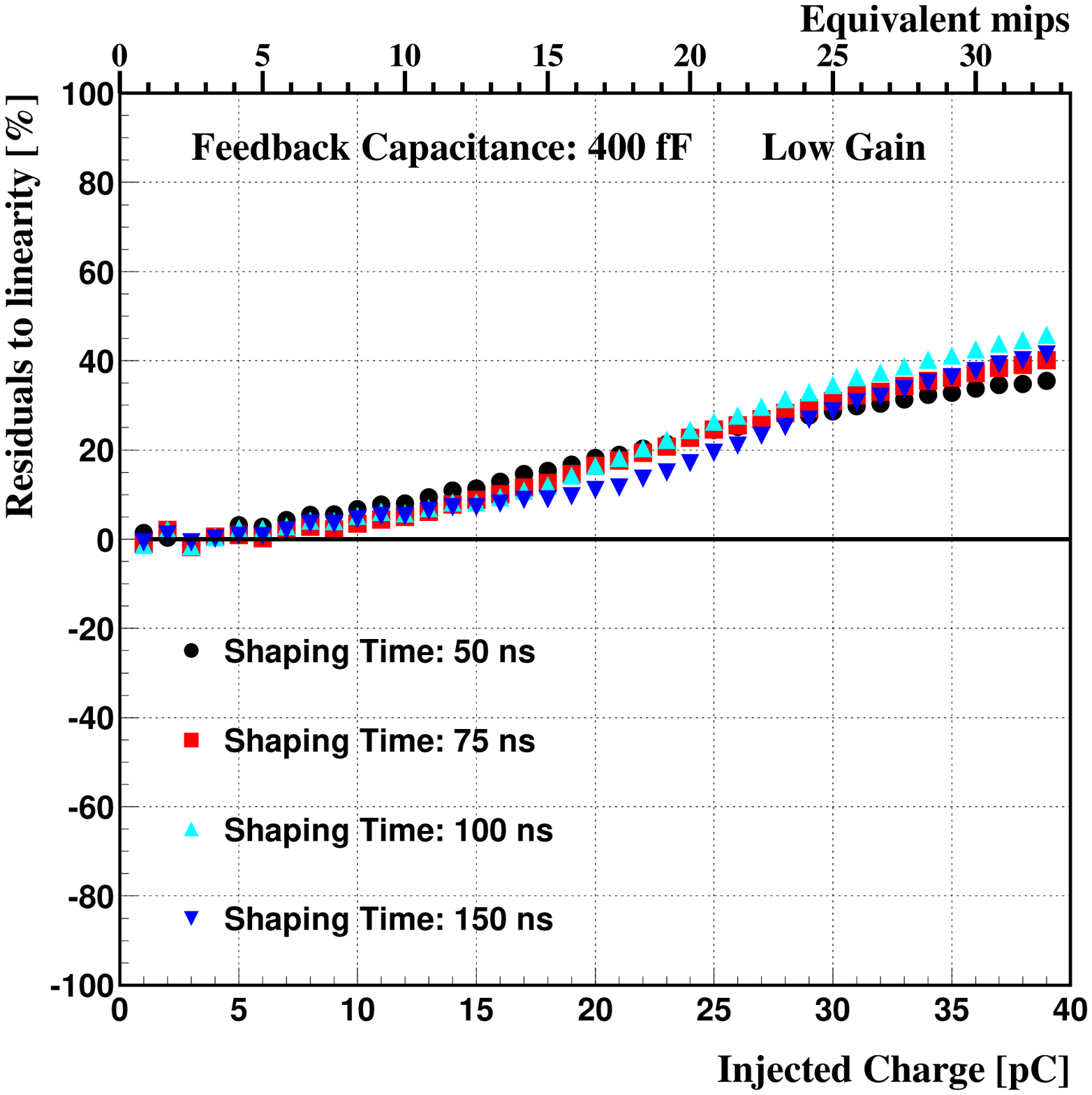}
  \vspace{-0.4cm}
  \caption{Upper panel: The output signal from the ASIC is measured in 
           low gain mode for $400$ fF feedback capacitance, and for 
           different values of injected charge and of shaping time.
           Bottom Panel: Residuals to linearity calculated as described 
           in the text.
           A $20$ dB attenuator ($10^{-1}$X) at the input line was used 
           during these measurements.}
  \label{fig:SHAPING_SCAN}
\end{center}
\end{figure}

The effect of the shaping time value on the measured signal peaking 
amplitude was investigated. The amplitude was measured operating the 
ASIC in low gain mode for different values of the shaping time and 
for the feedback capacitance value of $400$ fF, 
Fig.~\ref{fig:SHAPING_SCAN}.
Differently from what presented in Fig.~\ref{fig:LOW_GAIN_SCAN}, here 
a $20$ dB attenuator was used, to present results at small injected 
charge in finer binning. Also, results are presented for charge values 
down to $1$ pC. 

The signal peaking amplitude appears to increase with decreasing 
shaping time, in agreement with simulations. The sets of data were 
linearly fit in the range $1$-$4$ pC and the residuals to the linearity 
were calculated dividing the deviation
of the measured points from the fit results over the fit values.
A non linearity up to $40\%$ is observed for increasing injected
charge, bottom panel of Fig.~\ref{fig:SHAPING_SCAN}. In this region 
of injected charge values the non-linearity response (due to the track and
hold switch) appears to be weakly dependent on the chosen shaping time.
Note that a direct comparison with the results presented in the 
bottom panel of Fig.~\ref{fig:LOW_GAIN_SCAN} cannot be done, being 
different both the region considered in the linear fit and the smallest 
values of injected charge considered in these measurements.
%

\section{SiPM Single-Pixel Spectra}
%
\label{sec:single_pixel_spectra}
The possibility to operate the ASIC in auto-trigger mode during 
calibrations was shown to be feasible due to the 
low trigger jitter time walk, smaller than $1$ and $7$ ns, 
respectively (see Sec.~\ref{sec:trigger}). That possibility 
is confirmed by the following measurement of the single-pixel 
spectrum. 

A SiPM, flashed by an LED, was connected to the board, and the 
processed output signal was investigated. The device was operated 
at overvoltage larger ($+1.7$ V) than the nominal value, to increase
the cross-talk between the pixels (which is, on the other side,
not desirable for physics mode operations). 
The SiPM signal was measured using the auto-trigger mode described 
in Sec.7. After properly setting a discriminator threshold value, 
the generated trigger was used to trigger an external pulse 
generator (HP 8082A). One signal from the pulser was then used 
to open the gate of the external ADC module, and a second signal 
was input to the SPIROC board to hold the analogue signal from the 
SiPM at its peaking amplitude. 

\begin{figure}[t!]
\begin{center}
  \includegraphics[height=6.cm, width=11.0cm]
         {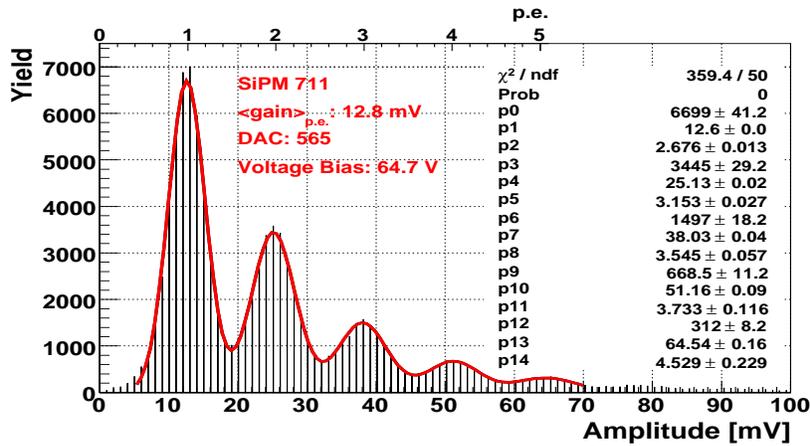} \\
  \vspace{-0.5cm}
  \caption{Single-pixel spectrum for the thermal noise obtained 
           by operating the SiPM in high gain ($13$ mV per pixel)
           and for the discriminator threshold level above the 
           pedestal. Superimposed is the gaussian fit to the
           peak structure of the spectrum.}
  \label{fig:ThermalNoise_HighGain}
\end{center}
\vspace{-0.5cm}
\end{figure}

When the discriminator threshold is set above the pedestal the 
contribution of the thermal noise dominates the measured spectrum, 
Fig.~\ref{fig:ThermalNoise_HighGain}, 
due to its large rate (typically around MHz).

\begin{figure}[b!]
\begin{center}
  \includegraphics[height=6.cm, width=11.0cm]
         {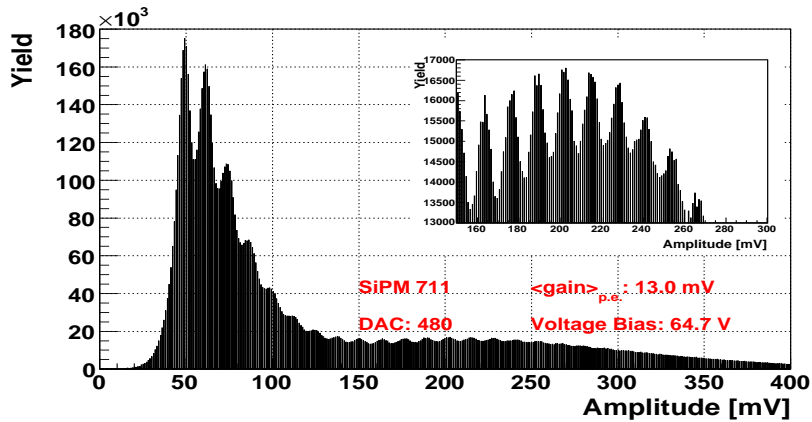} \\
  \vspace{-0.5cm}
  \caption{Single-pixel spectrum obtained by flashing the SiPM with LED
           light. The device was operated in high pixel gain ($13$ mV 
           per pixel) setting the SPIROC discriminator threshold level above 
           the third pixel peak, and running the ASIC in auto-trigger mode. 
           The contribution from the thermal noise and the LED events 
           dominates the first peaks and 
           the high tail of the spectrum, respectively.
           The peak structure of part of the spectrum for LED driven 
           events is also presented in the zoomed panel for better 
           visibility.}
  \label{fig:ThermalNoise_PlusSignal_HighGain}
\end{center}
\end{figure}

In order to observe the contribution of LED light induced events in the 
spectrum, the discriminator 
threshold was set to $480$ DAC units, corresponding to a value 
approximately above $3$ pixels for a measured gain of $13$ mV per pixel 
(operating the SiPM at $64.7$ V), 
Fig.~\ref{fig:ThermalNoise_PlusSignal_HighGain}. 
Also, the rate of the LED flashes was increased to $30$ kHz from the 
initial \mbox{$1$ kHz}.
Although the thermal noise contribution appears to be still sizable in 
this experimental configuration (dominating the spectrum first peaks), 
the LED signal appears in the spectrum 
at large amplitude values, and is also presented in the zoomed panel of 
the picture. The results show the single-pixel structure, and suggest 
that a fit of the spectrum (either the thermal noise or the signal 
contribution) might be possible, although its quality cannot be quantified 
at this stage of the analysis. 

\begin{figure}[t!]
\begin{center}
  \includegraphics[height=6.cm, width=11.0cm]
         {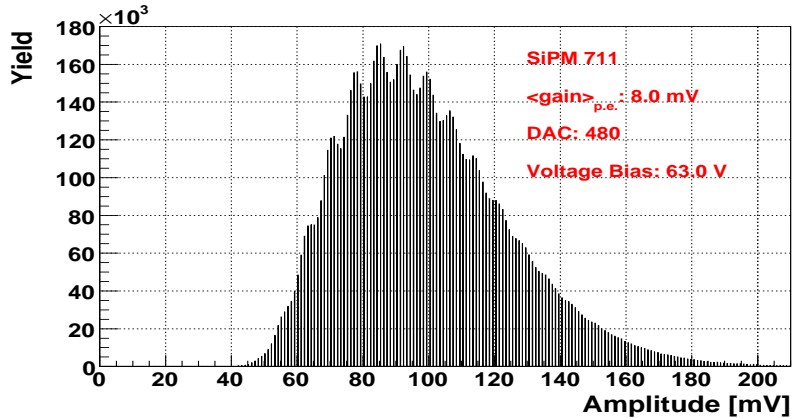} \\
  \vspace{-0.5cm}
  \caption{Single-pixel spectrum obtained by flashing the SiPM with LED
           light. The device was operated in low pixel gain ($8$ mV 
           per pixel) setting the SPIROC discriminator threshold level above 
           the fifth pixel peak, and running the ASIC in auto-trigger mode.
           Due to the lower device gain and to the high discriminator 
           threshold value the contribution from the thermal noise is 
           here strongly suppressed, and the spectrum is mainly populated
           by LED driven events.}
  \label{fig:ThermalNoise_PlusSignal_LowGain}
\end{center}
\vspace{-0.5cm}
\end{figure}

The above results were obtained for a quite large gain of the SiPM. To 
see the effects of a lower pixel gain value, the voltage bias applied to the 
device was decreased to $63.0$ V resulting in a gain around $8$ mV per pixel, 
which corresponds to approximately half a million electrons. 
The discriminator threshold value was kept unchanged, corresponding 
at about five pixels at this low pixel gain.  
The resulting spectrum is presented in 
Fig.~\ref{fig:ThermalNoise_PlusSignal_LowGain}.
Due to the lower device gain and to the high discriminator threshold 
value the contribution from the thermal noise is here strongly suppressed,
and the spectrum is mainly populated by LED driven events. 

\begin{figure}[t!]
\begin{center}
  \includegraphics[height=6.cm, width=11.0cm]
         {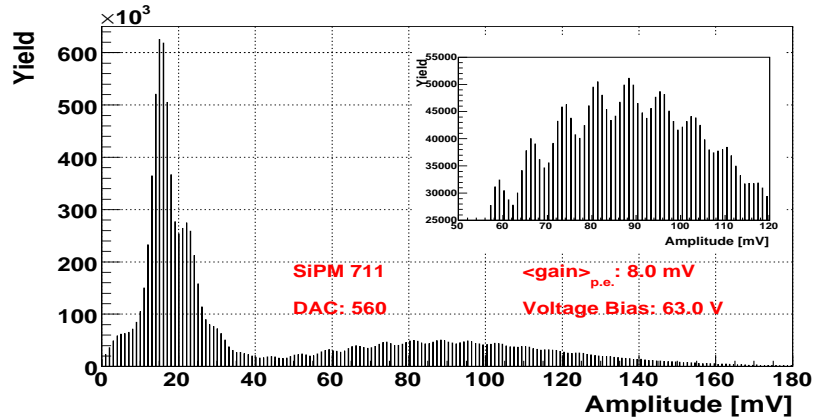} \\
  \vspace{-0.5cm}
  \caption{Single-pixel spectrum obtained by flashing the SiPM with LED
           light. The device was operated in low pixel gain ($8$ mV 
           per pixel) setting the SPIROC discriminator threshold level above 
           the pedestal, and running the ASIC in auto-trigger mode.
           The contribution from the thermal noise and the LED events 
           dominates the first peaks and 
           the high tail of the spectrum, respectively.
           The peak structure of part of the spectrum for LED driven 
           events is also presented in the zoomed panel for better 
           visibility.}
  \label{fig:ThermalNoise_PlusSignal_LowDAC_LowGain}
\end{center}
\end{figure}

To see the quality of the noise spectrum for this low SiPM gain, the 
threshold was set above the pedestal, 
Fig.~\ref{fig:ThermalNoise_PlusSignal_LowDAC_LowGain}.
At this low pixel gain value, the single-pixel structure of the thermal noise 
spectrum appears to be deteriorated, complicating the possibility to 
perform a fit of the pixel structure to calibrate the device. This 
complication might be overcome by operating the SiPM with a larger gain, 
as shown in Fig.~\ref{fig:ThermalNoise_HighGain}  . 

\begin{figure}[t!]
\begin{center}
  \includegraphics[height=6.cm, width=11.0cm]
         {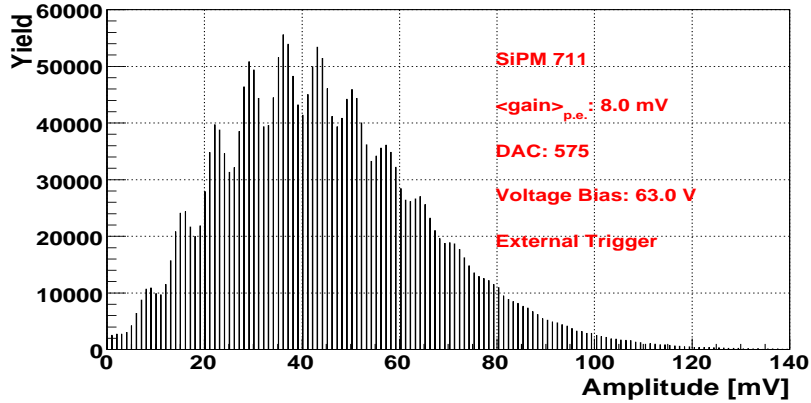} \\
  \vspace{-0.5cm}
  \caption{Typical SiPM single-pixel spectra obtained by flashing the SiPM
           with LED light. The device was operated in low pixel gain ($8$ mV
           per pixel) setting the SPIROC discriminator threshold level above
           the pedestal. 
           Here, the signal is held at its peaking amplitude 
           by an external trigger given by the main pulse generator.}
  \label{fig:EXT_TRIG_SiPM_Spectra}
\end{center}
\end{figure}

As a comparison, the measurement was repeated using an external 
trigger to hold the peaking amplitude and to open the gate of the 
ADC module, operating the SiPM at low pixel gain and setting the 
discriminator threshold value above the pedestal. The LED intensity
was decreased to have a lower number of pixels firing, similarly
to what is usually done calibrating a SiPM with an LED system. 
The single-pixel spectrum, now without any contribution from the 
thermal noise, is presented in Fig.~\ref{fig:EXT_TRIG_SiPM_Spectra}. 
The peaks appear to be well separated, allowing the calibration 
of the device using an external light source system.  
Potentially, the calibration of the photodetector might be also 
performed running the SPIROC in auto-trigger mode by analysing 
the thermal noise. In this case, during the calibration the SiPM should 
be operated at larger pixel gains, as shown above. 

\section{Towards Real Data Taking Conditions}
%
\label{sec:THRESHOLD_CUTS}
When the measurements presented in this note have been performed, 
the software to operate the chip in auto-trigger mode 
(see Sec.~\ref{sec:TRIGGER_EFF}) was not fully implemented. 
This limitation could be somehow bypassed using the following procedure, 
thus testing the capability and the efficiency of the chip to perform 
\begin{figure}[b!]
 \begin{center}
   \vspace{0.8cm}
   \includegraphics[height=5.5cm,width=8.5cm]{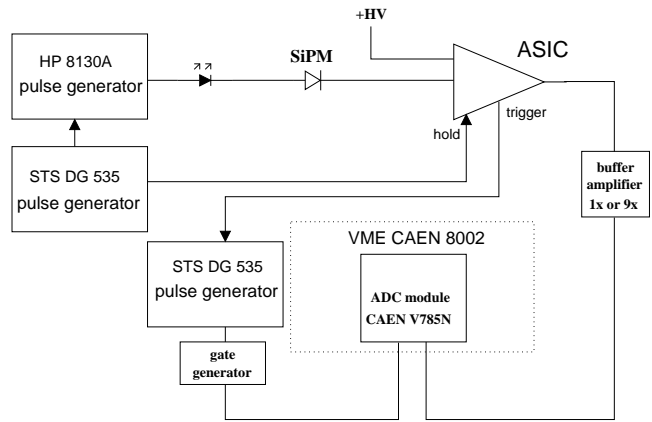}
   \caption{Test bench setup used at DESY for the investigation of the 
            efficiency of the SPIROC 1B chip to perform measurements in 
            physics and auto-trigger mode at different trigger
            threshold values.}
   \label{fig:TestBench2}
 \end{center}
\end{figure}
measurements in both physics and auto-trigger mode at different trigger 
threshold values, Fig.~\ref{fig:TestBench2}.

These measurements were performed using the signal from a SiPM 
(SiPM number 758), thus simulating, as much as possible, real data 
taking conditions. Spectra suitable for fitting the single-pixel 
structure of the signal were obtained operating the high voltage 
supply at $64.6$ V, corresponding to $59.4$ V voltage bias measured 
at the dynodes of SiPM by setting the input HV DAC tuning to zero units
(see Sec.~\ref{sec:InputDAC}).
Note that the nominal operation voltage for this SiPM is $59.3$ V, as 
declared in the SiPM database. 
The main pulse generator triggered the generation of a negative 
voltage pulse (via a Hewlett Packard $8130A$ pulse generator providing width, 
leading and trailing edges tunable down to the nanosecond level) to an LED, 
optically coupled to a SiPM whose signal was then directly input to the
ASIC board, using the same high voltage line, Fig.~\ref{fig:InputHvDAC}. 
The intensity of the LED, and in turns of the SiPM signal, could be then 
regulated by the chained pulser. 

During test-bench measurements the signal to open the ADC module gate 
is typically provided by the pulse generator, for every generated 
voltage pulse. 
Accessing the trigger allows instead to generate the open-gate signal 
only in case a processed signal (either from the thermal noise or from 
the LED) overshoots a chosen threshold value
in the discriminator. 
The generated trigger was therefore used as an external trigger to 
an additional pulse generator, which in turns generated the signal 
to open the ADC module gate at a suitable time.
The hold signal was provided by the main pulser used to switch the 
LED on.

\begin{figure}[t!]
\begin{center}
  \includegraphics[height=5.cm, width=6.0cm]{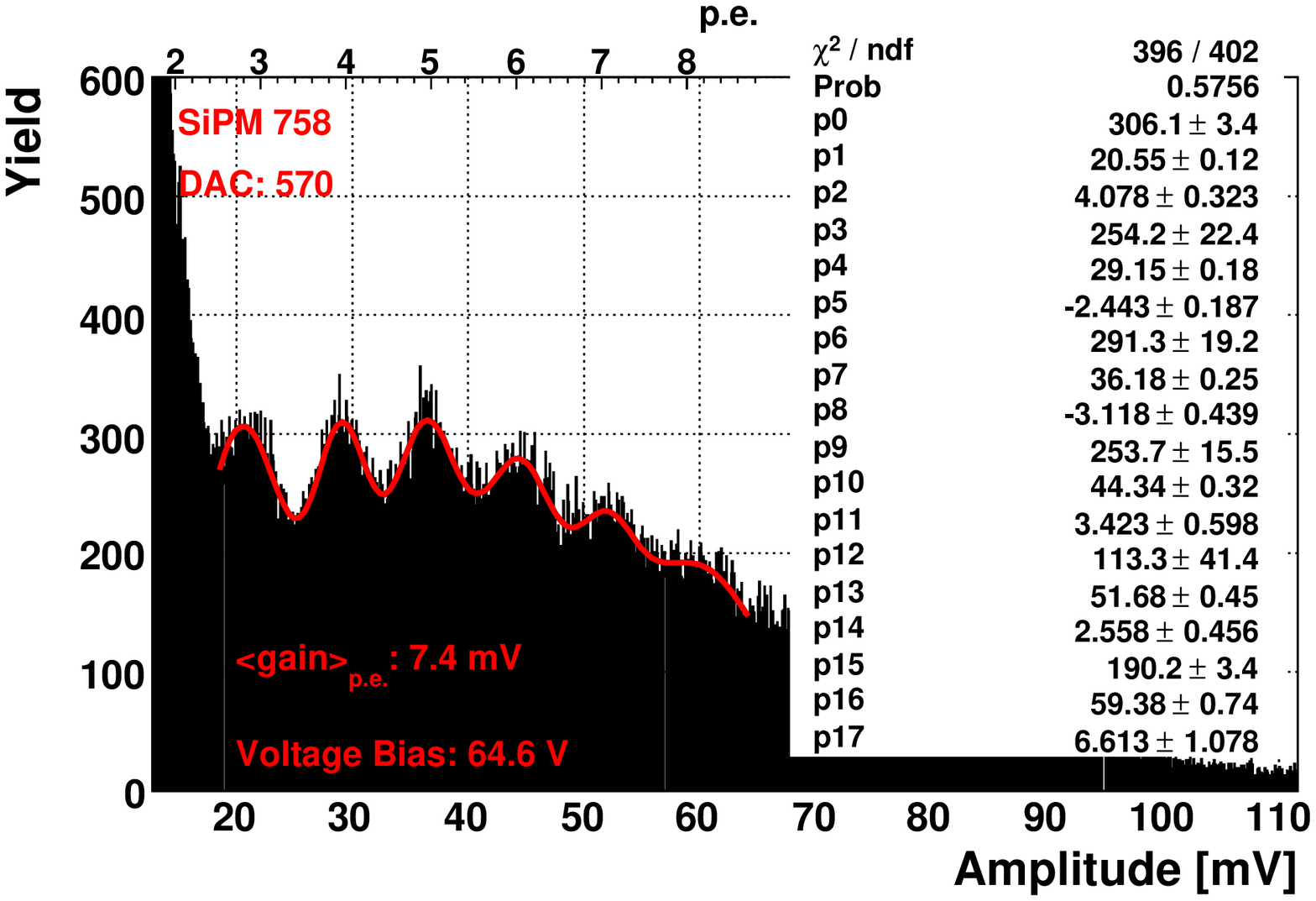} 
  \includegraphics[height=5.cm, width=6.0cm]{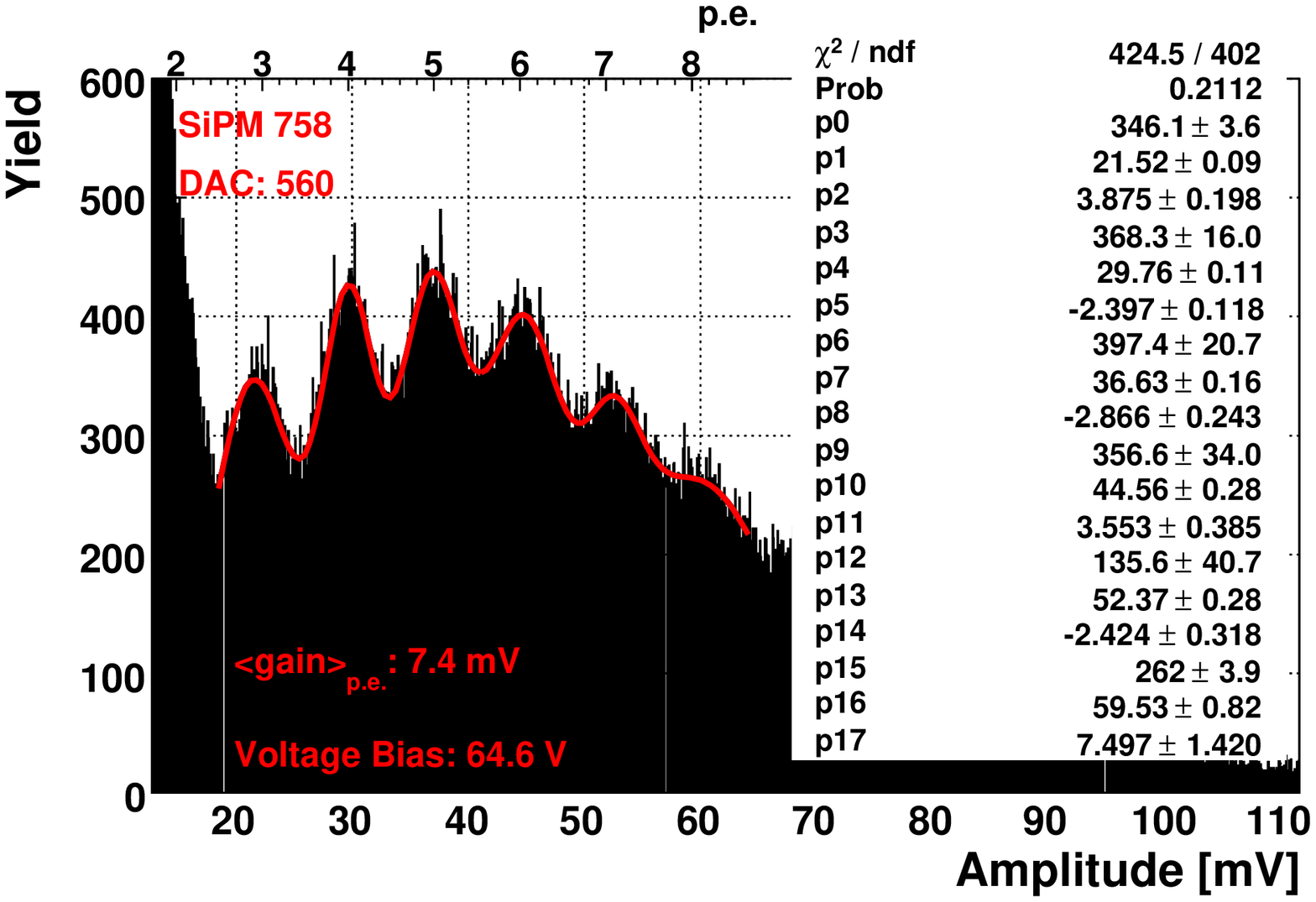} \\
  \includegraphics[height=5.cm, width=6.0cm]{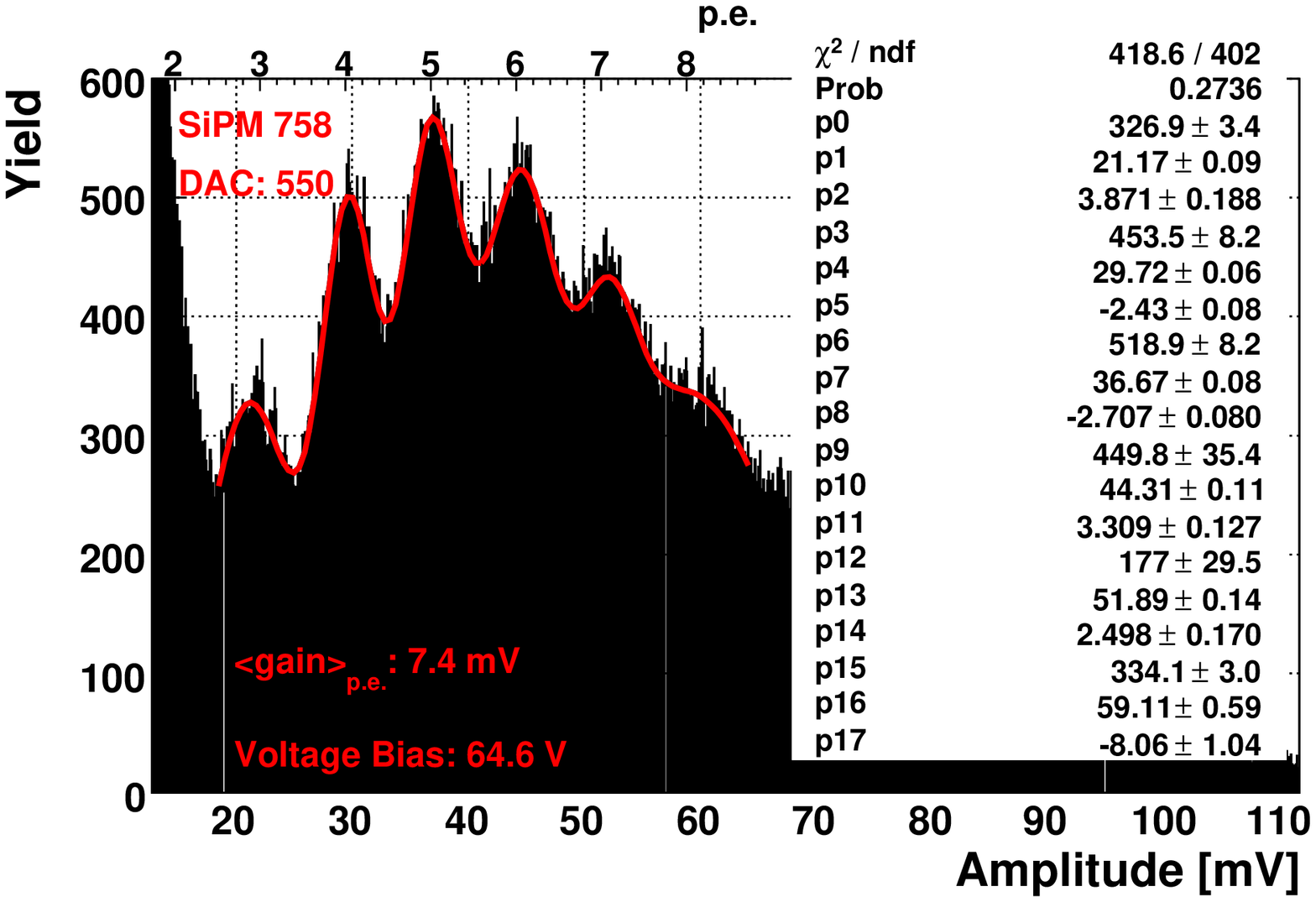} 
  \includegraphics[height=5.cm, width=6.0cm]{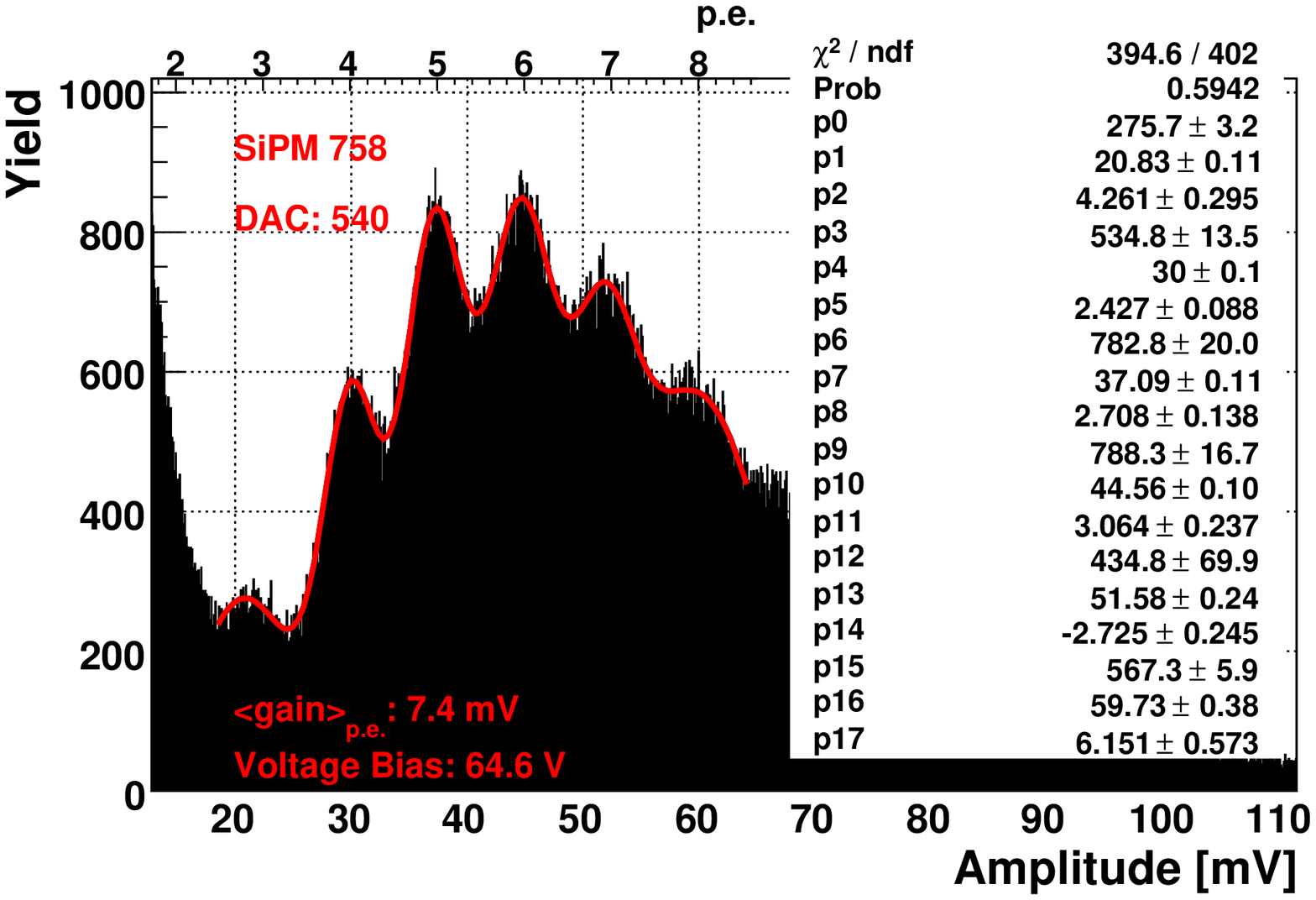} \\
  \includegraphics[height=5.cm, width=6.0cm]{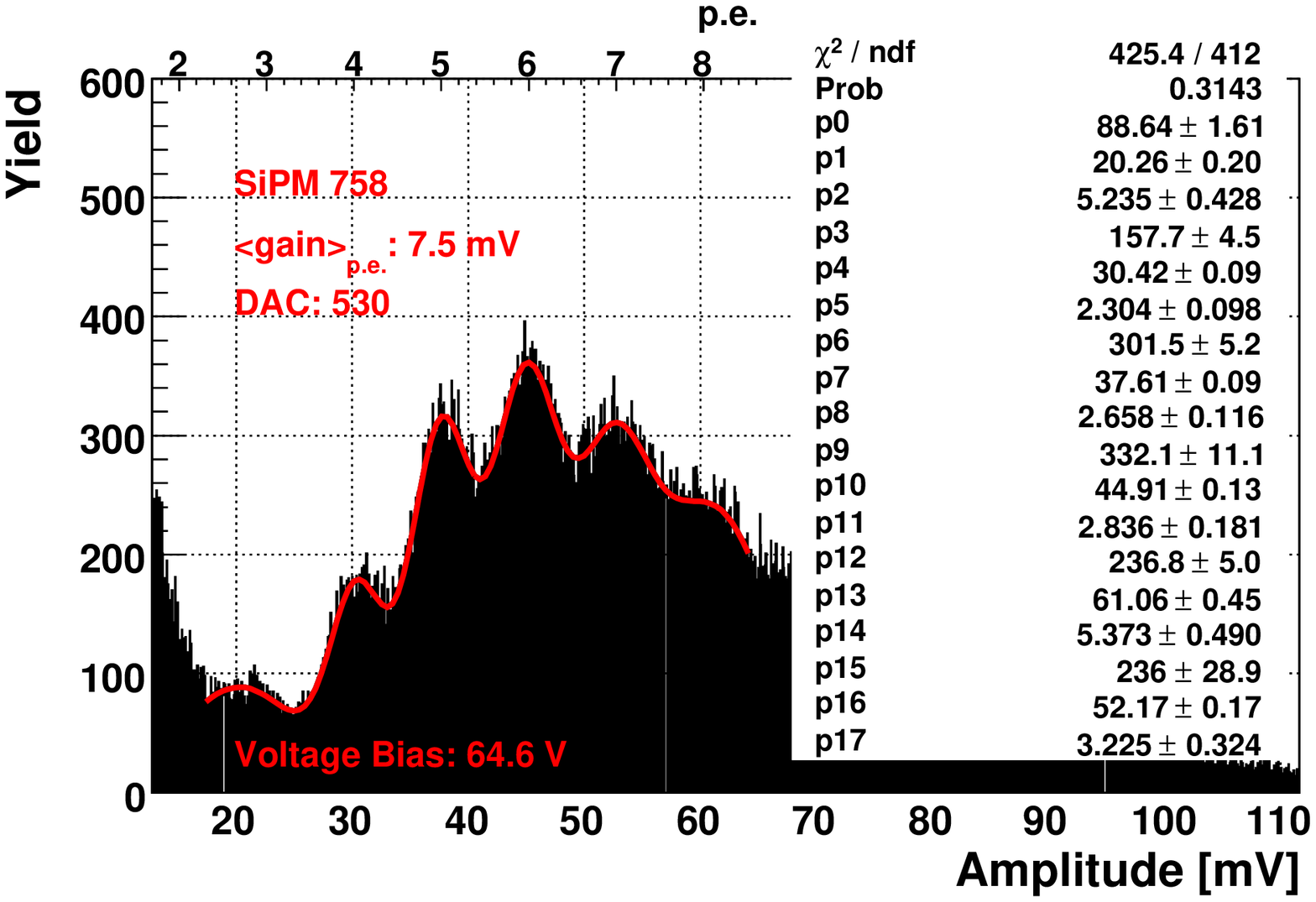} 
  \includegraphics[height=5.cm, width=6.0cm]{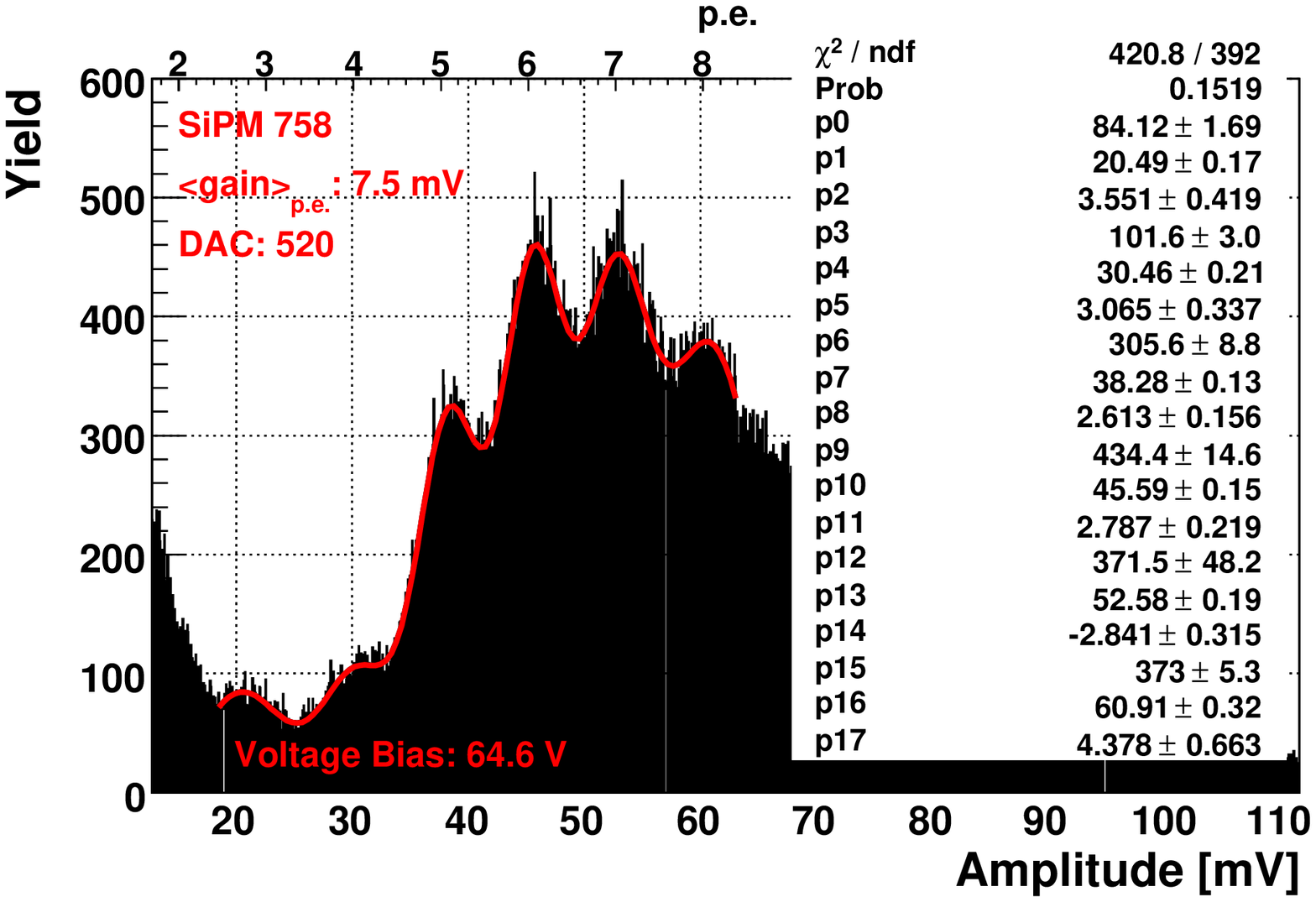} \\
  \vspace{0.25cm}
  \caption{The spectrum from SiPM nr.$758$ is measured in high gain mode
           applying different threshold DAC values at the signal discriminator
           located along the fast shaping line. The chip is operated in
           high gain mode at $25$ ns shaping time and $100$ fF feedback
           capacitance.}
  \label{fig:threshold_scan_HG1}
\end{center}
\end{figure}

\begin{figure*}[t!]
\begin{center}
  \includegraphics[height=5.cm,width=6.0cm]{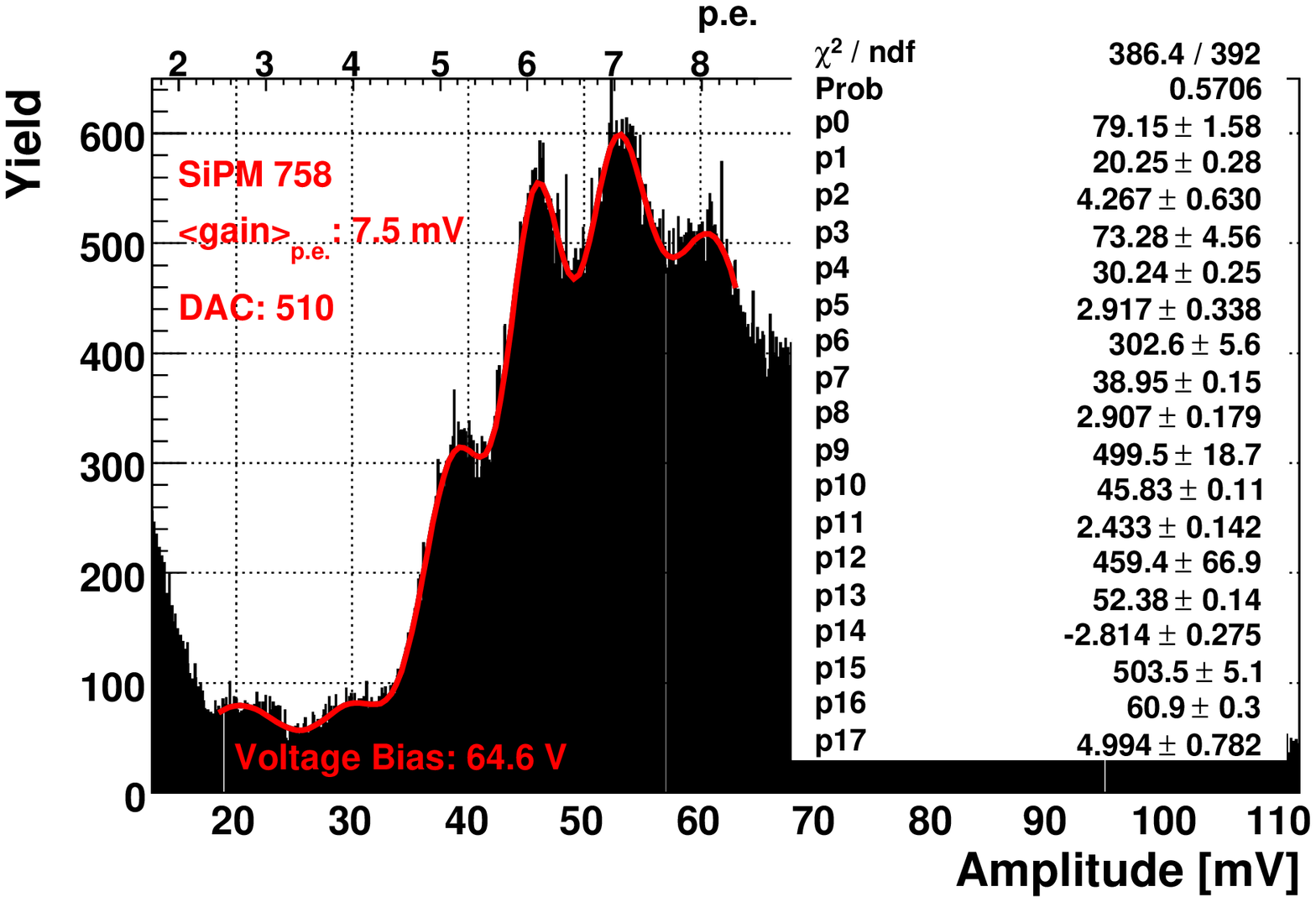} 
  \includegraphics[height=5.cm,width=6.0cm]{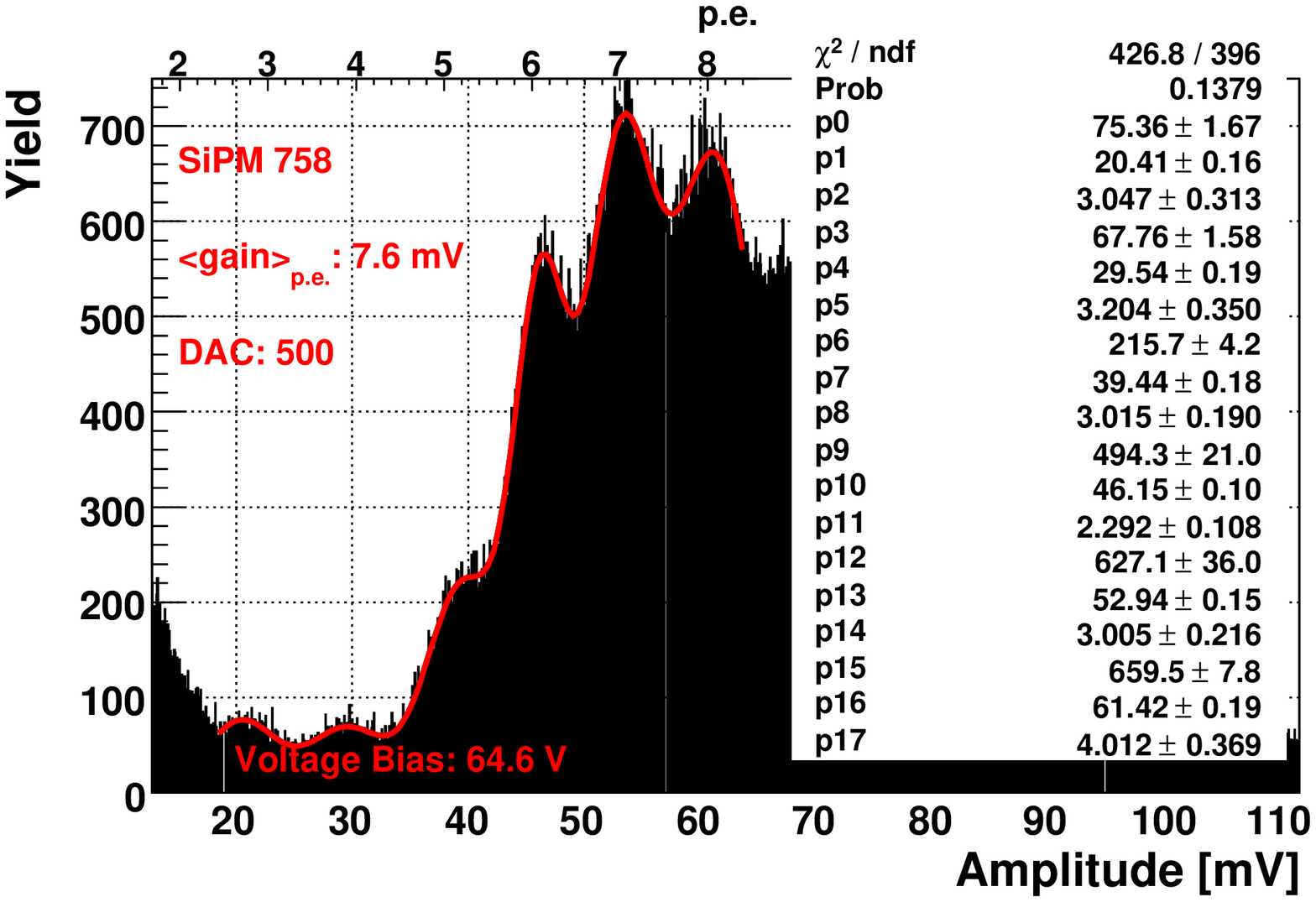} \\
  \includegraphics[height=5.cm,width=6.0cm]{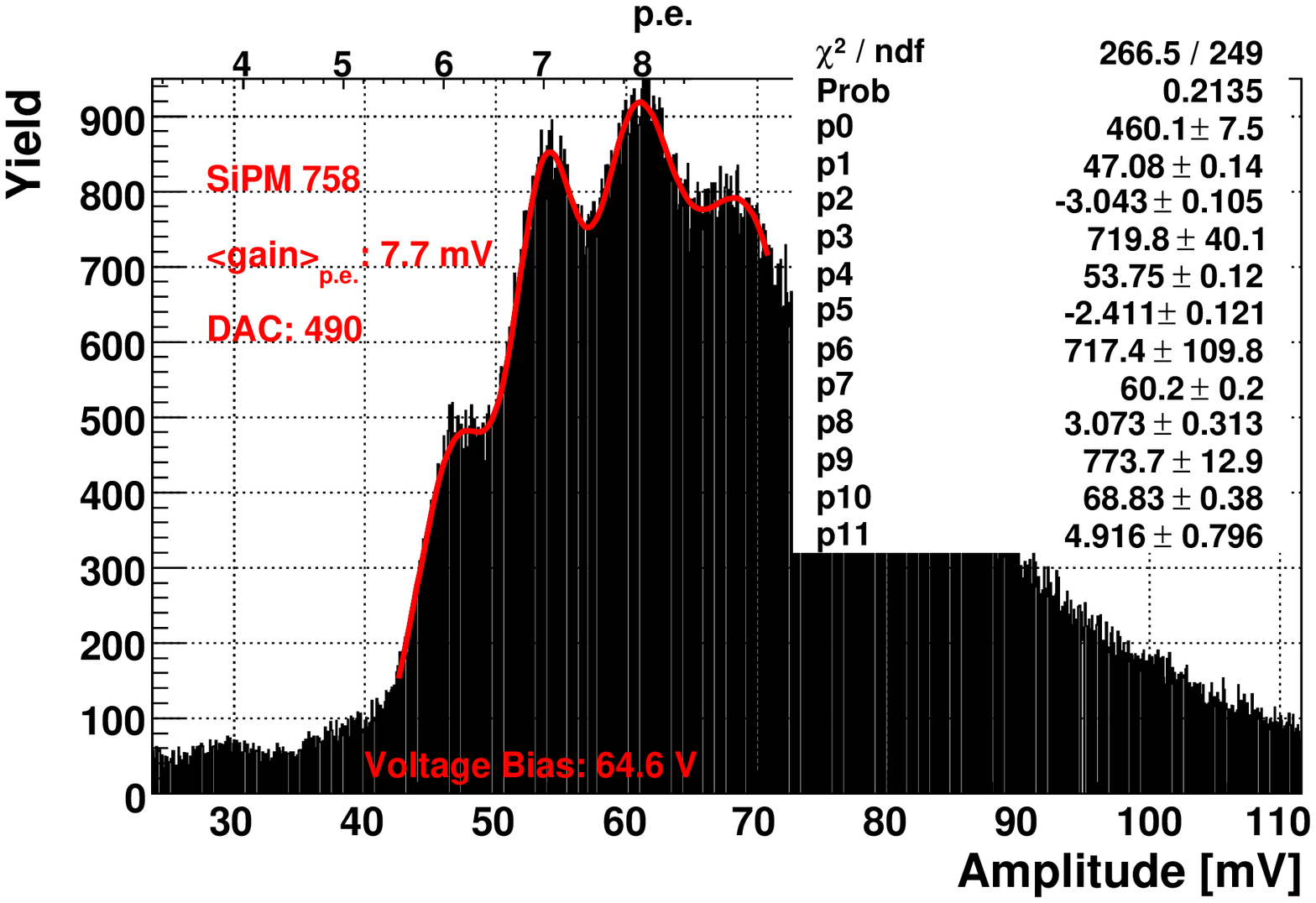} 
  \includegraphics[height=5.cm,width=6.0cm]{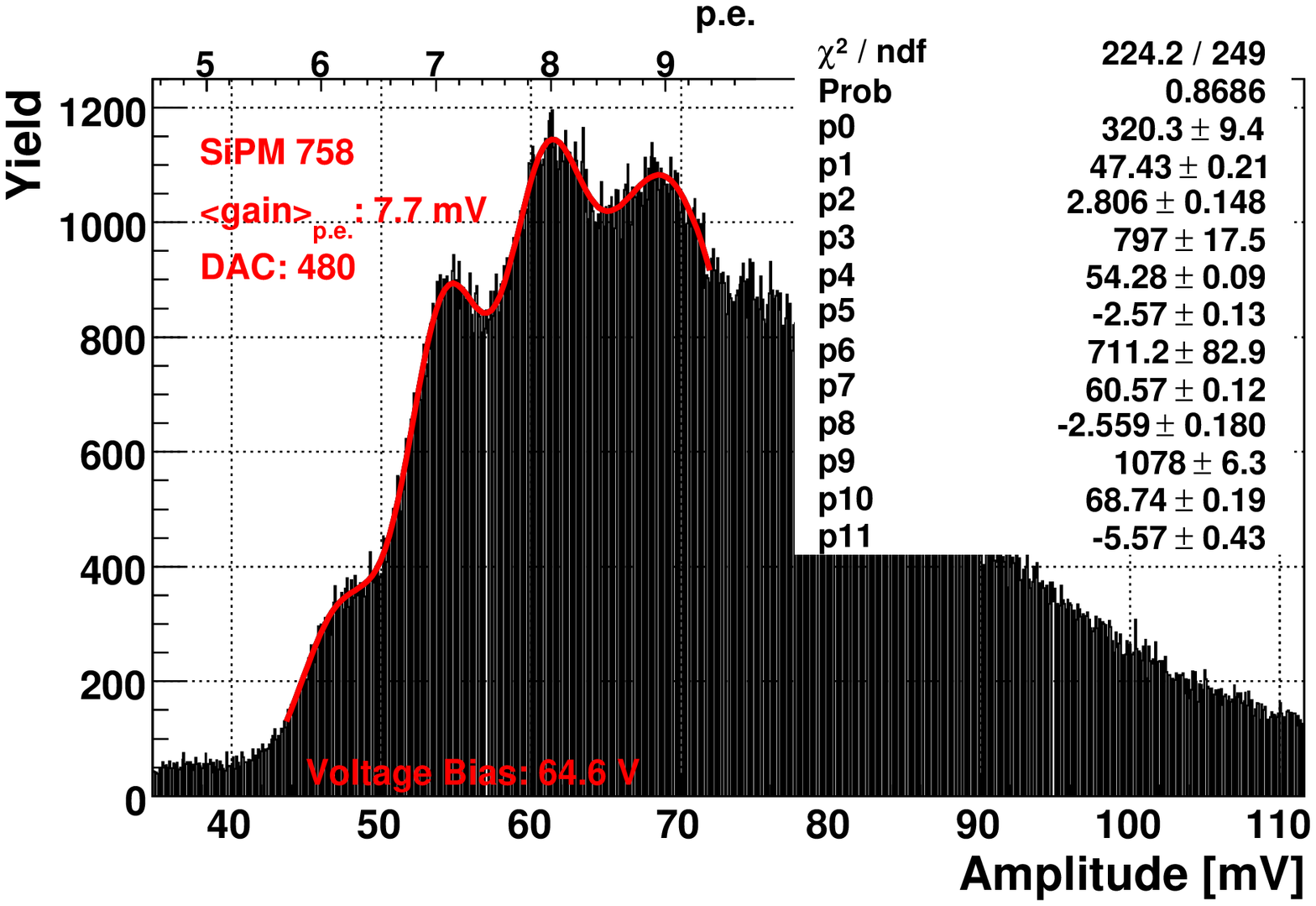} \\
  \includegraphics[height=5.cm,width=6.0cm]{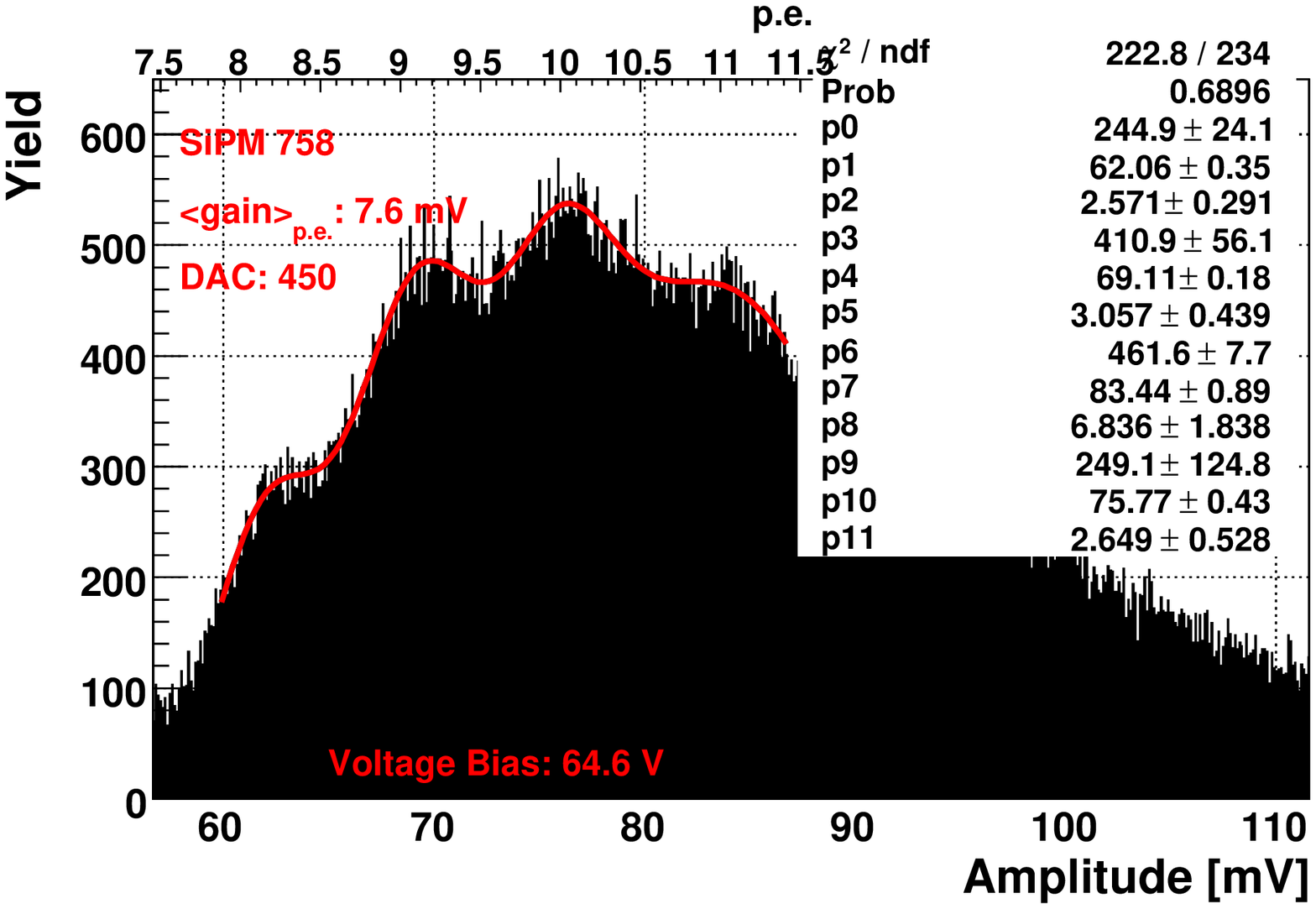} 
  \includegraphics[height=5.cm,width=6.0cm]{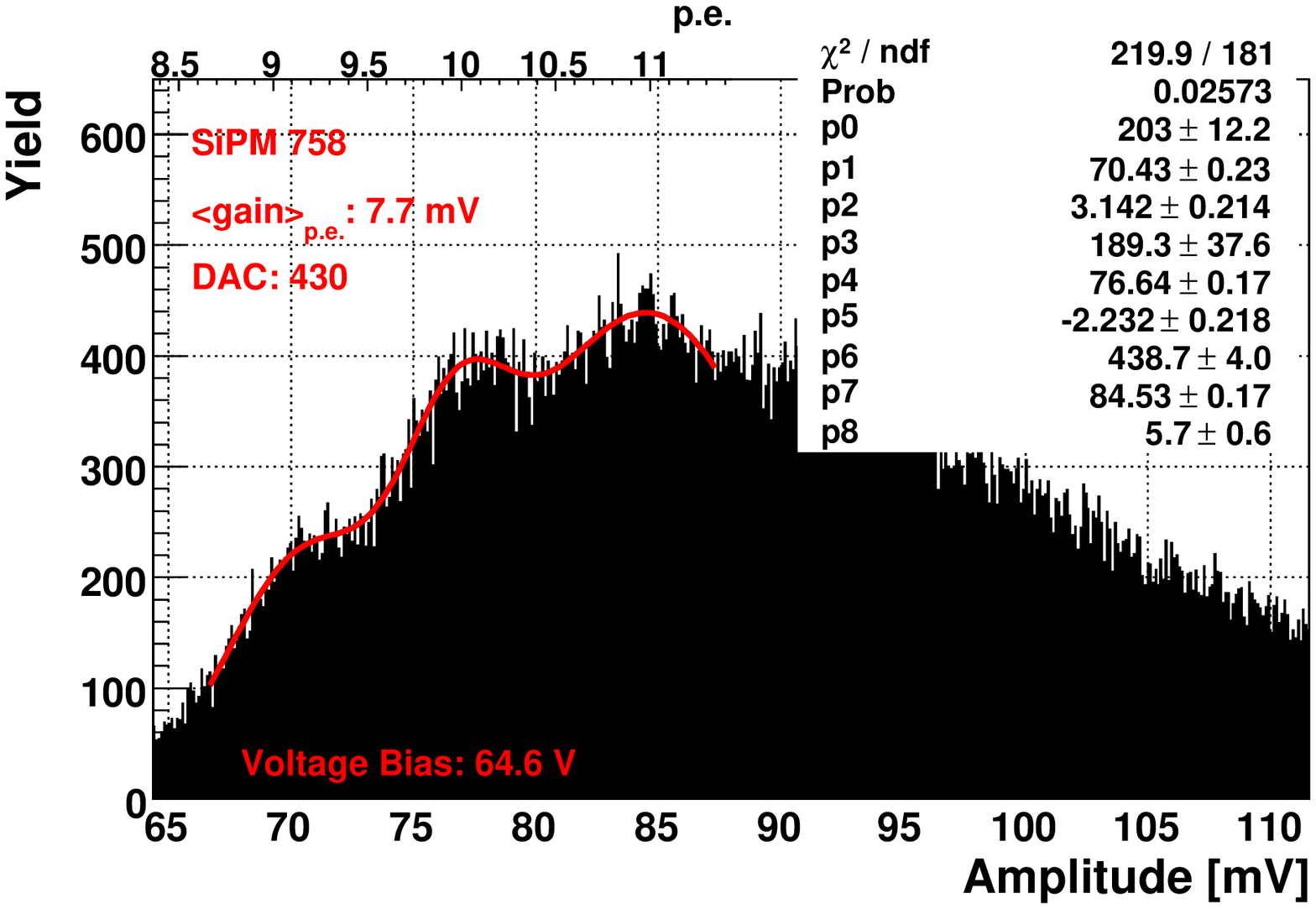} \\
  \vspace{0.25cm}
  \caption{The spectrum from SiPM nr.$758$ is measured in high gain mode
           applying different threshold DAC values at the signal discriminator
           located along the fast shaping line. The chip is operated in 
           high gain mode at $25$ ns shaping time and \mbox{$100$ fF}
           feedback capacitance.}
  \label{fig:threshold_scan_HG2}
\end{center}
\end{figure*}

As a first step, the chip was operated in high gain mode in parallel 
with low intensity LED light, while applying different threshold DAC 
values via the LabView interface (the larger is the DAC value set, 
the smaller is the applied threshold level value).
Using the trigger only to open the ADC gate, but holding the signal 
peaking amplitude via the external primary pulse generator 
allowed to easily fit the single-pixel structure of the spectrum 
corresponding to LED events, for the purpose of this measurement, 
\mbox{Fig.~\ref{fig:threshold_scan_HG1}-~\ref{fig:threshold_scan_HG2}}.
In case a signal induced by thermal noise overshoots the discriminator 
threshold level then the ADC gate is still opened by the generated trigger 
but the processed signal is not any more synchronised with the hold 
signal generated by the main pulser. This results in deteriorating the 
single-pixel structure of the thermal noise contribution, and 
populating the spectrum towards the pedestal region (the recorded signal
amplitude is lower than the real peaking amplitude).
This contribution from thermal noise to the spectrum is increasingly 
suppressed with increasing the threshold value, as clearly visible 
in Fig.~\ref{fig:threshold_scan_HG1}.
Although the number of suppressed peaks in the spectrum increases by 
increasing the threshold level, the measurements show that a remaining 
small contribution from the peaks expected to be suppressed is still 
present. This feature, possibly due to the experimental setup, has not
been furtherly investigated at this stage of the analysis, being not 
relevant for the measurement presented here.
 
From the above measurements the threshold values to cut specific 
spectrum peaks can be determined 
resulting in approximative $20$ DAC units per peak, 
Tab.~\ref{tab:threshold_values}. This corresponds to a step of 
$\approx 37$ mV in the fast shaping line (see Fig.~\ref{fig:DAC_calib}).
\begin{table}[t!]
  \begin{minipage}{6.5cm}
   \begin{tabular}{|c|c|c|}
      \hline
       DAC value & Suppressed peak number\\
      \hline
       & \\
       $530$ & $3$ \\
       $510$ & $4$ \\
       $490$ & $5$ \\
       $470$ & $6$ \\
       $450$ & $7$ \\
       $430$ & $8$ \\
       $410$ & $9$ \\
       $390$ & $10$ \\
      \hline
    \end{tabular}
  \end{minipage}
  \begin{minipage}{5.5cm}
    \caption{DAC values applied at the signal discriminator to suppress 
             specific peaks in the SiPM spectrum. 
             The obtained DAC values refer to the chip operated in high 
             gain mode at $25$ ns shaping time and $100$ fF feedback
             capacitance. One DAC units corresponds to $\approx 1.9$ mV
             in the fast shaping line.}
  \label{tab:threshold_values}
  \end{minipage}
\end{table}

\begin{figure*}[t!]
  \begin{minipage}{6.5cm}
       \includegraphics[height=6.5cm,width=6.5cm]{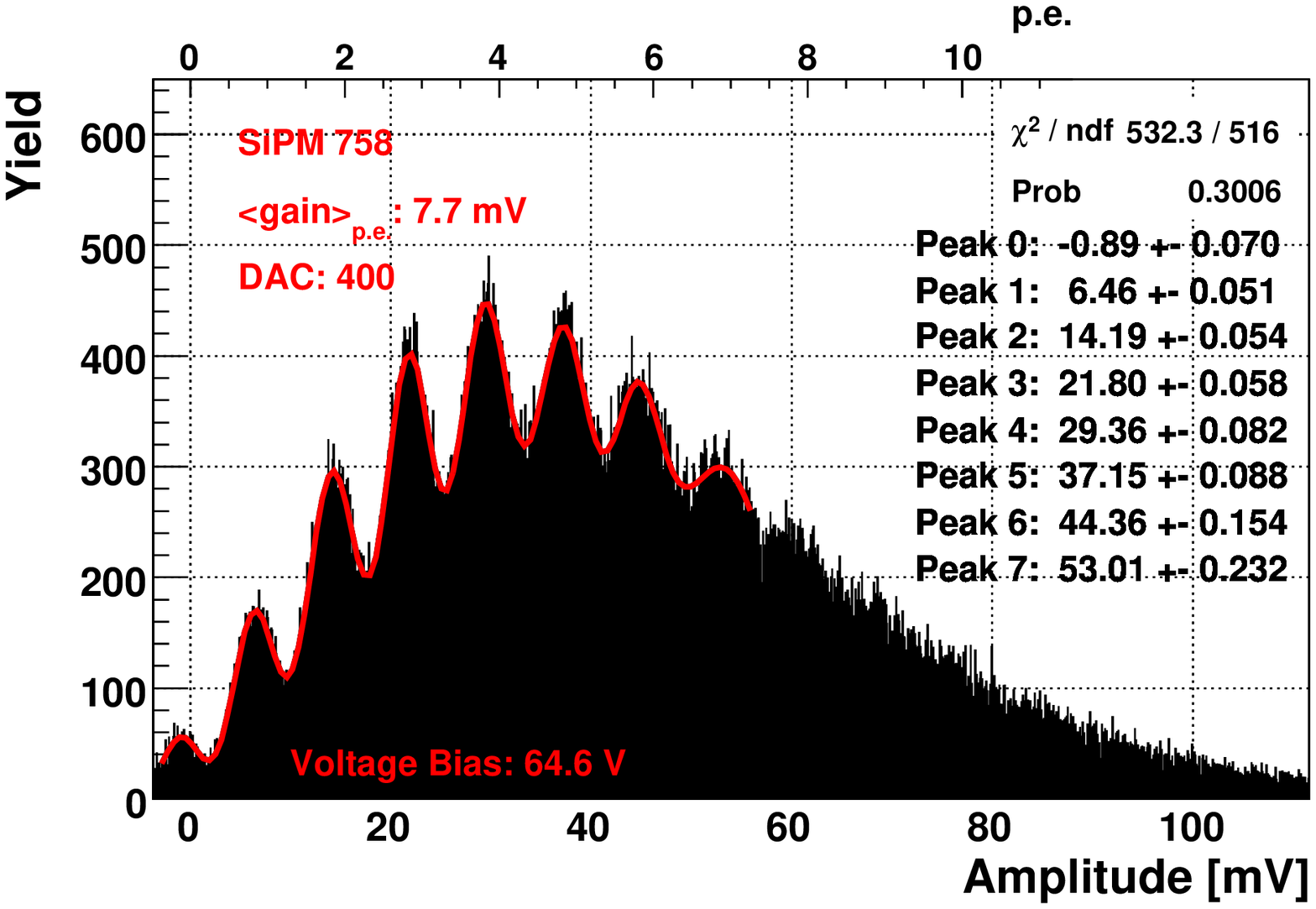}
  \end{minipage}
  \begin{minipage}{5.5cm}
  \caption{The SiPM single pixel spectrum obtained operating the chip in high 
           gain mode at $25$ ns shaping time and $100$ fF feedback 
           capacitance, and using the main external pulse generator to 
           open the ADC gate.}
  \label{fig:SiPM_SinglePixel_ExtTrigger}
  \end{minipage}
\end{figure*}

\begin{figure*}[t!]
  \begin{minipage}{6.5cm}
       \includegraphics[height=6.5cm,width=6.5cm]{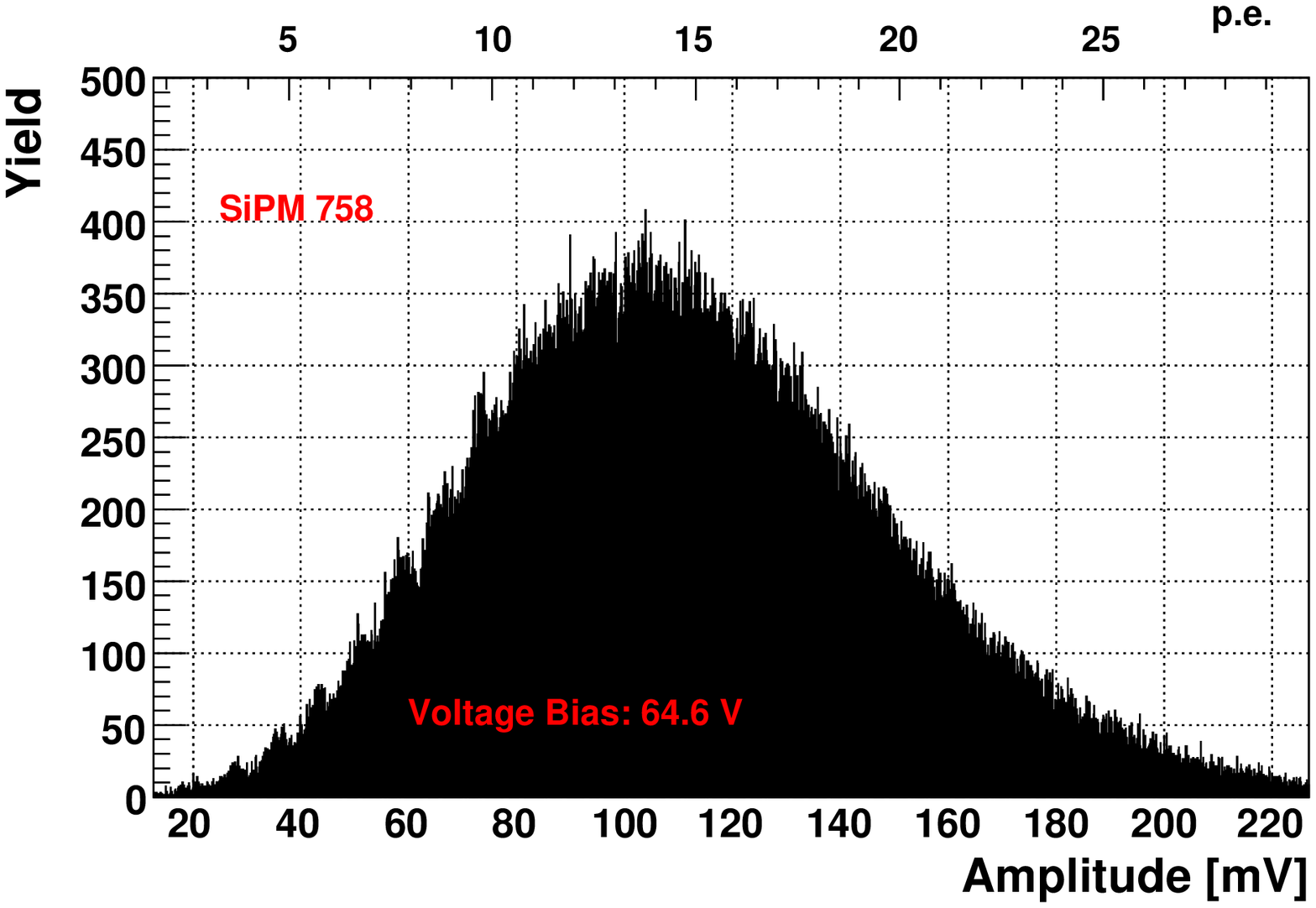}
  \end{minipage}
  \begin{minipage}{5.5cm}
  \caption{A mip-like signal measured operating the chip in high gain mode 
           at $25$ ns shaping time and \mbox{$100$ fF} feedback capacitance.}
  \label{fig:mip_HG}
  \end{minipage}
\end{figure*}

\begin{figure*}[t!]
   \includegraphics[height=6.cm,width=12.0cm]{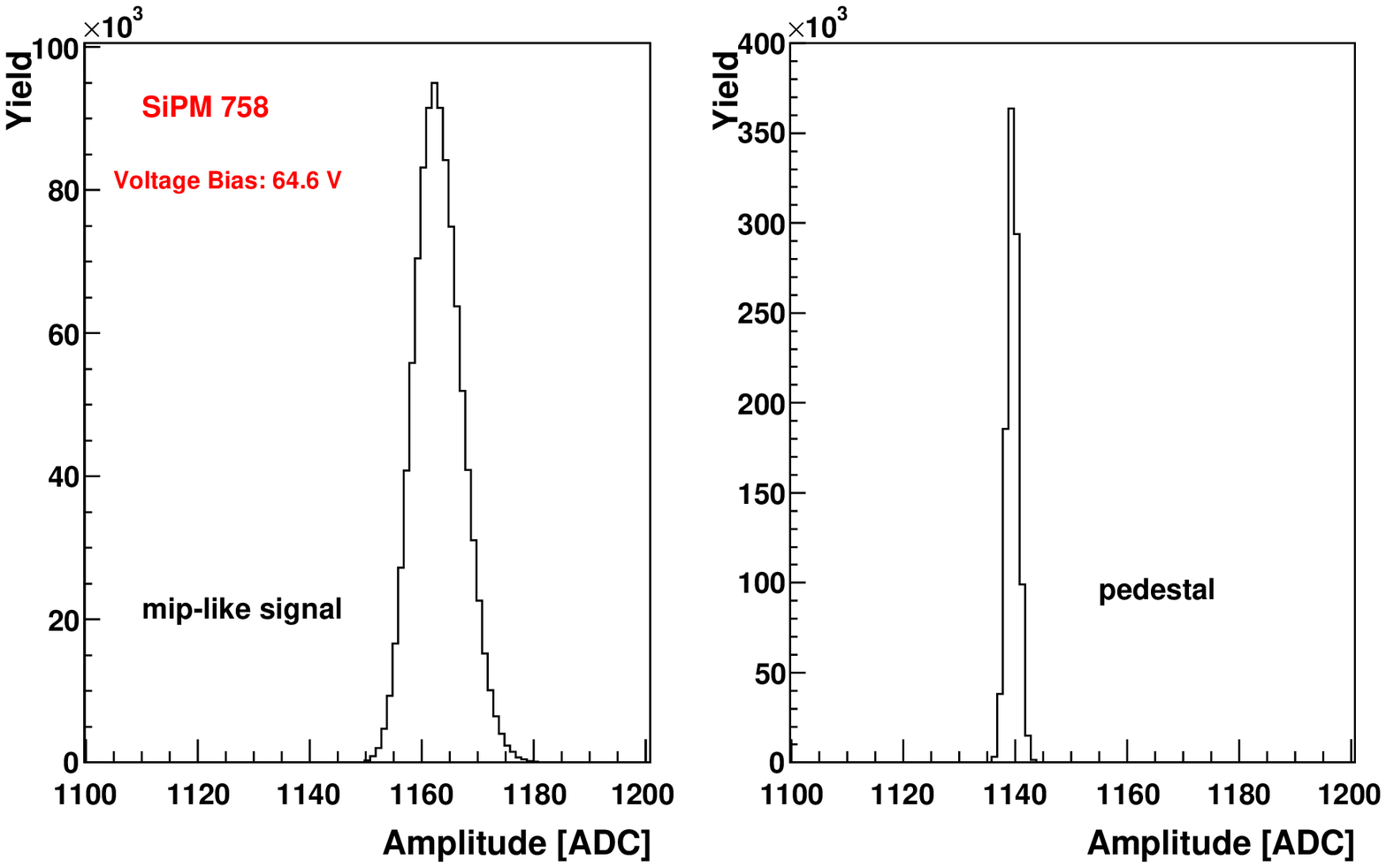}
  \caption{A mip-like signal (left panel) and the pedestal (right panel)
           are measured operating the chip in low gain mode at $50$ ns 
           shaping time and \mbox{$200$ fF} feedback capacitance.}
  \label{fig:mip_LG}
\end{figure*}

As a cross check, a similar spectrum was taken using the pulse generator to
open the ADC gate, obtaining the same single-pixel peak locations, 
Fig.~\ref{fig:SiPM_SinglePixel_ExtTrigger}.
With the same setup, a mip signal was tentatively obtained tuning the 
LED amplitude such that the SiPM spectrum shows a maximum around the 
$15$ pixel peak amplitude, Fig.~\ref{fig:mip_HG}. 
Keeping the same LED amplitude, the mip-like signal was then measured 
using the low gain mode of the chip, Fig.~\ref{fig:mip_LG}, for a
shaping time of $50$ ns.

\begin{figure*}[t!]
   \includegraphics[height=6.cm,width=6.0cm]{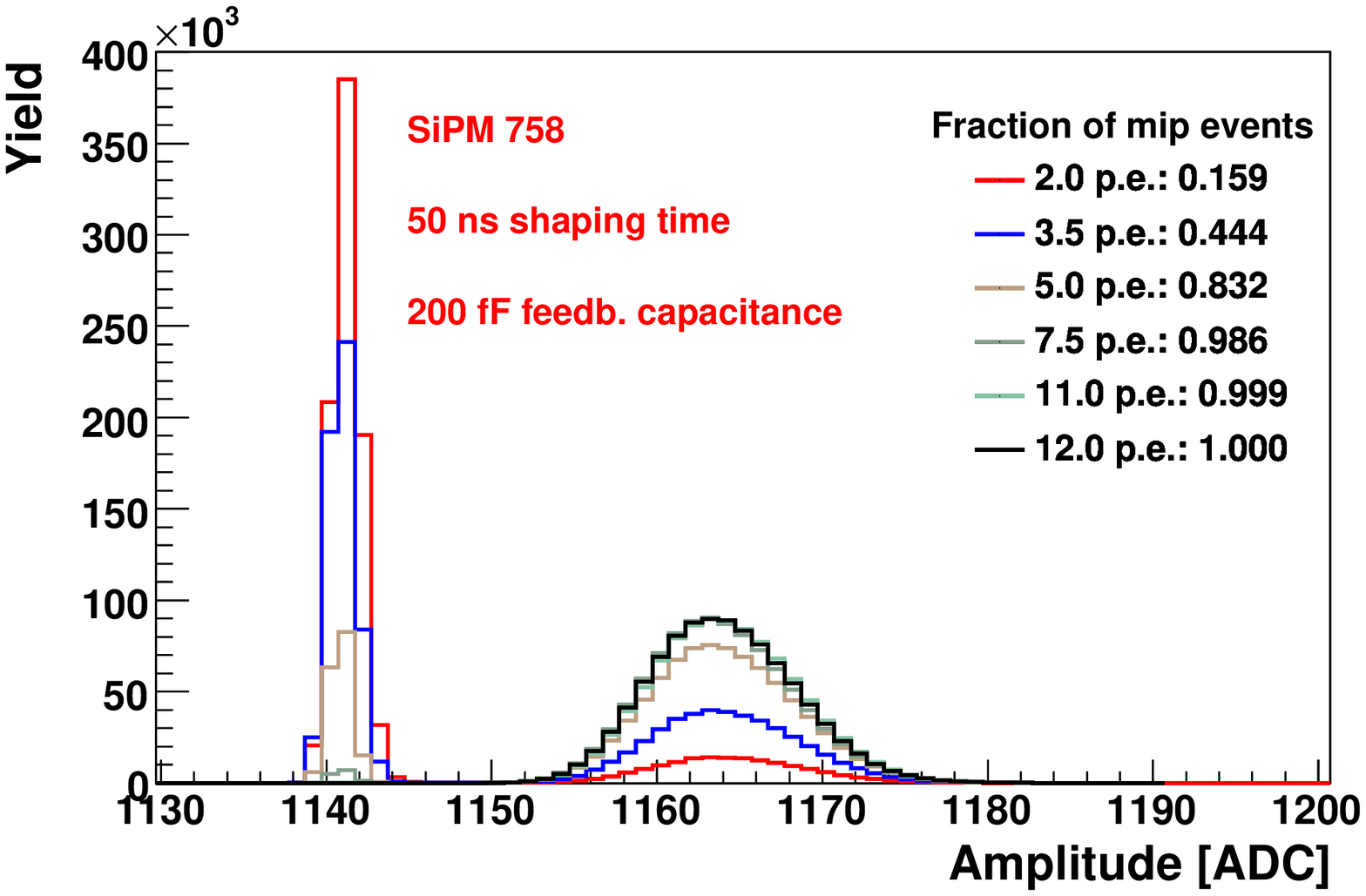}
   \includegraphics[height=6.cm,width=6.0cm]{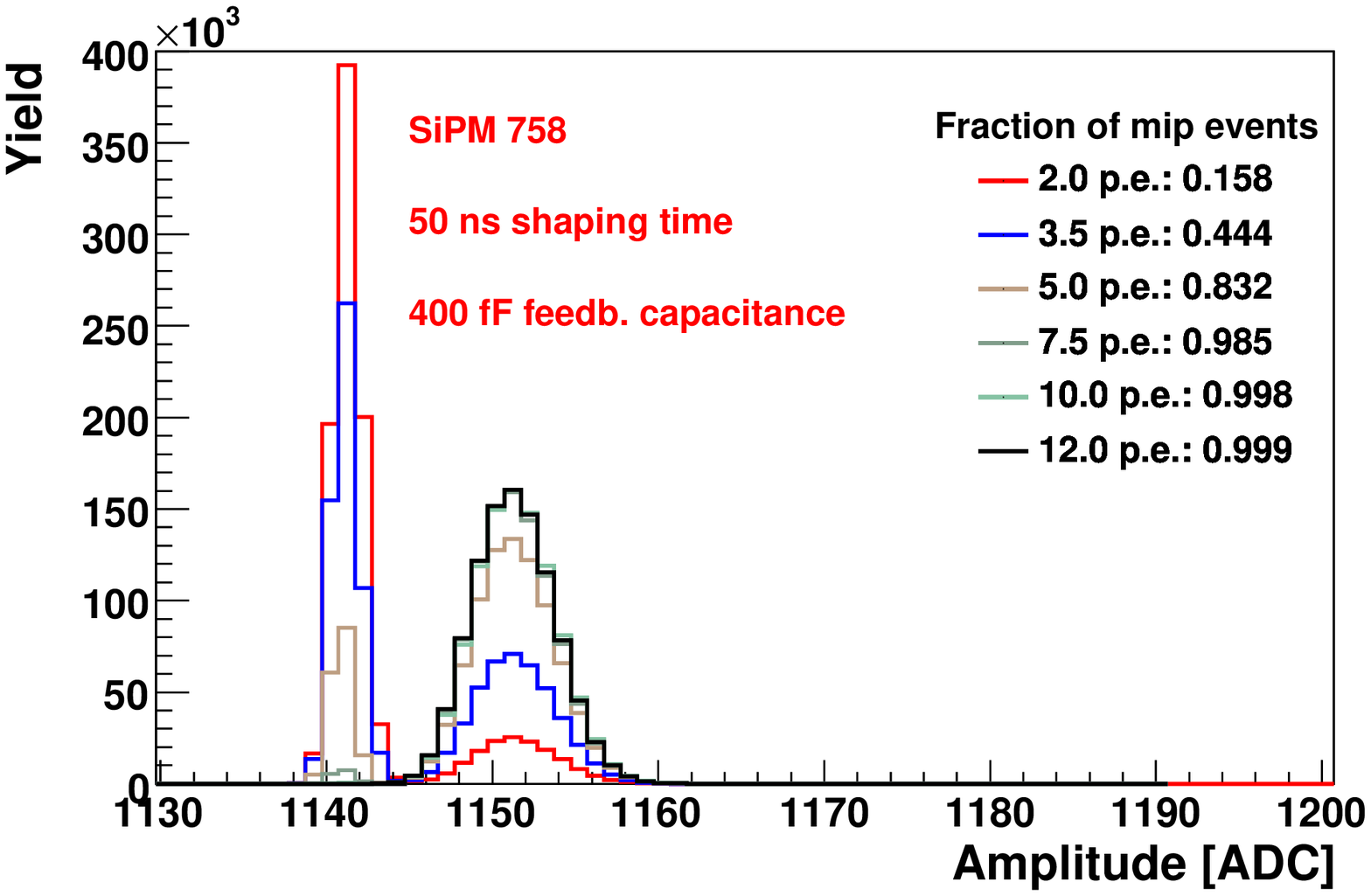}
  \caption{The efficiency in detecting a mip-like signal is measured
           at different threshold values (in SiPM photo-electron units)
           operating the chip in low gain mode at $50$ ns 
           shaping time, \mbox{$200$ fF} and \mbox{$400$ fF} feedback 
           capacitance, left and right panel, respectively.}
  \label{fig:mip_efficiency}
\end{figure*}

Using again the generated trigger to open the ADC gate,
and applying different threshold DAC values, the trigger efficiency 
with respect to the signal discrimination level
was investigated for two values of the feedback capacitance, 
Fig.~\ref{fig:mip_efficiency}.
In the plots, the threshold DAC values are presented in terms of 
number of photo-electron peaks which are cut by the selected threshold, 
according to Tab.~\ref{tab:threshold_values}. By increasing the DAC value 
(decreasing the trigger threshold) the number of triggers generated by 
the noise increases, resulting in a larger contribution of pedestal 
events to the overall accumulated events in the spectra.
Already at half a mip threshold the fractional contribution of SiPM 
signal events is almost $100\%$, for both the investigated ASIC 
configurations. 

The preliminary results presented here verify the possibility to operate 
the ASIC 
in auto-trigger mode (after the mentioned improvement of the track and hold
switch) processing the physics events at half a mip cut on the 
trigger threshold, as forseen in the ILC applications.

%
\section{Conclusions}
During 2008 and 2009 systematic studies have been performed 
to characterise the analogue component of the SPIROC ASIC, 
designed for the readout of the analogue hadronic calorimeter 
for the international linear collider project.
The more significant results are presented and described in 
this report showing the fulfilled requirements and the 
features which should be fixed in the next version of the
chip. These measurements have been done on the SPIROC version
SPIROC 1B, which has the analogue processing of signals almost 
fully implemented. 

The ASIC was shown to handle the dynamic range covered by the SiPMs which 
are expected to be readout in both calibration mode for the maximum 
amplification at $100$ fF feedback capacitance, and physics mode for 
values of feedback capacitance above $400$ fF, resulting non-saturated 
up to \mbox{$\approx 80$} and \mbox{$\approx 8$} mips, respectively.
The values reported here refer to a SiPM with pixel gain of order 
$0.5 \cdot 10^6$. 
In case of SiPMs with highly varying pixel gain (as for the SiPMs 
used in the past AHCAL test beam operations), the dynamic range 
covered by the photodetector changes, and the operation mode of the 
chip should be changed accordingly, resulting in a change of the
signal over noise ratio.  

The data acquisition is forseen to run in auto-trigger mode, 
being activated by the SiPM signals, and the chip appears to 
fulfil this requirement in the physics mode, reaching $100\%$ 
trigger efficiency for trigger threshold values above half a mip. 
For these values the trigger time walk and jitter were found to 
be less than $3$ and $1$ ns, respectively. Quite remarkable, 
is the observed possibility to fit the thermal noise, opening 
the potentiality of self-calibrating the device. 

The electronic noise was investigated and found to be 
typically within one millivolt, depending 
on the specific shaping time and amplification setting. 
The most sizable additional contribution to the 
noise is induced by switching on the track and hold component, 
approximately doubling the noise. When coupled to an external
signal source the noise dependence on varying the coupling capacitance 
(as for real SiPMs) was found to be within $20\%$ in the range 
\mbox{$10$-$100$ pF}.

Comparing the $36$ input channels, the pedestals showed
a spread of $2$ mV, and their deviation due to cross-talk 
was measured to be within $\pm 0.5$ mV in the investigated 
range from one to $33$ mips ($40$ pC).

Each input SiPM high voltage can be individually tuned via 
a DAC HV adjustment dedicated to each detector line. 
The DAC to volt calibration was performed for all channels 
showing a channel to channel average variation of \mbox{$100$ mV}, 
and an average maximum deviation to linearity of $80$ mV 
(for some channels up to $200$ mV).
The observed deviation from linearity appears to be systematically 
reproducible, and could be corrected for via a proper calibration. 

The high-low gain path coupling was studied and a sizable 
dependence (up to $10\%$) on the high gain amplification 
setting was measured for the physics mode readout.

Probably the most surprising feature of the chip was the observed 
increase of the signal peaking time with increasing injected charge, 
related to properties of the used track and hold switch. 
Investigation with simulations by the chip designers in Orsay is 
on-going, and a possible cure of the shown feature should be 
available in the next generations of the ASIC.

The investigation of the digital component of the ASIC, 
implemented in the version SPIROC 2, is on going and will 
be presented in a separate future note.

%
\begin{quotation}
      \vspace{0.5cm}
  \begin{center}
      {\bf Acknowledgments}
      \vspace{0.5cm}
  \end{center}
  We are deeply grateful to S.\ Callier, C.\ De La Taille, and 
  \mbox{L.\ Raux} of the OMEGA group at Orsay, and to F.\ Sefkow 
    and E.\ Garutti at DESY for useful discussions and suggestions.
\end{quotation}


\end{document}